\documentclass[preprint,10pt]{aastex}
\usepackage[]{epsfig,rotating}
\newcommand       \beq          {\begin{equation}}
\newcommand       \eeq          {\end{equation}}
\newcommand       \Angstrom     {\,{\rm \AA}}          

\newcommand       \cm           {\,{\rm cm}}
\newcommand       \erg          {\,{\rm erg}}
\newcommand       \eV           {\,{\rm eV}\,}
\newcommand       \g            {\,{\rm g}}

\newcommand       \K            {\,{\rm K}}

\newcommand	  \MJy		{\,{\rm MJy}}
\newcommand	  \mm		{\,{\rm mm}}
\newcommand	  \NC           {N_{\rm C}}
\newcommand	  \NH		{N_{\rm H}}
\newcommand       \nH           {n_{\rm H}}

\newcommand       \s            {\,{\rm s}}
\newcommand	  \sr		{\,{\rm sr}}

\newcommand       \simlt        {\lesssim}
\newcommand       \simgt        {\gtrsim}
\newcommand       \gtsim        {\gtrsim}
\newcommand       \ltsim        {\lesssim}

\newcommand	  \chiMMP	{\chi_{\rm MMP}}
\newcommand	  \fion		{\phi_{\rm ion}}	
\newcommand	  \qcont        {{q_{\rm gra}}}
\newcommand	  \bc		{{b_{\rm C}}}
\newcommand	  \bci		{{b_{{\rm C},i}}}
\newcommand	  \bcone	{{b_{{\rm C},1}}}
\newcommand	  \bctwo	{{b_{{\rm C},2}}}
\newcommand       \numatom	{N_{\rm atom}}
\newcommand	  \numC		{N_{\rm C}}	

\newcommand	  \emiss	{E_\lambda}
\newcommand	  \albedo	{{\rm albedo}}

\newcommand{\btdnote}[1]{}

\newcommand{\figwidth}{16.5cm}			
\newcommand{\figwidthland}{12.5cm}		

\countdef\decade=200
\advance\decade by \year
\countdef\hours=201
\advance\hours by \time
\divide\hours by 60
\countdef\mins=202
\advance\mins by \hours
\multiply\mins by 60
\multiply\hours by 100
\countdef\miltime=203
\advance\miltime by \hours
\advance\miltime by \time
\advance\miltime by -\mins

\shorttitle{Infrared Emission from Dust in the Diffuse ISM}

\begin{document}

\title{
	\vspace*{-3.0em}
	{\normalsize\rm {\it The Astrophysical Journal}, {\bf554}, 778-802\\
	with minor errors corrected in Table 6 and Fig. 16}\\
	\vspace*{1.0em}
	Infrared Emission from Interstellar Dust\\
	II. The Diffuse Interstellar Medium\\
	}

\author{Aigen Li and B.T. Draine}
\affil{Princeton University Observatory, Peyton Hall,
        Princeton, NJ 08544, USA;\\
       {\sf agli@astro.princeton.edu, 
	draine@astro.princeton.edu}}

\begin{abstract}
We present a quantitative model for the infrared emission from 
dust in the diffuse interstellar medium. The model consists of 
a mixture of amorphous silicate grains and carbonaceous grains,
each with a wide size distribution ranging 
from molecules containing tens of atoms
to large grains $\gtsim 1$ micron in diameter.
We assume that the carbonaceous grains
have polycyclic aromatic hydrocarbon (PAH)-like
properties at very small sizes, and graphitic properties 
for radii $a \gtrsim 50$\AA.
On the basis of recent laboratory studies and 
guided by astronomical observations, we propose
``astronomical'' absorption cross sections for use in modeling 
neutral and ionized PAHs from the far ultraviolet to the far infrared.
We also propose modifications to the far-infrared emissivity of
``astronomical silicate''.

We calculate energy
distribution functions for small grains undergoing ``temperature
spikes'' caused by stochastic absorption of starlight photons,
using realistic heat capacities and optical properties.
Using a grain size distribution consistent with the observed
interstellar extinction,
we are able to reproduce the near-IR to submillimeter 
emission spectrum of the diffuse interstellar medium, 
including the PAH emission features at 3.3, 6.2, 7.7, 8.6, and 11.3$\micron$.
The model is compared with the observed emission at
high Galactic latitudes as well as in the Galactic plane, 
as measured by the COBE/DIRBE, COBE/FIRAS, IRTS/MIRS, and 
IRTS/NIRS instruments. 
The model has $60\times 10^{-6}$ of C 
(relative to H) locked up in PAHs, with $45\times 10^{-6}$ of C 
in a component peaking at $\sim 6$\AA\ ($N_{\rm C}\approx 100$ carbon 
atoms) to account for the PAH emission features,
and with $15\times 10^{-6}$ of C in a component
peaking at $\sim 50$\AA\ to account for the 60$\mu$m flux. 
The total infrared emission is in excellent agreement with COBE/DIRBE
observations at high galactic latitudes, just as the albedo for our
grain model is in accord with observations of the diffuse galactic light.
The aromatic absorption features 
at 3.3$\mu$m and 6.2$\mu$m predicted by our dust model are consistent 
with observations.

We calculate infrared emission spectra for our dust model heated by
a range of starlight intensities, from 0.3 to 10$^4$ times the
local interstellar radiation field, and we tabulate the intensities
integrated over the SIRTF/IRAC and MIPS bands.  
We also provide dust opacities
tabulated from the extreme ultraviolet to submillimeter wavelengths.
\end{abstract}


\section{Introduction
	\label{sec:introduction}}

It has been 70 years since Trumpler's discovery of color excesses 
provided the first definitive proof of the existence of 
interstellar dust (Trumpler 1930); however, many aspects of the 
nature of interstellar dust still remain unclear. Our knowledge 
of the composition of interstellar grains is based mainly on
their spectroscopic absorption/emission features and observed 
elemental depletions. The most generally accepted view is
that interstellar grains consist of amorphous silicates and some
form of carbonaceous materials; the former inferred from the 
9.7$\mu$m Si-O stretching mode and 
18$\mu$m O-Si-O bending mode absorption features in interstellar
regions as well as the fact that the cosmically abundant heavy 
elements such as Si, Fe, Mg are highly depleted; the latter 
mainly inferred from the strongest interstellar absorption feature
-- the 2200\AA\ hump -- and the fact that silicates alone are not 
able to provide enough extinction. 

The dust sizes, inferred from the wavelength-dependent interstellar 
extinction and polarization curves, may be separated into 
two domains -- 1) radii $a > 0.02\micron$, which includes 
the ``classical'' grains
(with $a \simgt 0.1\mu$m), which are primarily responsible for the 
extinction, polarization and scattering at visible wavelengths; 
and 2) the ``very small grain'' component 
(with $a <0.02\mu$m) which contribute importantly to the
extinction in the
vacuum-ultraviolet.
While the size distribution for the 
``classical grain'' component is relatively well constrained 
by fitting the observed interstellar extinction curve for an 
assumed dust composition, our knowledge of the size distribution 
for the ``very small grain'' component is much poorer,
due to the fact that, for $\lambda \simgt 0.1\mu$m,
these very small grains are in the Rayleigh limit and their 
extinction cross sections per unit volume are independent of size, 
so that the observed UV/far-UV extinction curve only constrains
the total volume (mass) of this grain component.  

Forty-five years ago, Platt (1956) proposed that very small grains 
or large molecules with radii $\simlt 10$\AA\ may be present in 
interstellar space. Donn (1968) further proposed that 
polycyclic aromatic hydrocarbon-like ``Platt particles'', 
may be responsible for the UV interstellar extinction. 
However, the inability to determine the detailed optical 
properties of ``Platt particles'' and the limited observational 
information on the UV/far-UV interstellar extinction forestalled
further development of these ideas (Greenberg 1960).

Since the 1980s, an important new window on the ``very small grain component''
has been opened by
infrared (IR) observations.
The near-IR continuum emission 
of reflection nebulae (Sellgren, Werner, \& Dinerstein 1983) 
and the 12 and 25$\mu$m ``cirrus'' emission detected by the
{\it Infrared Astronomical Satellite} (IRAS) 
(Boulanger \& P\'{e}rault 1988) explicitly indicated the 
presence of a very small interstellar dust component
since large grains (with radii $\sim 0.1\mu$m) heated by diffuse starlight
emit negligibly at such short wavelengths, whereas very small grains
(with radii $\simlt 0.01\mu$m) can be transiently heated to 
very high temperatures ($\simgt 1000$\ K depending on grain 
size, composition, and photon energy) as first pointed out 
by Greenberg (1968). Subsequent measurements by the 
{\it Diffuse Infrared Background Experiment} (DIRBE) instrument 
on the {\it Cosmic Background Explorer} (COBE) satellite confirmed 
this and detected additional broadband emission at 
3.5 and 4.9$\mu$m (Arendt et al.\ 1998). 

More recently, spectrometers aboard the {\it Infrared Telescope in Space} 
(IRTS) (Onaka et al.\ 1996; Tanaka et al.\ 1996) and the {\it Infrared Space 
Observatory} (ISO) (Mattila et al.\ 1996) have shown that the diffuse 
interstellar
medium radiates strongly in emission features at 3.3, 6.2, 7.7, 8.6, and 
11.3$\mu$m. These emission features, first seen in the spectrum of the 
planetary nebulae NGC 7027 and BD+30$^{\rm o}$3639 (Gillett, Forrest, \& 
Merrill 1973), have been observed in a wide range of astronomical
environments including planetary nebulae, protoplanetary nebulae, 
reflection nebulae, HII regions, circumstellar envelopes, and
external galaxies (see Tielens et al.\ 1999 for a review). 
Often referred to as 
``unidentified infrared'' (UIR) bands, these emission
features are now usually attributed to polycyclic aromatic hydrocarbons
(PAHs) which are vibrationally excited upon absorption of a single 
UV/visible photon (L\'{e}ger \& Puget 1984; Allamandola, Tielens, 
\& Barker 1985) although other carriers have also been proposed such 
as hydrogenated amorphous carbon (Duley \& Williams 1981;
Borghesi, Bussoletti, \& Colangeli 1987; Jones, Duley, \& Williams 1990), 
quenched carbonaceous composite (Sakata et al.\ 1990), 
coal (Papoular et al.\ 1993), fullerenes (Webster 1993), 
and interstellar nanodiamonds with $sp^3$ surface atoms 
reconstructed to $sp^2$ hybridization (Jones \& d'Hendecourt 2000).
The emission mechanism proposed for the UIR bands
-- UV excitation of gas-phase PAHs followed by 
internal conversion and IR fluorescence --
is supported by laboratory measurements of the IR {\it emission} 
spectra of gas-phase PAH molecules (Cherchneff \& Barker 1989; 
Brenner \& Barker 1989; Kurtz 1992; Cook et al.\ 1998)
and by theoretical investigations of the heating and cooling processes 
of PAHs in interstellar space (Allamandola, Tielens, \& Barker 1989; 
Barker \& Cherchneff 1989; d'Hendecourt et al.\ 1989;
Draine \& Li 2001).

By far the most extensively modelled dust property is the
interstellar extinction curve. Existing grain models for 
the diffuse interstellar medium (see Witt 2000 for a recent 
review) are mainly based on an analysis of extinction 
(Mathis, Rumpl, \& Nordsieck 1977; Greenberg 1978; 
Hong \& Greenberg 1980; Draine \& Lee 1984; 
Duley, Jones, \& Williams 1989; Mathis \& Whiffen 1989; 
Kim, Martin, \& Hendry 1994; Mathis 1996; Li \& Greenberg 1997; 
Zubko 1999; Weingartner \& Draine 2001a). 
The near-IR (1--5$\mu$m), mid-IR (5--12$\mu$m) emission spectrum 
along with the far-IR ($>$12$\mu$m) continuum emission of the 
diffuse Galactic medium yields further insights into the 
composition and physical nature of interstellar dust; 
in particular, the PAH emission features allow us to place 
new constraints on the size distribution of the very small dust 
component.    

To model this IR emission, we require realistic calculations of 
the excitation and deexcitation rates of small grains subject
to stochastic heating by absorption of starlight, and 
cooling by spontaneous emission of infrared photons. We require detailed
knowledge of the optical and thermal properties of interstellar 
dust materials to derive the probability distribution for the 
vibrational energy of a grain of specified size and composition.  

Attempts to model the IR emission of interstellar dust have
been made by various previous workers. Following the initial
detection of 60 and 100$\mu$m cirrus emission (Low et al.\ 1984),
Draine \& Anderson (1985) calculated the IR emission from 
a graphite/silicate grain model with grains as small as 3\AA\
and argued that the 60 and 100$\mu$m emission could be accounted
for. When further processing of the IRAS data revealed 
stronger-than-expected 12 and 25$\mu$m emission from interstellar
clouds (Boulanger, Baud, \& van Albada 1985), Weiland et al.\ (1986) 
showed that this emission could be explained if very large numbers
of 3--10\AA\ grains were present. A step forward was taken by 
D\'{e}sert, Boulanger, \& Puget (1990), Siebenmorgen \& Kr\"{u}gel 
(1992), Schutte, Tielens, \& Allamandola (1993), and Dwek et al.\ (1997) 
by including PAHs as an essential grain component. 
Early studies were limited to the IRAS observation 
in four broad photometric bands, but Dwek et al.\ (1997) were able 
to use DIRBE and FIRAS data. No calculations have been carried out 
to ascertain whether {\it both} the PAH features {\it and} the 
far-IR continuum can be quantitatively reproduced. 

There has been considerable progress in both experimental 
measurements and quantum chemical calculations of the optical 
properties of PAHs (Allamandola, Hudgins, \& Sandford 1999a;
Langhoff 1996; and references therein).
There is also an improved understanding of the heat capacities 
of dust candidate materials (Draine \& Li 2001a, hereafter DL01)
and the stochastic heating of very small grains 
(Barker \& Cherchneff 1989; d'Hendecourt et al.\ 1989; DL01)
the interstellar dust size distributions 
(Weingartner \& Draine 2001a, hereafter WD01a), and
the grain charging processes (Weingartner \& Draine 2001b).
In the present paper we make use of these advances 
to model the full emission spectrum,
from near-IR to submillimeter, 
of dust in the diffuse interstellar medium (ISM).
The model is compared to  
IRTS, DIRBE, and FIRAS observations.
We consider dust in three distinct regions: 
high-latitude regions ($|b|\simgt 25^{\rm o}$; hereafter HGL); 
and two regions in the Galactic plane --
the ``MIRS'' region ($44^{\rm o}\le l \le 44^{\rm o}40^{'}$, 
$-0^{\rm o}40^{'}\le b \le 0^{\rm o}$) for which the 
4.5--11.7$\mu$m spectrum was obtained by the mid-IR 
spectrometer on board IRTS;
and the ``NIRS'' region ($47^{\rm o}30^{'}\le l \le 48^{\rm o}$, 
$|b|\le 15^{'}$) for which the 2.8--3.9$\mu$m spectrum was 
obtained by the near-IR spectrometer on board IRTS.

In the present work we consider a dust model which is a natural
extension of the original graphite-silicate model for interstellar dust
(Mathis, Rumpl, \& Nordsieck 1977; Draine \& Lee 1984, hereafter DL84).
Our dust model
consists of 
two grain types -- amorphous silicate grains and carbonaceous 
grains -- both with size distributions ranging 
from ultrasmall grains/large molecules a few angstroms in size
(in the molecular domain) to large grains $\gtsim 1\micron$ in 
diameter\footnote{%
	To avoid confusion, we emphasize here 
	that ``very small grains'', ``large molecules'', 
        and ``ultrasmall grains'' are synonymous.}.
We assume that the carbonaceous grains have
graphitic properties at large sizes and PAH-like properties 
at very small sizes. For PAH-like grains, we distinguish between 
the optical properties of neutral PAHs and PAH ions.
The current work assumes spherical grains and therefore 
does not yet address issues related to polarization.

This paper is organized as follows. We first explore in 
\S\ref{sec:optc} the optical properties of interstellar 
dust materials, with emphasis on the absorption cross 
sections of both ionized and neutral PAHs from far-UV to 
far-IR. We then briefly summarize the heat capacities of 
PAHs, graphite and silicates in \S\ref{sec:enthalpy}.
In \S\ref{sec:isrf} we describe the radiation field 
adopted for the diffuse ISM.
We calculate in \S\ref{sec:T_eq} the equilibrium 
temperatures for large grains ($\simgt 200$\AA). 
In \S\ref{sec:T_spike} we describe the methods 
developed in DL01 to calculate the energy
distribution functions for very small ($a\simlt 200$\AA) grains heated by
starlight.
Selected results are presented for illustration, and we demonstrate
the importance of stochastic heating for grains
as large as $a=200\Angstrom$.

The adopted dust size distribution functions are summarized in 
\S\ref{sec:sizedb}. The PAH charging computation is described 
in \S\ref{sec:charging}, and \S\ref{sec:ir_em} details the IR 
emission modelling.

In \S\ref{sec:results} we  discuss our choices of model parameters,
and compare our model with observations of infrared emission 
from high galactic latitudes as well as two galactic plane regions.
We find our model to be in very good agreement with
observations from the near-IR (including the PAH emission features)
to the submillimeter. We compute emission spectra for our ``standard'' dust
heated by starlight with intensities ranging from 0.3 to $10^4$ times the
local value, and we tabulate the predicted intensities
averaged over the SIRTF IRAC and MIPS photometric bands.
The total infrared emission from our grain model is seen to be in excellent
agreement with DIRBE observations, just as the albedo for our grain
model is in accord with observations of the diffuse galactic
light.

In \S\ref{sec:discussion} we show the extinction curve for our model,
from the far-UV to submillimeter wavelengths, as well as tabulated
values of the dust opacity.
We discuss the elemental depletion, the aromatic absorption 
features, and the variations of the relative abundance of PAH 
in various regions. 
We also summarize 
the upper limits for the abundance of ultrasmall silicate grains 
(both amorphous and crystalline) estimated in Li \& Draine (2001). 
Finally, we summarize our conclusions in \S\ref{sec:summary}.

\section{Optical Properties of Interstellar Dust Materials
	\label{sec:optc}}

To calculate the energy (temperature) distribution functions 
of very small grains (undergoing ``temperature fluctuations''), 
or the ``equilibrium temperatures'' of larger ``classical'' 
grains, and the resulting IR emission spectrum, an accurate 
knowledge of the absorption and emission properties of 
interstellar dust materials over a broad range of wavelengths 
is required.

\subsection{Silicate Grains\label{sec:optc_sil}}

We use Mie theory (see, e.g., Bohren \& Huffman 1983) 
to calculate the absorption and scattering cross
sections for amorphous silicate spheres. We adopt the dielectric 
functions of DL84 with two modifications:
1) the sudden steep rise of the imaginary part in the far-UV 
   has been smoothed (see WD01a); and 
2) the dielectric function has been modified at $\lambda > 250\mu$m
   to better match the average high Galactic latitude dust emission
   spectrum measured by FIRAS (Wright et al.\ 1991; Reach et al.\ 1995; 
   Finkbeiner, Davis, \& Schlegel 1999) -- for $\lambda >250\mu$m we 
   take the imaginary part of the silicate dielectric function to be
\begin{equation}\label{eq:cabs_sil}
\epsilon_2(\lambda) = \epsilon_2^{\rm DL}(\lambda)\times
\left\{\begin{array}{lr} 
\left[1 + \beta\frac{\rm ln\left(\lambda/250\mu m\right)\times
     ln\left(\lambda/850\mu m\right)}{\rm ln(850/250)}\right],
     & 250 \le \lambda \le 850{\rm \mu m},\\
\left(\lambda/850{\rm \mu m}\right)^{\beta}, 
& 850 \le \lambda \le 10^{4}{\rm \mu m},\\ 
\left(10^4/850\right)^{\beta}, 
& \lambda \ge 10^{4}{\rm \mu m}.\\ 
\end{array}\right.
\end{equation}
where $\beta$=0.4, and $\epsilon_2^{\rm DL}$ is the 
imaginary part of the DL84 silicate dielectric function. 
Using the Kramers-Kronig relation (see DL84), we have recomputed
the real part of the silicate dielectric function from the newly
modified $\epsilon_2$.
The modification to $\epsilon_2(\lambda)$ and the absorption cross sections
$C_{\rm abs}(\lambda)$ for silicates
is very slight except at very long wavelengths: there is no change at all
for $\lambda \leq250\micron$, and for $250< \lambda \leq 1100\micron$
the new $\epsilon_2$ is within $\pm12\%$ of $\epsilon_2^{\rm DL}$.
Only for $\lambda > 4.8\mm$ is the silicate emissivity altered by more than
a factor of 2.

We stress that we do {\it not} intend to suggest that the
dielectric function for ``astronomical silicate'' has been determined
with the precision that
might be suggested by the modest adjustments represented by 
eq.\ (\ref{eq:cabs_sil}) -- i.e., we do {\it not} believe that $\epsilon_2$
is known to an accuracy of $\pm$12\% for $\lambda < 1100\micron$.
Our objective is simply to demonstrate that with an entirely reasonable
dielectric function, our carbon-silicate 
grain model can accurately reproduce the
available observations of the far-infrared emission spectrum.
The original DL84 astro-silicate had $C_{\rm abs}\propto\lambda^{-2.0}$
for $\lambda \gtsim 100\micron$.
The above modification has $C_{\rm abs} \propto\lambda^{-1.6}$ for
$800 \ltsim \lambda < 1\cm$, similar to the $\lambda^{-1.6}$ frequency
dependence observed by Agladze et al.\ (1996) 
over 700--2500$\micron$ for amorphous 2MgO$\cdot$SiO$_2$ 
grains at $T=20\K$ [although note that at $\lambda=1000\micron$
we have a silicate opacity 
$\kappa = C_{\rm abs}/(4\pi\rho a^3/3)=0.33\cm^2 \g^{-1}$,
whereas the material studied by Agladze et al.\ (1996)
had $\kappa=1.25\cm^2 \g^{-1}$ at this frequency].

\subsection{Carbonaceous Grains\label{sec:optc_gra}}

We assume a continuous distribution for the optical properties of 
interstellar carbonaceous grains with graphite-like properties for
large sizes ($\simgt 50$\AA) and PAH-like properties in
the small size limit ($\simlt 20$\AA). 
The absorption cross sections $C_{\rm abs}^{\rm gra}(a,\lambda)$
of graphite grains are calculated from Mie theory using
the ``1/3--2/3 approximation'' (Draine \& Malhotra 1993) and
the graphite dielectric functions of DL84.
The absorption cross sections of PAHs will be discussed below 
(\S\ref{sec:optc_pah}). Finally, we make a transition from PAH 
optical properties to graphite properties at $a=50$\AA\ by taking 
the absorption cross sections of carbonaceous grains to be
\begin{equation}\label{eq:cabs_gra}
C_{\rm abs}^{\rm carb}(a,\lambda) = 
\xi_{\rm PAH} C_{\rm abs}^{\rm PAH}(a,\lambda) + 
       (1-\xi_{\rm PAH}) C_{\rm abs}^{\rm gra}(a,\lambda)
\end{equation}
\begin{equation}\label{eq:xi_pah}
\xi_{\rm PAH} (a) = \left(1 - \qcont\right) \times 
                    {\rm min}\left[1, (a_{\xi}/a)^3\right],
                                ~~~~ a_{\xi} = 50{\rm \AA}, 
				~~~~ \qcont = 0.01
\end{equation}
where $C_{\rm abs}^{\rm PAH}$, $C_{\rm abs}^{\rm gra}$ are, 
respectively, the absorption cross sections of PAHs and graphite grains 
of radius\footnote{
	We adopt a mass density
	${\rm \approx 2.24\ g\ cm^{-3}}$ for graphite. 
	The ``radius'' $a$ of a PAH containing $\numC$ C atoms 
	is {\it defined} to be the
	radius of a sphere with the carbon density of graphite
	containing the same number of C atoms,
	i.e., $a = 1.286 N_{\rm C}^{1/3}\Angstrom$.}
$a$ at wavelength $\lambda$; 
$a_{\xi}$ is the grain radius where the transition from PAH properties
to graphite properties begins; 
$0 < \qcont \ll 1$ so that even for small carbonaceous grains
we introduce a small amount of ``continuum'' absorption 
(see \S\ref{sec:optc_pah} below);
$\xi_{\rm PAH}$, the PAH ``weight'', 
drops from $1-\qcont$ to 0 as $a$ increases from $a_{\xi}$ to infinity.
Our choice of the functional form of $\xi_{\rm PAH}$ is quite arbitrary,
but the resultant IR spectrum is insensitive to the detailed behavior of 
$\xi_{\rm PAH}$ provided $\xi_{\rm PAH}=(1-\qcont)$ for $a\simlt 30$\AA\ 
and $\xi_{\rm PAH} \simlt 0.5$ for $a\simgt 100$\AA. 
For example, the much flatter 
function,
$\xi_{\rm PAH} (a) = \left(1 - \qcont\right) \times 
{\rm min}\left\{1, 1/\left[1 + (a/a_{\xi}-1)^3\right]\right\}$,
results in a model spectrum (not shown here) which 
is nearly indistinguishable from that of eq.(\ref{eq:xi_pah}).

\subsection{Absorption Cross Sections of PAHs\label{sec:optc_pah}}

In the present model, a PAH molecule is characterized by the 
number of carbon atoms ($\numC$), the hydrogen to carbon ratio (H/C), 
and charge state (neutral or charged). The H/C ratio is determined 
by the aromatic structure. Since it seems likely that astronomical 
PAHs are relatively pericondensed, as favoured by stability 
considerations (van der Zwet \& Allamandola 1985; Omont 1986; 
Allamandola et al.\ 1989; Jochims et al.\ 1994) 
and the fact that large compact PAHs provide a better match 
to interstellar PAH band positions (Langhoff 1996), we assume 
\begin{equation}\label{eq:nh2nc}
{\rm H/C} = 
\left\{\begin{array}{lr} 
0.5, & \numC \le 25,\\
0.5/\sqrt{\numC/25}, & 25 \le \numC \le 100,\\
0.25, & \numC \ge 100,\\
\end{array}\right.
\end{equation}
which for $\numC < 100$ approximates those of compact, 
symmetric PAHs (series 1, 3 in Stein \& Brown 1991). 
But we have also considered a higher H/C
ratio which is appropriate for PAHs with more open, uneven 
structures (see \S\ref{sec:h2c}).

Following previous workers (L\'{e}ger, d'Hendecourt, \& D\'{e}fourneau 
1989a; D\'{e}sert et al.\ 1990; Schutte et al.\ 1993),
we construct the absorption cross sections for both 
neutral and ionized PAHs from far-UV to far-IR
based on available laboratory measurements and guided by the 
astronomical observations (see Appendix A for details).
The resulting cross sections are characterized by a set of 
Drude profiles: the $\sigma$-$\sigma^{*}$ transition with a peak at 
$\lambda_0^{-1}\simeq 14\mu$m$^{-1}$; the $\pi$-$\pi^{*}$ transition 
with $\lambda_0^{-1}\simeq 4.6\mu$m$^{-1}$;
the C-H stretching mode ($\lambda_0 = 3.3\mu$m); 
two C-C stretching modes ($\lambda_0 = 6.2, 7.7 \mu$m);
the C-H in-plane bending mode ($\lambda_0 = 8.6\mu$m); 
three C-H out-of-plane bending modes 
($\lambda_0 = 11.3, 11.9, 12.7\mu$m);
and a few weak features probably due to C-C bending 
modes ($\lambda_0 = 16.4, 18.3, 21.2, 23.1\mu$m). 
In addition to these discrete features, there is also 
far-UV, near-UV/visible and far-IR continuum absorption.  

The absorption cross section per C atom 
$C_{\rm abs}^{\rm PAH}/\numC$ is taken to be
\begin{eqnarray}
\frac{C_{\rm abs}^{\rm PAH}(\lambda)}{\numC}
&=& C_{\rm abs}^{\rm gra}(\lambda)/\numC
	\hspace*{22em}
    	x > 17.25;			\label{eq:cabs_pah_1}
\\
&=& \left( 126.0-6.4943 x\right)\times 10^{-18}\cm^2/{\rm C},
	\hspace*{10em}
	15 < x < 17.25;			\label{eq:cabs_pah_2}
\\
&=& S_1(\lambda) + \left(-3.0+1.35x\right)\times10^{-18}\cm^2/{\rm C},
	\hspace*{9em}
    	10 < x < 15;			\label{eq:cabs_pah_3}
\\
&=& \left(66.302 - 24.367 x + 
                   2.950 x^2 - 
                   0.1057x^3\right)\times10^{-18}\cm^2/{\rm C}, 
	\hspace*{1em}
    	 7.7 < x < 10;		\label{eq:cabs_pah_4}
\\
&=& S_2(\lambda) + \Big[1.8687 + 0.1905x + 0.4175\left(x-5.9\right)^2 +
                   				\nonumber
\\
&&	0.04370\left(x-5.9\right)^3\Big]
		\times10^{-18}\cm^2/{\rm C},
	\hspace*{10em}
    	5.9 < x < 7.7;		\label{eq:cabs_pah_5}
\\
&=& S_2(\lambda) + (1.8687+0.1905x)\times10^{-18}\cm^2/{\rm C},
	\hspace*{7em}
	3.3 < x < 5.9;		\label{eq:cabs_pah_6}
\\
&=& 34.58\times10^{-18-3.431/x}\times{\rm cutoff}(\lambda,\lambda_{\rm c})
\cm^2/{\rm C} + \sum_{j=3}^{14} S_j(\lambda),
	\hspace*{4em}
	x < 3.3;			\label{eq:cabs_pah_7}
\end{eqnarray}
where  $x\equiv (\lambda/\micron)^{-1}$.
The ultraviolet
       $\sigma$-$\sigma^{*}$  and
       $\pi$-$\pi^{*}$ transitions
       ($j$=1,2),
       C-H stretching ($j$=3), C-C stretching ($j$=4,5),
       C-H in-plane bending ($j$=6), C-H out-of-plane 
       bending ($j$=7,8,9), C-C bending ($j$=10,11,12,13) modes,
       and far-IR continuum ($j$=14)
are represented by Drude profiles, with the absorption cross section
per carbon atom contributed by feature $j$ given by
\begin{equation}\label{eq:drude}
S_j(\lambda) \equiv \frac{2}{\pi} 
	\frac{\gamma_j \lambda_j \sigma_{{\rm int},j}}
	{(\lambda/\lambda_j-\lambda_j/\lambda)^2+\gamma_j^2}, 
\end{equation}
where Table \ref{tab:drude_parameters} lists values of the
central wavelength $\lambda_j$, the broadening parameter $\gamma_j$,
and the integrated absorption strength\footnote{
	The cross section integrated over wavelength 
	$\int S_j(\lambda)d\lambda = \lambda_j^2 \sigma_{{\rm int},j}$.
	}
\begin{equation}\label{eq:int_strength}
\sigma_{{\rm int},j}\equiv\int S_j (\lambda) d\lambda^{-1} 
	= \frac{\pi}{2} S_j(\lambda_j)\gamma_j\lambda_j^{-1}.
\end{equation}
The function ${\rm cutoff(\lambda, \lambda_c)}$
and the cutoff wavelength ${\rm \lambda_c}$, determined by 
the PAH size (the number of fused benzenoid rings),
are defined in eqs.\ (\ref{eq:pah_cutoff_func},\ref{eq:pah_cutoff_wave}). 

\begin{table}
\caption{Drude profile parameters for model PAHs.
	\label{tab:drude_parameters}}
\begin{tabular}{c c c c c c c c}
\hline \hline
 & & & &\multicolumn{2}{c}{ FWHM}&\multicolumn{2}{c}{$\sigma_{{\rm int},j}\ \equiv\int S_j(\lambda)d\lambda^{-1}$}\\ 
$j$ & $\lambda_j$ & $\lambda_j^{-1}$ &$\gamma_j$& $\gamma_j \lambda_j$ & 
	$\gamma_j \lambda_j^{-1}$ & neutral & ionized\\
 & ($\mu$m) & (cm$^{-1}$) && ($\micron$) & (cm$^{-1}$) & ($10^{-20}$cm/C) 
 & ($10^{-20}$cm/C) \\
\hline
1 & .0722          & 138500 &0.195& .0141 & 27000 & $7.97\times10^{7}$
                                            & $7.97\times10^{7}$\\
2 & .2175          & 46000  &0.217& .0473 & 10000 & $1.23\times10^{7}$
                                            & $1.23\times10^{7}$\\
\hline
3 & 3.3            & 3030   &0.012& 0.04  &    37 & $197\times{\rm H/C}$
                   & $44.7\times{\rm H/C}$\\
4\tablenotemark{a} & 6.2  & 1610   &0.032 & 0.20  &    52 
                   & $19.6\times E_{6.2}$ & $157\times E_{6.2}$\\
5\tablenotemark{a} & 7.7  & 1300   &0.091 & 0.70  &   118 
                   & $60.9\times E_{7.7}$ & $548\times E_{7.7}$\\
6\tablenotemark{a} & 8.6  & 1161   &0.047& 0.40  &   54 
                   & $34.7\times E_{8.6}\times{\rm H/C}$ 
                   & $242\times E_{8.6}\times {\rm H/C}$\\
7\tablenotemark{b} &11.3   &  886   &0.018& 0.20 &   16 
                   & $427\times (1/3){\rm H/C}$ & $400\times (1/3){\rm H/C}$\\ 
8\tablenotemark{b} &11.9   &  840   &0.025& 0.30  & 21 
                   & $72.7\times (1/3){\rm H/C}$ 
                   & $61.4\times (1/3){\rm H/C}$\\
9\tablenotemark{b} & 12.7  &  787   &0.024& 0.30 &   19 
                   & $167\times (1/3){\rm H/C}$
                   & $149\times (1/3){\rm H/C}$\\
\hline
10\tablenotemark{c}& 16.4 &  610   &0.010& 0.16  &    6& $5.52$
				            & $5.52$ \\ 
11\tablenotemark{d}& 18.3   &  546   &0.036& 0.66  &    20 & $6.04$
                                            & $6.04$ \\
12\tablenotemark{d}& 21.2   &  472   &0.038& 0.81  &    18 & $10.8$
                                            & $10.8$ \\
13\tablenotemark{d}& 23.1   &  433   &0.046& 1.07  &    20 & $2.78$
                                            & $2.78$ \\
\hline
14\tablenotemark{e}& 26.0  &  385   &0.69& 18.0  &   266 & $15.2$ 
                                            & $15.2$ \\
\hline
\end{tabular}\\[2mm]
\tablenotetext{a}{With $E_{6.2}$=1, 
              $E_{7.7}$=1 and $E_{8.6}$=1 one obtains the
              laboratory measured band strengths. 
              The observed IR spectrum of the diffuse
              ISM is best fitted by our dust 
              model with $E_{6.2}$=3, $E_{7.7}$=2 and $E_{8.6}$=2.}
\tablenotetext{b}{For the C-H out-of-plane bending modes, the $\frac{1}{3}$
             factor arises from the assumption that PAHs have equal 
             numbers of H in solo, duet, and trio adjacent CH units 
             (see \S\ref{app:NIR_to_MIR}).}
\tablenotetext{c}{Band width from Moutou et al.\ (2000).}
\tablenotetext{d}{Band widths from Moutou et al.\ (1996).}
\tablenotetext{e}{To reproduce the weak $\lambda > 14\mu$m continuum 
	absorption measured by Moutou et al.\ (1996).}
\end{table}

\begin{figure}			
\epsfig{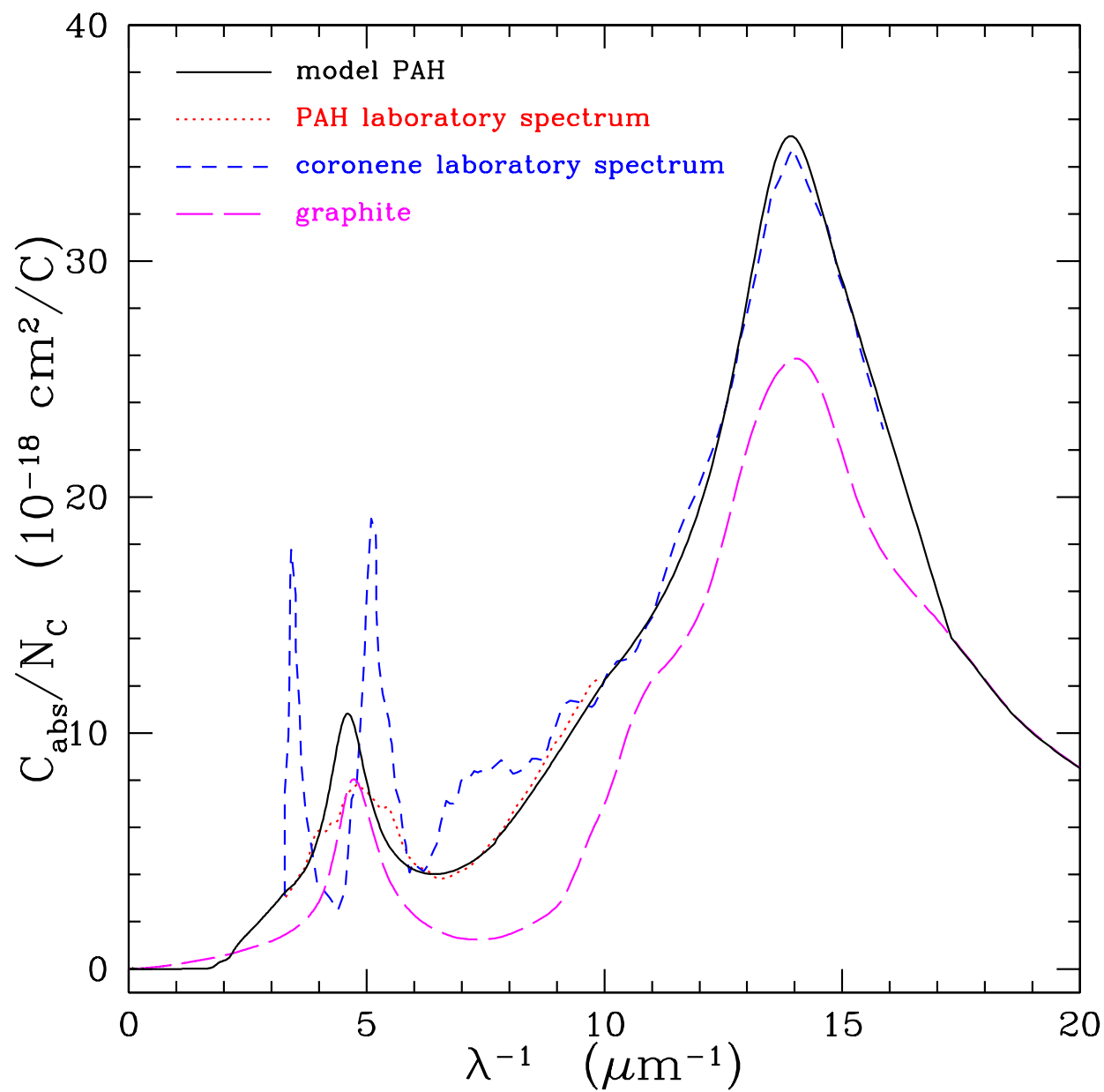}
\caption{
	\footnotesize
        \label{fig:pahcs_uv}
        The ultraviolet and far ultraviolet 
        absorption cross section per C atom of PAHs.
        }
\end{figure}

\begin{figure}			
\epsfig{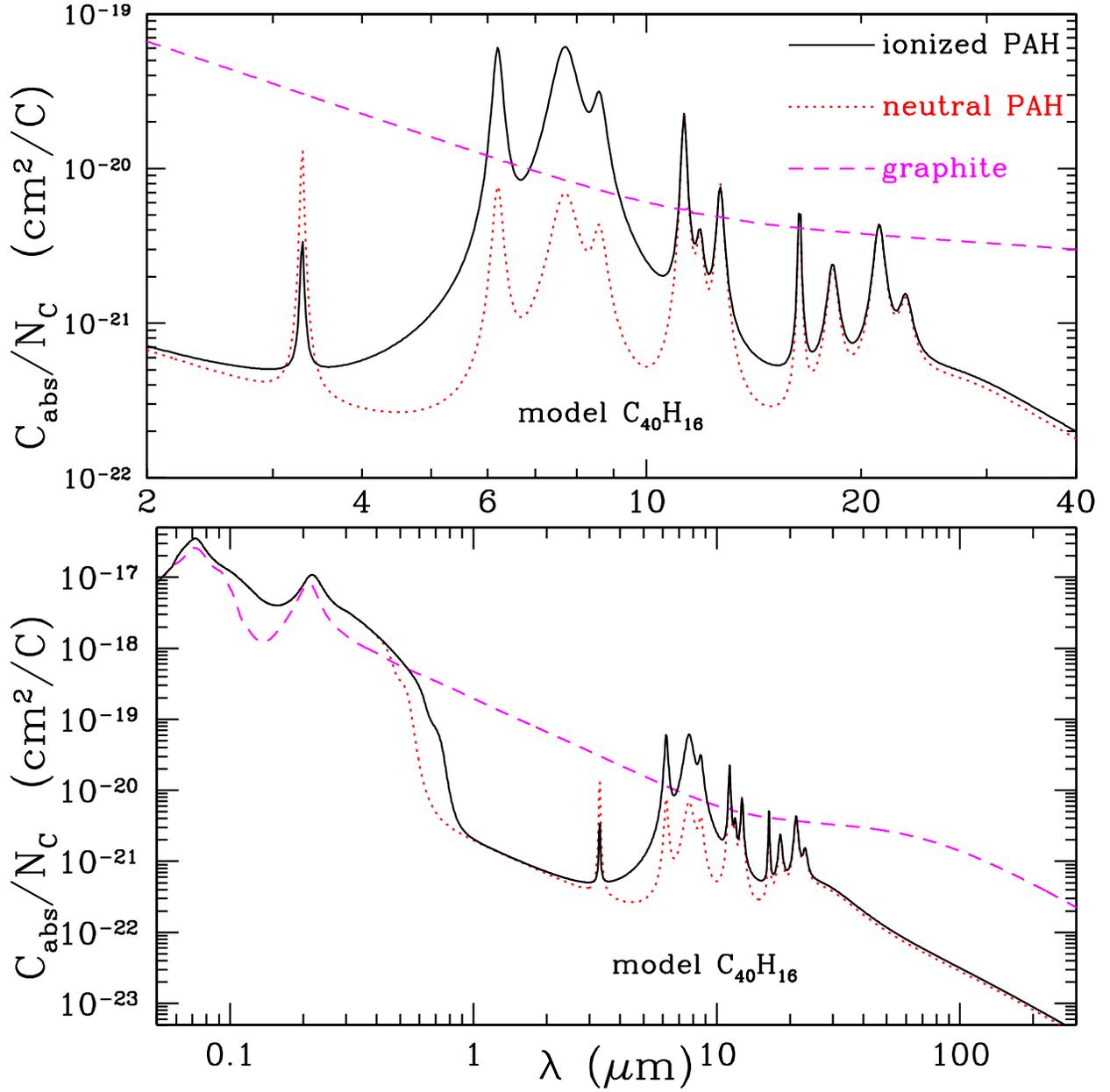}
\caption{
	\footnotesize
        \label{fig:pahcs_ir}
        Absorption cross section per C atom of neutral and 
        ionized PAHs (with $E_{6.2}$=3, $E_{7.7}$=2, $E_{8.6}$=2). 
	Ionization enhances the C-C stretching modes
        (6.2, 7.7$\mu$m) and C-H in-plane bending mode (8.6$\mu$m),
	and weakens the C-H stretching mode
        (3.3$\mu$m).
        Ionization shifts the visual absorption edge 
        (which is also size-dependent) redward.
        The absorption cross section per C atom of 
        small graphite spheres is shown for comparison. 
        }
\end{figure}

Figure \ref{fig:pahcs_uv} 
plots the absorption spectrum of PAHs in the UV/far-UV. 
Also plotted are the experimental spectra of coronene 
(C$_{24}$H$_{12}$) (L\'{e}ger et al.\ 1989b), PAH mixtures 
(L\'{e}ger et al.\ 1989b; Joblin, L\'{e}ger, \& Martin 1992), 
and the cross section estimated for graphite (DL84).
Figure \ref{fig:pahcs_ir} displays the absorption cross sections
of neutral and ionized PAHs for a wide wavelength range with 
emphasis on the C-H, C-C IR vibrational modes (see following
and \S\ref{sec:pahircont}, \S\ref{app:NIR_to_MIR},
\S\ref{app:FIR} for details).

We have included enhancement factors $E_{6.2}$, $E_{7.7}$, 
and $E_{8.6}$ in the strengths of the 6.2, 7.7, and 8.6$\micron$ 
bands\footnote{%
	Absolute band strengths for some PAH species 
	vary considerably among different laboratory groups 
	and between experimental results and theoretical calculations. 
	For example, the integrated cross sections of naphthalene, 
	anthracene, tetracene, phenanthrene, benzanthracene, pyrene, 
	and coronene cations can differ by a factor of 1.5--20 between 
	different experimental studies, or between experimental and 
	theoretical studies (see Langhoff 1996).
	}. 
For $E_{6.2}=E_{7.7}=E_{8.6}=1$ we recover the laboratory values, 
but we are unable to satisfactorily reproduce astronomical spectra unless
we take
$E_{6.2}$, $E_{7.7}$, $E_{8.6}>1$.
For the $7.7$ and $11.3\micron$ features,
laboratory data (see \S\ref{app:NIR_to_MIR}) indicate 
a relative band strength 
$\sigma_{{\rm int},7.7}/\sigma_{{\rm int},11.3}\approx 1.3$ 
for neutral PAHs, 
and $\approx 12$ for ionized PAHs (assuming ${\rm H/C}=1/3$). 
Let $P(\lambda_j)\equiv\int I_\lambda d\lambda$ be the total power
integrated over the profile of feature $j$.
Values of $P(7.7\micron)/P(11.3\micron)$
as large as 11 are observed in some objects, such as compact HII 
regions (Roelfsema et al.\ 1996). DL01 have modelled 
the emission expected from PAHs in various radiation fields, 
and find that for $E_{7.7}$=2, fully ionized PAHs with 
$\numC<$100 in a region of very intense UV radiation 
(up to $10^{6}\times$ the Habing [1968] UV intensity) achieve
$P(7.7\micron)/P(11.3\micron)\approx 21$.
However, since the observed 
spectra include emission from larger PAHs with smaller values of
$P(7.7\micron)/P(11.3\micron)$,
it is necessary to take $E_{7.7}\approx 2$ to reproduce the strength
of the strongest observed $7.7\micron$ features. 

In addition, a wide variety of objects show a 
$P(6.2\micron)/P(7.7\micron)$
ratio exceeding the maximum predicted from the 
laboratory PAH feature strengths.
For example, 
$P(6.2\micron)/P(7.7\micron)\approx$ 0.3 -- 0.6
in a recent study of 
820 mid-IR spectra of the diffuse interstellar 
medium (Chan et al.\ 2001), and $\approx 0.54$ for 
an average of planetary nebulae, reflection
nebulae, and HII regions (Cohen et al.\ 1989).
Thermal emission has
a strict upper limit 
$P(6.2\micron)/P(7.7\micron) < (7.7/6.2)^2
\sigma_{{\rm int},6.2}/\sigma_{{\rm int},7.7} \approx 0.44 E_{6.2}/E_{7.7}$
(see Table \ref{tab:drude_parameters}),
and detailed modelling of stochastic heating finds
$P(6.2\micron)/P(7.7\micron)\ltsim 0.35 E_{6.2}/E_{7.7}$
(see Figure 16 in DL01).
Therefore, we take $E_{6.2}=1.5E_{7.7}\approx 3$.

For most objects, the strength of the $8.6\micron$ C-H in-plane
bending mode is consistent with the strength of the $7.7\micron$
C-C bending mode, so we take $E_{8.6}=E_{7.7}\approx 2$.

\subsection{Continuum Opacity of Polycyclic Aromatic Hydrocarbons\label{sec:pahircont}}

Even when $a<a_\xi$, eqs.\ (\ref{eq:cabs_gra},\ref{eq:xi_pah}) cause
$C_{\rm abs}$ to include a fraction $\qcont$ of the opacity per
C atom of graphite, resulting in ``continuum'' opacity extending into
the infrared.
This is required
to account for
the near-IR continuum emission at 1--5$\mu$m underlying the 3.3$\mu$m 
C-H stretching feature detected in a number of astronomical objects,
including 
the Orion Bar 
	(Joblin et al.\ 1996a; 
	Bregman et al.\ 1989;
	Geballe et al.\ 1989), 
various reflection nebulae 
	(Sellgren, Werner, \& Allamandola 1996), 
the Red Rectangle
	(Geballe et al.\ 1989),
normal galaxies 
	(Helou et al.\ 2000), 
and the starburst galaxies M82 and NGC 253 
	(Sturm et al.\ 2000). 
This continuum emission
may arise from the electronic fluorescence or 
a quasi-continuum of overlapping overtone and combination bands 
from vibrational fluorescence in PAHs (L\'{e}ger \& Puget 1984; 
Allamandola et al.\ 1985, 1989; Sellgren et al.\ 1996).
Electronic intraband transitions can also contribute an optical-infrared
continuum due to the closing of the band gap as the PAH size increases
and/or upon ionization of the PAH (Allamandola et al.\ 1989; 
Tielens et al.\ 1999).

We represent 
this continuum by the parameter $\qcont$ in eq.\ (\ref{eq:xi_pah}).
High resolution spectroscopic observations of NGC 1333 (Joblin et al.\ 
1996a), the Orion Bar (Geballe et al.\ 1989; Joblin et al.\ 1996a), and 
the Red Rectangle (Geballe et al.\ 1989) resolving the 3.3$\mu$m feature
with FWHM$\approx 0.04\mu$m (same as that adopted here)
show a feature peak-to-continuum ratio as high as $\sim 30$.
If we require a $3.3\micron$ peak-to-continuum ratio of 30, then 
PAHs with $\NC\ltsim50$ (small enough to be
heated to $T\gtsim1500\K$ by a UV photon -- see DL01)
must have 
\begin{equation}
\qcont \approx \frac{1}{30}\times
\frac{2\times 3.3\mu {\rm m}\times\sigma_{{\rm int},3.3\mu {\rm m}}}
{\pi\gamma_{3.3} C_{\rm abs}^{\rm gra}(3.3\micron)} \approx 0.01 ~,
\end{equation}
where we have taken 
$C_{\rm abs}^{\rm gra}(3.3\micron)\approx 3\times10^{-20}\NC\cm^2$,
H/C$\approx 1/3$, and have taken $\sigma_{\rm int,3.3\mu m}$
to be the average of the values in Table \ref{tab:drude_parameters} for
neutral and ionized PAHs. 
The resulting PAH infrared cross sections are shown in
Figure 2, for both neutral and ionized PAHs. 

Finally, there is the question of the value of $a_\xi$.
The observed PAH emission features are at $\lambda\ltsim 25\micron$,
and this will be contributed primarily by particles with
$a\ltsim 40\Angstrom$ which can be heated to $\gtsim100\K$ by
a photon with $hc/\lambda<13.6\eV$.
Therefore, the observed PAH emission features only require
$a_\xi\gtsim 25\Angstrom$.
We will take $a_\xi\approx50\Angstrom$ so that the 6.2$\micron$
feature in {\it absorption} is in agreement with observations,
as discussed in \S\ref{sec:summary},
but this value -- as well as the form of the transition from 
PAH-like to graphitic optical properties (\S\ref{sec:optc_gra})
-- is poorly determined, as the calculated emission spectrum is relatively
insensitive to variations in $a_\xi$ in the range 25 -- 100$\Angstrom$.
IR spectra calculated with $a_{\xi}=25$\AA\ or
$a_{\xi}=75$\AA\ (not shown here) 
are almost identical to that for $a_\xi=50$\AA.

\section{Enthalpies\label{sec:enthalpy}}

As discussed in DL01, the vibrational energy of 
a PAH molecule containing $\numC$ C and 
${\rm (H/C)}\times N_{\rm C}$ H atoms
can be approximated as 
\begin{equation}\label{eq:E_pah}
E_{\rm PAH}(T) = \frac{\rm H}{\rm C}\NC 
             \sum\limits^{3}_{j=1}\frac{\hbar \omega_j}
             {{\rm exp}(\hbar \omega_j/kT)-1} + 
             (\numC-2)\left[k \Theta_{op} f_2(T/\Theta_{op}) +
             2k \Theta_{ip} f_2(T/\Theta_{ip}) \right] ~~~;
\end{equation}
the summation index $j=1-3$ runs 
over the C-H out-of-plane bending ($\omega_1/2\pi c$=886\ cm$^{-1}$), 
in-plane bending ($\omega_2/2\pi c$=1161\ cm$^{-1}$), 
and stretching ($\omega_3/2\pi c$=3030\ cm$^{-1}$) modes; the term containing 
$\numC-2$ is a two-dimensional Debye model with Debye 
temperatures $\Theta_{op} \approx 863$\ K for the out-of-plane C-C modes, 
$\Theta_{ip} \approx 2504$\ K for the in-plane C-C modes;  
$k$ is the Boltzman constant; $T$ is the ``vibrational temperature''; 
and
\begin{equation}\label{eq:fn_debye}
f_n(x) = \frac{1}{n} \int^{1}_{0} \frac{y^n dy}{{\rm exp}(y/x)-1}.
\end{equation}

For graphitic grains, we use eq.\ (\ref{eq:E_pah}) 
with ${\rm H/C}=0$ (see DL01).

For amorphous silicate grains, DL01 recommend
\begin{equation}\label{eq:E_sil}
E_{\rm sil}(T) = (\numatom-2)\left[2k \Theta_2 f_2(T/\Theta_2) +
             k \Theta_3 f_3(T/\Theta_3) \right] ,
\end{equation}
where $\numatom$ is the number of atoms contained in the silicate 
grain (we assume a mass density 3.5${\rm g\ cm^{-3}}$ for MgFeSiO$_4$ 
silicate material); 
$\Theta_2 \approx 500$\ K, $\Theta_3 \approx 1500$\ K are Debye
temperatures for the two- and three-dimensional 
Debye models, respectively.

\section{Radiation Field
	\label{sec:isrf}}

We adopt the solar neighbourhood interstellar radiation field (ISRF)
(Mathis, Mezger, \& Panagia 1983; hereafter MMP) for the diffuse 
ISM 
\begin{equation}\label{eq:mmpisrf}
u_{\lambda} = \chiMMP \left\{ u_{\lambda}^{\rm UV_{\odot}} + 
              \sum\limits_{i=2}^{4}W_{i} 
              \frac{4\pi}{c}B_{\lambda}(T_i)\right\} 
              + \frac{4\pi}{c}B_{\lambda}(2.9 {\rm K})
\end{equation}
where $\chiMMP$ describes the enhancement of the starlight
component 
(relative to the solar neighborhood)\footnote{
	Note that $\chiMMP=1$ corresponds to $\chi=1.23$, 
	where $\chi$ is the ratio of the intensity relative to 
        the Habing (1968) estimate at 1000\AA,
	or $G_0=1.14$, where $G_0$ is the 6--13.6~eV
	energy density relative to the 6--13.6~eV energy density
	in the Habing (1968) radiation field.
	};
$u_{\lambda}^{\rm UV_{\odot}}$ is the 
ultraviolet component; $(W_2, W_3, W_4)=(10^{-14}, 10^{-13}, 
4\times 10^{-13})$ and $(T_2, T_3, T_4)=(7500, 4000, 3000)$\ K 
are dilution factors and corresponding black-body temperatures.
The last term is the cosmic background radiation.

\section{Thermal Equilibrium Temperatures of Large Grains
	\label{sec:T_eq}}

\begin{figure}			
\epsfig{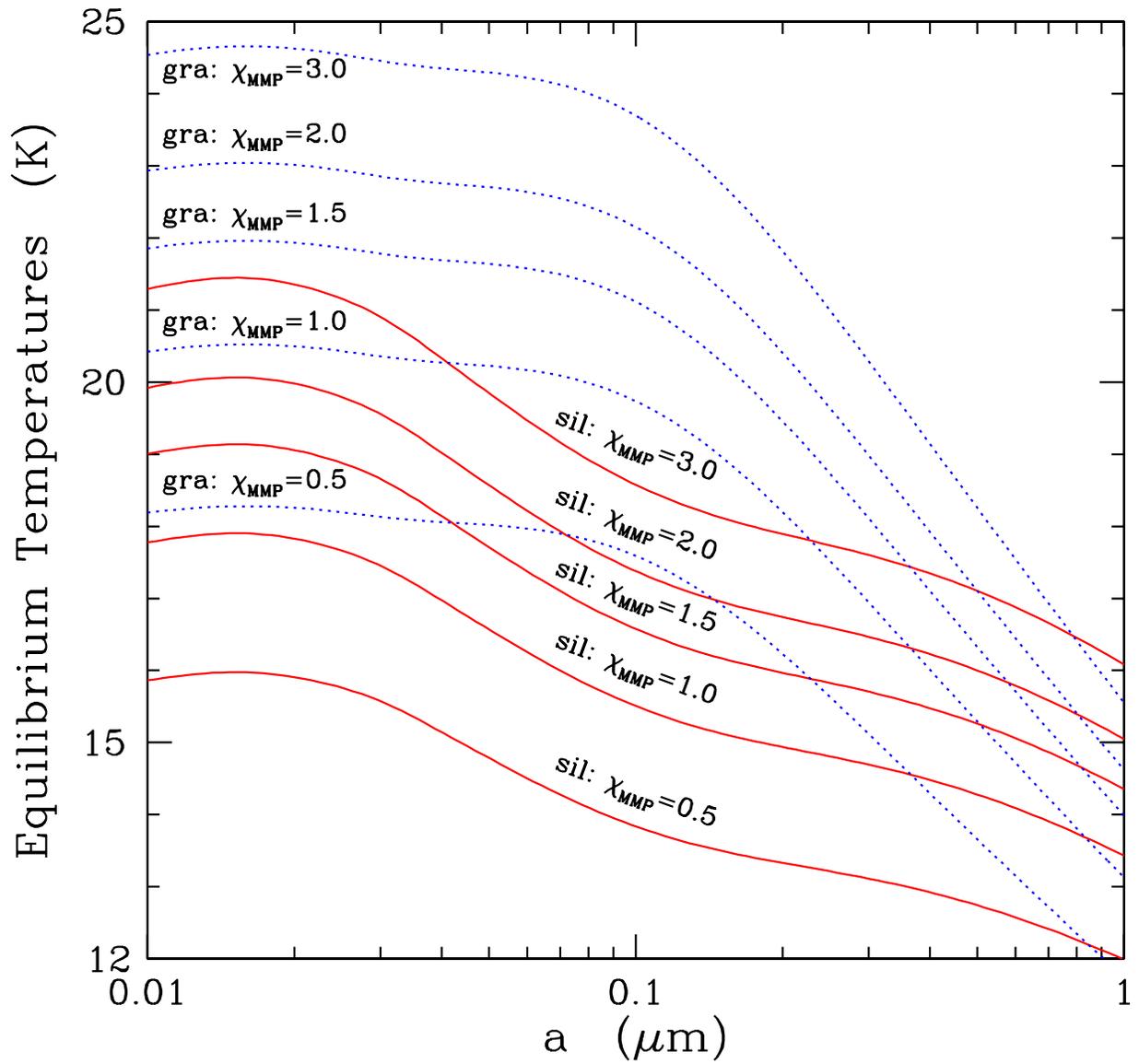}
\caption{
	\footnotesize
        \label{fig:temp_equil}
	Equilibrium temperatures for graphite (dotted lines) 
        and silicate grains (solid lines) in environments with
        various starlight intensities.
        }
\end{figure}

Large grains maintain a
nearly constant
temperature $\bar{T}$ determined by balancing
absorption and emission,
\begin{equation}\label{eq:cal_Teq}
\int^{\infty}_{0} C_{\rm abs}(a,\lambda)  c u_{\lambda} d\lambda
= \int^{\infty}_{0} C_{\rm abs}(a,\lambda) 4\pi B_{\lambda}(\bar{T})d\lambda
~~~,
\end{equation}
where $C_{\rm abs}(a,\lambda)$ is the absorption cross section for a grain
with size $a$ at wavelength $\lambda$, $c$ is the speed of light,
$B_{\lambda}(T)$ is the Planck function at temperature $T$, and
$u_{\lambda}$ is the energy density of the radiation field 
(see \S\ref{sec:isrf}). In Figure \ref{fig:temp_equil} we display
these ``equilibrium'' temperatures for graphitic and silicate grains
as a function of size in environments with various UV intensities. 
 
\section{Transient Heating of Very Small Grains
	\label{sec:T_spike}}

Since Greenberg (1968) first noted that very 
small grains heated by starlight would experience temperature 
spikes, there have been a number of theoretical studies of 
this process. 
DL01 have recently carried 
out a detailed study in which they compared the energy
distribution functions and resulting emission spectra 
calculated ``exactly'' (the ``{\it exact-statistical}'' method) 
versus two approximations: the ``{\it thermal-discrete}'' 
approximation and the ``{\it thermal-continuous}'' approximation.
The {\it exact-statistical} method involves calculation 
of the vibrational density of states.
In contrast, both the {\it thermal-discrete} method and 
the {\it thermal-continuous} method assume that at any 
given time, the vibrationally excited grain has a ``vibrational 
temperature'', and the deexcitation process is assumed to
correspond to a thermal emission spectrum.  
In the {\it thermal-discrete} method both the excitation and 
deexcitation processes are treated as discrete events; however,
in the {\it thermal-continuous} method, the deexcitation process
is approximated as continuous rather than discrete.

\begin{figure}			
\epsfig{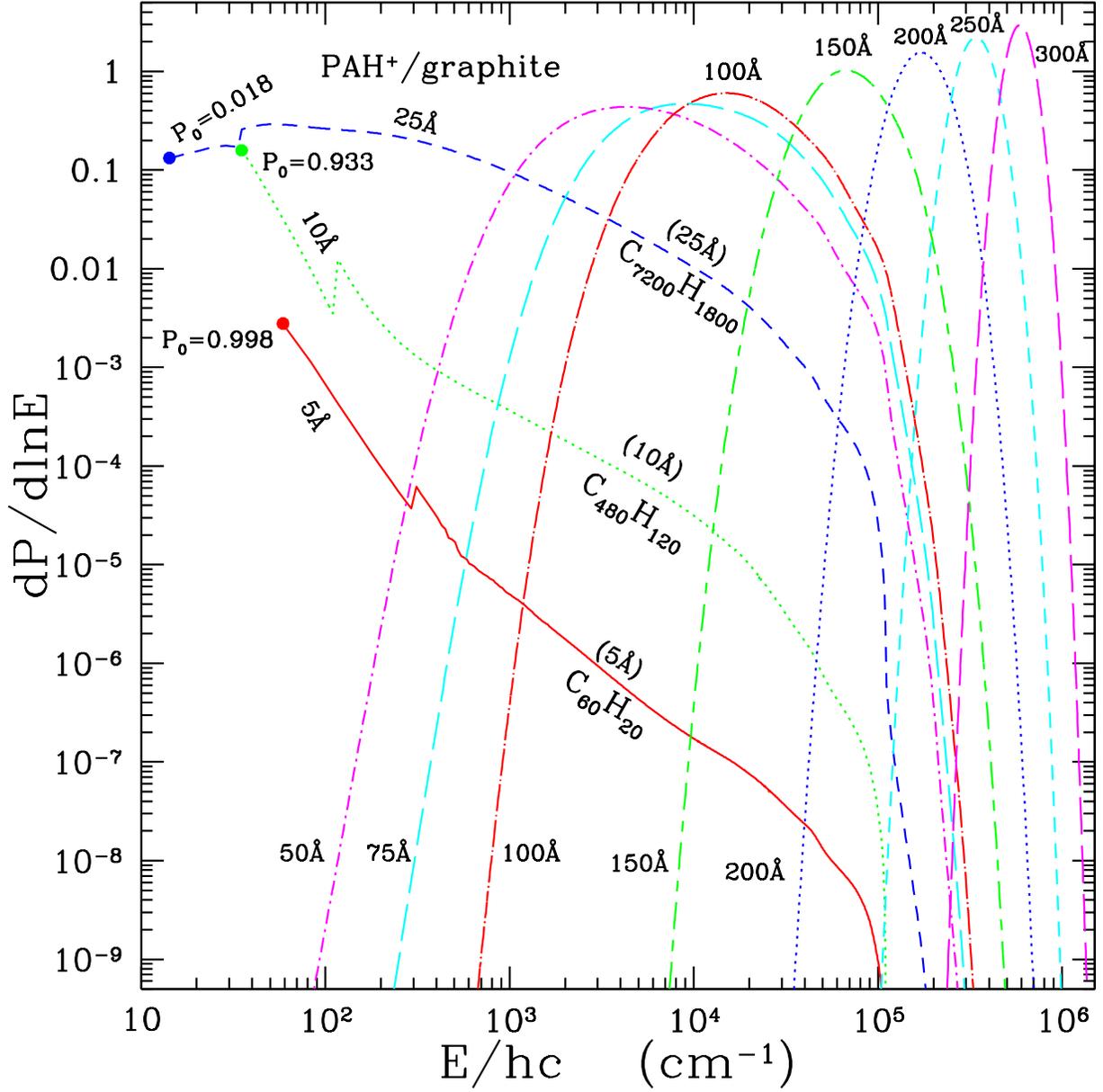}
\caption{
	\footnotesize
        \label{fig:P_E_pah}
        The energy distribution functions for charged carbonaceous
        grains ($a=5\ [{\rm C_{60}H_{20}}]$, 10 [${\rm C_{480}H_{120}}$], 
        25 [${\rm C_{7200}H_{1800}}$], 50, 75, 100, 150, 200, 250,
	300\AA) for $\chiMMP=1$. 
        The discontinuity in the 5, 10, and 25\AA\ curves 
	is due to the change of the estimate 
        for grain vibrational ``temperature'' at the 20th vibrational
        mode (see DL01). For 5, 10, and 25\AA\ a dot
        indicates the first excited state, and $P_0$ is
        the probability of being in the ground state.
        }
\end{figure}

\begin{figure}			
\epsfig{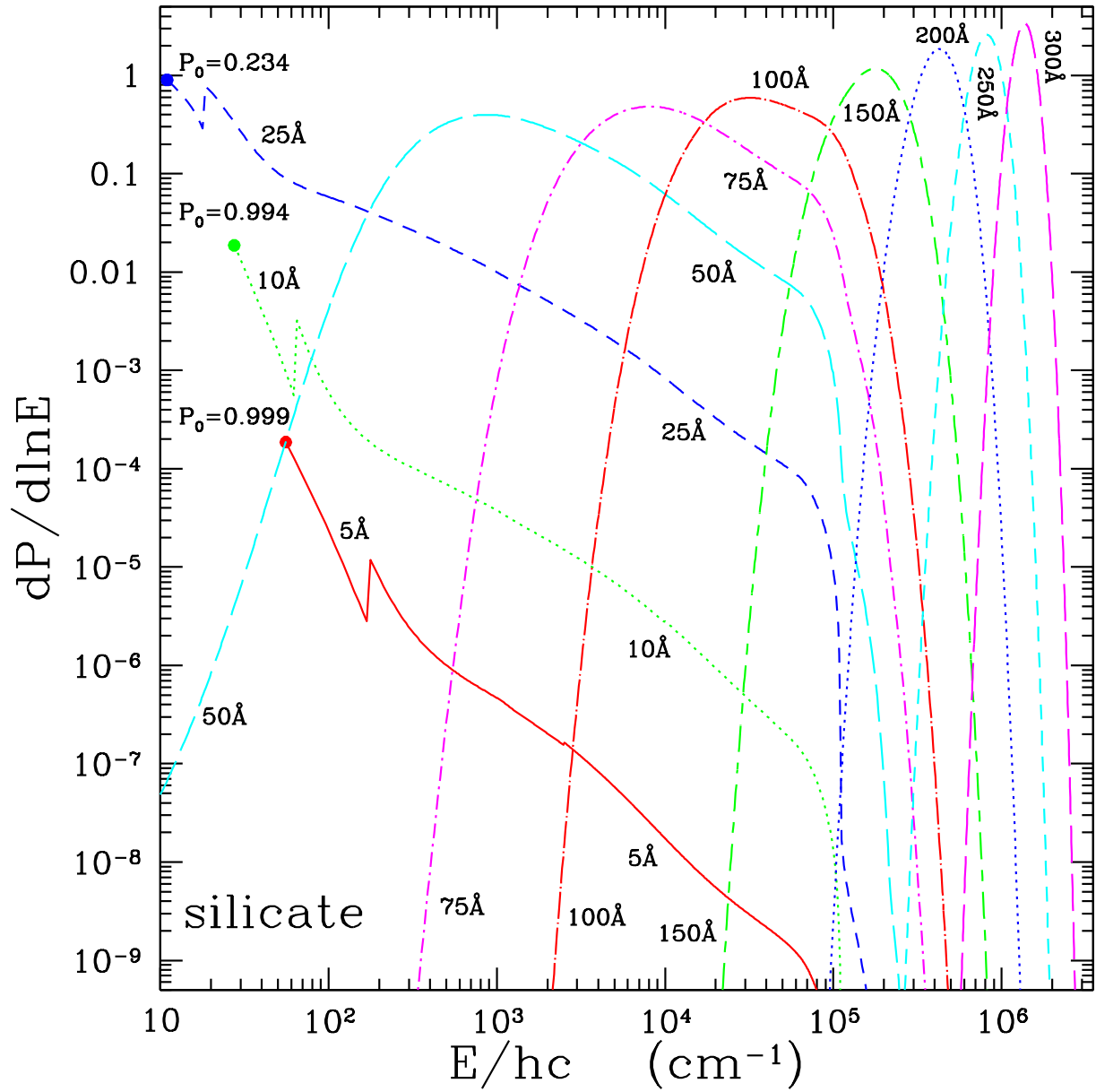}
\caption{
	\footnotesize
        \label{fig:P_E_sil}
	As in Figure \ref{fig:P_E_pah}, but for silicate grains.
	}
\end{figure}

\begin{figure}			
\epsfig{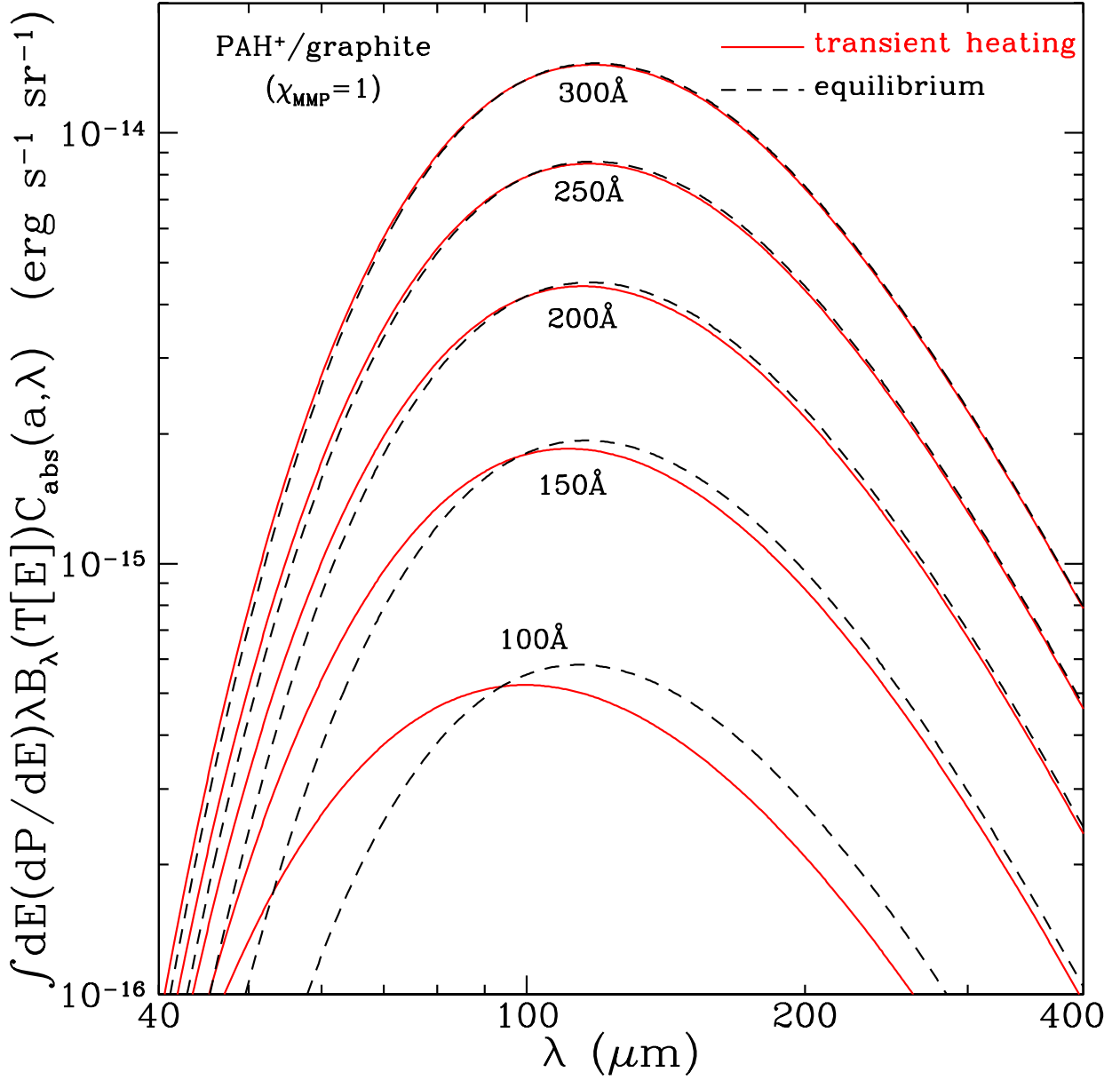}
\caption{
	\footnotesize
	\label{fig:emsn_spike_equil}
	Infrared emission spectra for small carbonaceous grains
	of various sizes heated by starlight with $\chiMMP=1$,
	calculated using the full energy distribution function 
        $P(E)$ (solid lines); also shown (broken lines) are spectra
	computed for grains at the
	 ``equilibrium'' 
        temperature $\bar{T}$. 
	Transient heating effects lead to
        significantly more short wavelength emission for 
        $a \ltsim 200\Angstrom$.
	}
\end{figure}

Although the thermal methods are not exact, 
DL01 have shown that the overall IR emission spectrum calculated
using the {\it thermal-discrete} and the {\it thermal-continuous}
methods is very close to that of the exact treatment (see also
Barker \& Cherchneff 1989; d'Hendecourt et al.\ 1989;
Allamandola et al.\ 1989; Schutte et al.\ 1993).
Since the {\it exact-statistical} method involves heavy 
computations of the grain vibrational mode spectrum and 
density of states, we adopt the {\it thermal-discrete} method. 
Figures \ref{fig:P_E_pah} and \ref{fig:P_E_sil} 
show the energy distribution 
functions found for PAHs and silicate grains with radii
$a=5, 10, 25, 50, 75, 100, 150, 200, 300$\AA\ 
illuminated by the average ISRF ($\chiMMP=1$).
Very small grains ($a \ltsim 100\Angstrom$) have a very broad $P(E)$, and
the smallest grains ($a \ltsim 30\Angstrom$) have an appreciable probability
$P_0$
of being found in the vibrational ground state $E=0$.
As the grain size increases, $P(E)$ becomes narrower, 
so that for $\chiMMP=1$ it can be approximated by a delta 
function for $a > 250\Angstrom$. 
However, for radii as large as $a = 200\Angstrom$, both carbonaceous
and silicate grains have energy distribution functions which are broad 
enough that the emission spectrum deviates noticeably from the emission 
spectrum for grains at a single temperature $\bar{T}$, as shown in 
Figure \ref{fig:emsn_spike_equil}. For accurate computation of infrared 
emission spectra it is therefore important to properly calculate the 
energy distribution function $P(E)$, including grain sizes which are 
large enough that the average thermal energy content exceeds a few eV.

\section{Grain Size Distributions\label{sec:sizedb}}

Following Weingartner \& Draine (2001a), we adopt size 
distribution functions for silicate grains and carbonaceous grains of
the following form:
\begin{equation}\label{eq:sizedb_sil}
\frac{1}{\nH} \frac{dn_{\rm sil}}{da} = \frac{C_{\rm sil}}{a}
\left( \frac{a}{a_{\rm t,sil}} \right)^{\alpha_{\rm sil}} 
F(a; \beta_{\rm t,sil},a_{\rm t})\times G(a; a_{\rm t,sil},a_{\rm c,sil}) ~,
\end{equation}
\begin{equation}\label{eq:sizedb_gra}
\frac{1}{\nH} \frac{dn_{\rm carb}}{da} = D(a) + \frac{C_{\rm carb}}{a} 
\left( \frac{a}{a_{\rm t,carb}} \right)^{\alpha_{\rm carb}} 
F(a; \beta_{\rm carb},a_{\rm t})\times G(a; a_{\rm t,carb},a_{\rm c,carb}) ~,
\end{equation}
\begin{equation}
F(a; \beta,a_{\rm t}) \equiv
\cases{1 + \beta a/a_{\rm t}, ~~~ &$\beta\ge 0$
\cr
1/(1-\beta a/a_{\rm t}), ~~~ &$\beta < 0$} ~,
\end{equation}
\begin{equation}
G(a; a_{\rm t}, a_{\rm c}) \equiv
\cases{1, &$3.5 \Angstrom < a < a_{\rm t}$
\cr
\exp \left\{ - [(a - a_{\rm t})/a_{\rm c}]^3 \right\},
&$a > a_{\rm t}$\cr} ~,
\end{equation}
where $n_{\rm H}$ is the hydrogen number density, and
$D(a)$ is an additional population of very small ($a\ltsim 100\Angstrom$)
carbonaceous grains
which is introduced in order for the grain model to reproduce
the observed infrared emission.
The grains in the distribution $D(a)$ are all in the Rayleigh
limit at the wavelengths $\lambda > 1000\Angstrom$ where 
interstellar extinction
has been measured, and therefore extinction measurements 
constrain only 
$\bc$, the number of carbon atoms per total H atoms
in the population $D(a)$.
For the PAH ultraviolet absorption cross section discussed in
\S\ref{sec:optc_pah}, WD01a showed that
values of $\bc \leq 60$ppm are allowed for the average diffuse ISM
extinction law ($R_V=3.1$).
For a given choice of $\bc$ within the allowed range, 
the parameters
$a_{\rm t,sil}$, $a_{\rm c,sil}$, $\alpha_{\rm sil}$,
$\beta_{\rm sil}$, $C_{\rm sil}$, $a_{\rm t,carb}$, $a_{\rm c,carb}$, 
$\alpha_{\rm carb}$, $\beta_{\rm carb}$, $C_{\rm carb}$ 
can be found in WD01a.

The population of very small
grains required to explain the IR emission of the cirrus
and translucent molecular clouds is
larger than could be obtained from simple extrapolation of
the MRN power law ($dn/da \propto a^{-3.5}$) to very small sizes
(Weiland et al.\ 
1986; D\'{e}sert et al.\ 1990; Dwek et al.\ 1997;
Verter et al.\ 2000),
and therefore $D(a)$ must include such very small grains.
The observed infrared emission requires a substantial population of
grains in
the 5 -- $10\Angstrom$ size range so that single-photon heating can
raise them to high enough temperatures to emit in the $3-12\micron$
range (see DL01).  
A log-normal distribution was used
by Draine \& Lazarian (1998) to represent the size distribution of
ultrasmall carbonaceous grains producing both
$3-12\micron$ infrared vibrational emission and microwave rotational emission,
and we adopt such a distribution here.

We will see below (see Figure \ref{fig:HGL}) that the 
$a\gtsim 100\Angstrom$ grains in the present grain
model, with the adopted optical properties of graphite and silicate
grains, produce less emission than observed near $60\micron$ for
$a \gtsim 100\Angstrom$ grains, as the steady-state grain temperatures
$\bar{T}$ are slightly too cool.
This could be because the grain properties may be in error -- perhaps
$C_{\rm abs}$ is too low in the optical and ultraviolet, or too large
in the far-IR, or perhaps the starlight radiation field has
been underestimated.  However, we find that the grain model gives
excellent agreement with the total radiated power (suggesting that
the product of starlight intensity and 
optical-UV absorption cross sections are about right), and excellent
agreement with the $\lambda \gtsim 100\micron$ emission (suggesting that
the far-IR emission cross sections are about right).

As previously suggested by Draine \& Anderson (1985), the shortfall at
$60\micron$ could be made up by emission from a population of very
small grains which single-photon heating can raise to $\sim 40\K$, so
that much of the absorbed energy will be radiated near $60\micron$.
The optimal grain size for this purpose is $a\approx50\Angstrom$.
We use a second log-normal component to add a population
of $a\approx50\Angstrom$ grains to $D(a)$:
\begin{equation}\label{eq:sizedb_pah1}
D(a) = \sum_{i=1}^{2}\frac{B_i}{a} \exp \left\{ - \frac{1}{2} 
       \left[ \frac{\ln (a/a_{0i})}{\sigma} \right]^2 \right\},
       ~~~~~~~a > 3.5 \Angstrom 
\end{equation}
\begin{equation}\label{eq:sizedb_pah2}
B_i = \frac{3}{(2\pi)^{3/2}}\frac{\exp(-4.5\sigma^2)}
      {a_{0i}^3\rho\sigma}\frac{m_{\rm C}\bci}
        {
        \left\{ 1 + {\rm erf}[3 \sigma / \sqrt{2} + 
        \ln (a_{0i} / 3.5 \Angstrom) / \sigma \sqrt{2}] \right\}
        }~~~,~~~
\sum_{i=1}^{2}\bci = \bc,     
\end{equation}
where $m_{\rm C}$ is the mass of a C atom; 
      $\rho=2.24\g\cm^{-3}$ is the mass density of graphite;
      $a_{0i}$ determines the location of the peak,
      $\bci$ is the amount of C atoms 
      relative to H consumed by the $i$-th 
      log-normal component. 
      The width of each of the log-normal components is determined by
      $\sigma$. 
The lower cutoff is set at $a$=3.5\AA\ (corresponding to 
$\numC\approx 20$ for PAHs) since smaller grains are 
photolytically unstable (Omont 1986; Allamandola et al.\ 1989; 
Guhathakurta \& Draine 1989). Component $i$ of $D(a)$
has $a^3dn/d\ln a$ peaking at $a_{{\rm peak},i}=a_{0i}\exp(3\sigma^2)$.

As we will show below, we obtain a good fit to the observed
interstellar emission spectrum with
$b_{\rm C1}=45$ppm, $a_{01}=3.5\Angstrom$,
$b_{\rm C2}=15$ppm, $a_{02}=30\Angstrom$,
and $\sigma=0.4$.
The resulting size distribution, with $a^3 dn/d\ln a$ having three
separate peaks at $\sim 6\Angstrom$, $\sim 50\Angstrom$, and $3000\Angstrom$
(see Figure 2 of WD01a) seems somewhat artificial, but this may be
attributable in part to the procedure used for fitting the UV extinction.

\section{PAH Charging\label{sec:charging}}

Due to their low ionization potential (6--7\eV), 
PAH molecules can be ionized through photoelectric emission 
(Allamandola et al.\ 1985; van der Zwet \& Allamandola 1985; 
Lepp \& Dalgarno 1988; Bakes \& Tielens 1994). 
On the other hand, PAHs may also acquire charge through collisions
with electrons and ions (Draine \& Sutin 1987). The grain charging 
process has been recently reanalyzed by Weingartner \& Draine (2001b);
we use their rates for photoelectric emission and electron capture
to study PAH charging in three idealized interstellar environments 
characterizing the diffuse ISM: cold neutral medium 
(CNM), warm neutral medium (WNM), and warm ionized medium (WIM). 
The parameters adopted in the PAH charge distribution computation 
are displayed in Table \ref{tab:cnmwnmwim}. 

We estimate $\fion(a)$ -- the probability of finding
a PAH molecule of radius $a$ in a non-zero charge state
-- for CNM, WNM, and WIM respectively. 
In Figure 5 we present the PAH ``ionization fraction'' 
$\fion(a)$ as a function of size for
CNM, WNM, and WIM conditions, and in addition the
average $\fion(a)$ for
the mass fractions listed in Table \ref{tab:cnmwnmwim}. 

\begin{table}
\caption{Idealized phases for the diffuse ISM
	\label{tab:cnmwnmwim}.}
\begin{tabular}{c c c c}
\hline \hline
Item & CNM & WNM & WIM \\
\hline
Mass fraction         &0.43   &0.43   &0.14   \nl
$\nH (\cm^{-3})$      &30     &0.4    &0.1    \nl
$T (\K)$              &100    &6000   &8000   \nl
$n_{\rm e}/n_{\rm H}$ &0.0015 &0.1    &0.99   \nl
\hline
\end{tabular}
\end{table}

\begin{figure}			
\epsfig{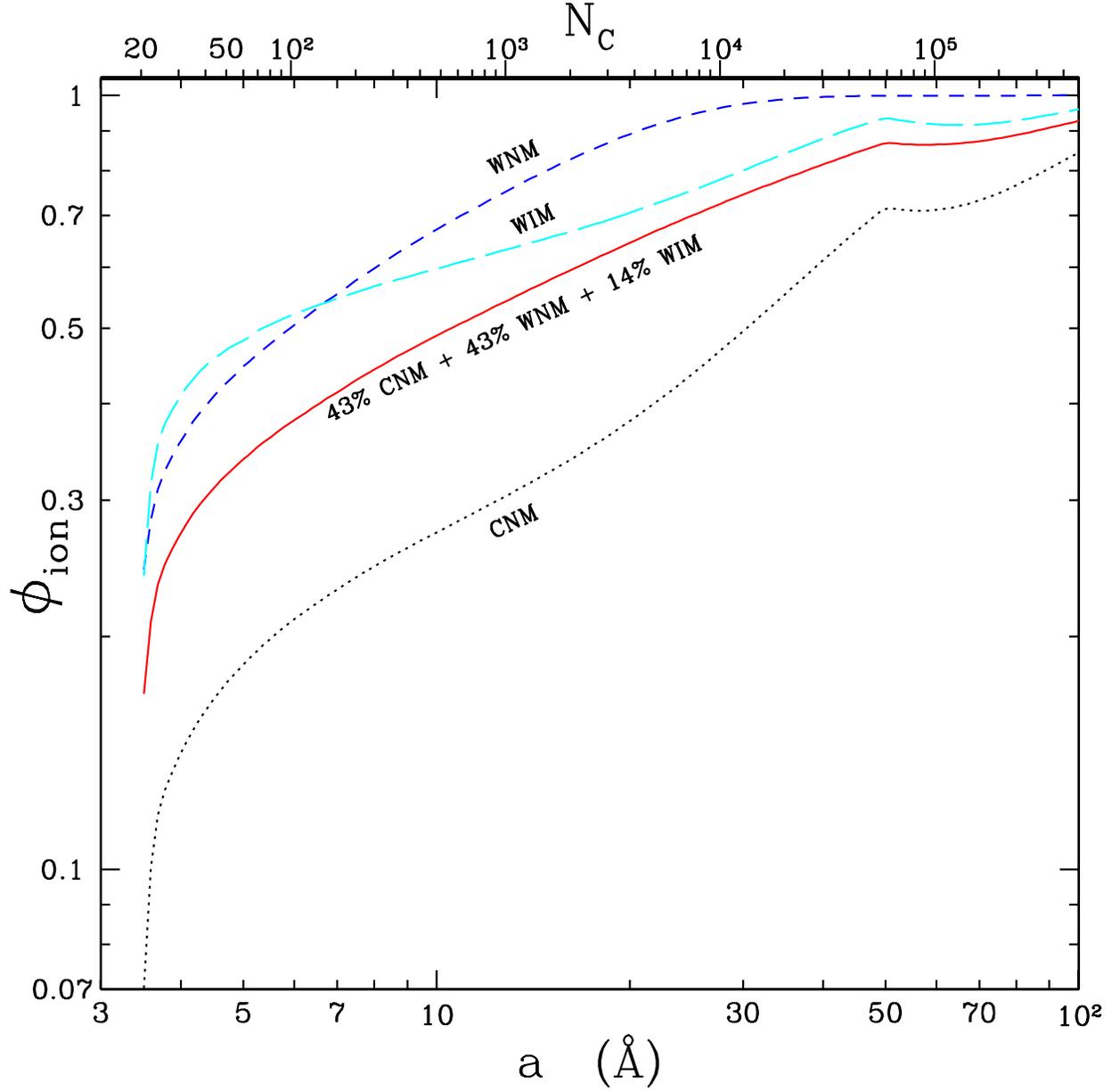}
\caption{
	\footnotesize
        \label{fig:f_ion}
        The ``ionization fraction'' $\fion$ -- the probability
	of finding a PAH molecule in a non-zero charge state --
	as a function of size for
        three idealized phases of the 
        diffuse ISM: CNM (dotted), WNM (short-dashed), 
        and WIM (long-dashed).
        Solid line shows the averaged ionization fraction weighted by 
        the mass fractions of Table \ref{tab:cnmwnmwim}. 
        The slope discontinuity at $a=50$\AA\ is due to the 
        transition from PAH optical properties
        to graphite properties above $a_{\xi}=50\Angstrom$. 
        The upper axis label give the number of carbon atoms
	$\numC = 0.468(a/\Angstrom)^3$.
        }
\end{figure}

\section{Modeling the IR Emission Spectrum\label{sec:ir_em}}

With the absorption cross sections $C_{\rm abs}(a,\lambda)$ 
(\S\ref{sec:optc}), the UV radiation field strength $\chiMMP$ 
(\S\ref{sec:isrf}), the grain size distribution
(\S\ref{sec:sizedb}), the energy 
distribution functions $P(E)$ 
for very small grains (\S\ref{sec:T_spike}), the equilibrium 
temperature $\bar{T}$ for large grains (\S\ref{sec:T_eq}), and the 
ionization fraction $\fion(a)$ of PAHs (\S\ref{sec:charging}),
we can now calculate the IR emission spectrum from 
a mixture of silicate and carbonaceous grains.
The IR emissivity per H nucleon, for dust illuminated by
starlight of strength $\chi$, is
\begin{eqnarray}
\emiss(\chi)
&=& 
  \Bigg\{ \int_{250\Angstrom}^{a_{\rm max}^{\rm sil}}
  da \frac{1}{\nH}\frac{dn_{\rm sil}}{da}
  C_{\rm abs}^{\rm sil}(a,\lambda) B_{\lambda}(\bar{T}_{\rm sil}[a,\chi])
\nonumber
\\
&& + \int^{a_{\rm max}^{\rm gra}}_{250\Angstrom}
  da \frac{1}{\nH}\frac{dn_{\rm carb}}{da}
  C_{\rm abs}^{\rm gra}(a,\lambda) B_{\lambda}(\bar{T}_{\rm gra}[a,\chi]) 
\nonumber
\\
&& + \int^{250\Angstrom}_{a_{\rm min}}
  da \frac{1}{\nH}\frac{dn_{\rm sil}}{da}
  \int^{\infty}_{hc/\lambda} dE \left(\frac{dP}{dE}\right)_{{\rm sil}}
   C_{\rm abs}^{\rm sil}(a,\lambda) 
   B_{\lambda}(T_{\rm sil}[E,a]) 
\nonumber
\\
&~& + \int^{250\Angstrom}_{a_{\rm min}} 
   da \frac{1}{\nH}\frac{dn_{\rm carb}}{da}
   \int^{\infty}_{hc/\lambda} dE 
	\Big[
   \fion \left(\frac{dP}{dE}\right)_{\rm carb^{+}}
   C_{\rm abs}^{\rm carb^+}(a,\lambda) B_{\lambda}(T_{\rm carb^+}[E,a])
\nonumber
\\
&&  + (1-\fion) \left(\frac{dP}{dE}\right)_{\rm carb^{0}}
   C_{\rm abs}^{\rm carb^0}(a,\lambda) B_{\lambda}(T_{\rm carb^0}[E,a])
	\Big]
	\Bigg\}
   \times   \left[1 + \frac{u_{\lambda}}{8\pi hc/\lambda^5} \right] ~,
\label{eq:calc_irem}
\end{eqnarray}
where $C_{\rm abs}^{\rm sil}$, $C_{\rm abs}^{\rm gra}$, 
$C_{\rm abs}^{\rm carb^+}$, and $C_{\rm abs}^{\rm carb^0}$
are the absorption cross sections for silicate grains, graphitic grains,
and neutral and ionized carbonaceous grains, respectively; 
$\bar{T}_{\rm sil}$, $\bar{T}_{\rm gra}$ are the equilibrium temperatures 
of silicate and graphite grains, respectively; $dP(E)$ is the probability 
that the grain vibrational energy will be in $[E, E+dE]$;
and $T_{\rm x}(E,a)$ is the effective vibrational temperature for 
a grain with composition ${\rm x}$, vibrational energy $E$, and 
radius $a$ (see DL01). The first term on the right represents the 
contribution of big silicate grains ($a\ge 250$\AA; hereafter 
``${\rm B_{sil}}$''); the second term represents the big 
carbonaceous grain component ($a\ge 250$\AA; hereafter 
``${\rm B_{carb}}$''); the third term represents the 
very small silicate component (3.5\AA $\le a <$ 250\AA; 
hereafter ``${\rm S_{sil}}$''); the fourth and fifth terms 
represent the charged and neutral very small carbonaceous 
grains (including PAHs) for 3.5\AA $\le a <$ 250\AA; 
hereafter ``${\rm S_{carb}}$''.
The $u_{\lambda}/(8\pi hc/\lambda^5)$ factor is 
a (usually negligible) correction for stimulated emission.
In regions where dust is optically thin to its own thermal 
emission, the resulting IR intensity is simply
\begin{equation}
\label{eq:I_thin_dust}
(I_\lambda)_{\rm dust} = \emiss(\chi)	\times \NH
\end{equation}
where $\NH$ is hydrogen column density. 
Dust self-absorption should be taken into account  
in regions with large $\NH$:
\begin{equation}
\label{eq:I_dust}
(I_\lambda)_{\rm dust} =
\emiss(\chi)	\times 
	\frac{1-{\rm exp}\left[-\NH \Sigma_{\rm abs}(\lambda)\right]}
        {\Sigma_{\rm abs}(\lambda)}
\end{equation}
where $\Sigma_{\rm abs}(\lambda)$ is the total 
absorption\footnote{
	If the radiation field in the emitting region is
	approximately isotropic, scattering does not alter the intensity.}
cross section per H nucleon for the dust model.

\section{Model Emission Spectra and Observations\label{sec:results}}

\subsection{Overview of Model Parameters\label{sec:parameters}}

If we choose a value for $\sigma$ (we take $\sigma=0.4$), then
the grain model has a total of 18 adjustable 
parameters: 
\begin{enumerate}
\item $\qcont$, determining the PAH near-IR continuum 
   (1--5$\mu$m) opacity; we take $\qcont=0.01$ 
   (see \S\ref{sec:optc_gra}, \S\ref{sec:pahircont});
\item $a_{\xi}$, determining the transition 
   from PAH optical properties to graphite properties;
   we take $a_\xi=50\Angstrom$ 
   (see \S\ref{sec:optc_gra});
\item $a_{\rm t,sil}$, $a_{\rm c,sil}$, $\alpha_{\rm sil}$, 
   $\beta_{\rm sil}$, $C_{\rm sil}$, $a_{\rm t,carb}$, $a_{\rm c,carb}$, 
   $\alpha_{\rm carb}$, $\beta_{\rm carb}$, $C_{\rm carb}$,
   characterizing the size distributions and abundances of silicate 
   and carbonaceous grains 
   (\S\ref{sec:sizedb});
\item $a_{01}$, $\bcone$, $a_{02}$, 
   $\bctwo$ ($=\bc-\bcone$), characterizing
   the size distributions and abundances of the two log-normal
   components of very small carbonaceous grains (\S\ref{sec:sizedb});
\item $\chiMMP$, the starlight intensity (\S\ref{sec:isrf});
\item $\NH$, the total column density of material.
\end{enumerate}
In addition, the spectrum depends on the ionized fraction 
$\fion(a)$ for PAHs, determined by $\chiMMP/n_e$ ($n_e$ is 
the electron number density) and the gas temperature $T$.

For the UV properties of PAHs adopted here, WD01 show that
the observed interstellar extinction puts an 
upper limit on $\bc=\bcone+\bctwo\ltsim60~$ppm
for the average extinction in diffuse regions (WD01).
For a given choice of $\bc\leq60~$ppm,
ten of the parameters ($a_{\rm t,sil}$, $a_{\rm c,sil}$, 
$\alpha_{\rm sil}$, $\beta_{\rm sil}$, $C_{\rm sil}$, 
$a_{\rm t,carb}$, $a_{\rm c,carb}$, $\alpha_{\rm carb}$, 
$\beta_{\rm carb}$, $C_{\rm carb}$) are determined by fitting
the interstellar extinction curve
(see WD01a). 

The strength of the 5--12$\mu$m IRTS spectrum
relative to the total far infrared emission in the MIRS region
requires a population of PAHs with $\bcone\approx 45\,$ppm
with a mass distribution peaking near $\sim6\Angstrom$.
Thus $a_{01}\approx 6\Angstrom \exp(-3\sigma^2)$; adopting
$\sigma=0.4$ gives $a_{01}\approx 3.5\Angstrom$.

For the high galactic latitude dust we take $\chiMMP=1$, since the
starlight heating this nearby, optically-thin diffuse material should
be very close to the local ISRF estimate of Mathis, Mezger, 
\& Panagia (1983).

The dust at high galactic latitudes has substantially more 60$\micron$
emission than would be produced by graphite and silicate grains
at their ``thermal equilibrium'' temperatures $\bar{T}$.
Within the context of the graphite-silicate model, we can account for
this extra emission with a population of
very small carbonaceous grains which undergo single-photon heating 
to $T\approx 40\K$, which corresponds to
$a\approx50\Angstrom$ grains.
We therefore 
add $\bctwo\approx15$ppm 
in a population of grains with 
$a_{02}\approx50\Angstrom\exp(-3\sigma^2)\approx30\Angstrom$. 

This completes specification of the properties of our
``standard'' grain model,
with parameters summarized in 
Table \ref{tab:model_parameters}.
To compute the emission, it remains only to specify the intensity 
of the starlight heating the grains, and the total column density $\NH$.

\begin{table}
\caption[]{Parameters for standard dust model for $R_V=3.1$
           \label{tab:model_parameters}}
\begin{tabular}{ccc}
\hline \hline
PAHs & carbonaceous\tablenotemark{a} & silicate\tablenotemark{a} \\
\hline
$\bcone$=45ppm, $a_{01}$=3.5\AA~ ($\sigma$=0.4) 
	&$a_{\rm t,carb}=0.0107\mu$m
	&$a_{\rm t,sil}=0.164\mu$m \\

$\bctwo$=15ppm, $a_{02}$=30\AA~ ($\sigma$=0.4) 
	&$a_{\rm c,carb}=0.428\mu$m
	&$a_{\rm c,sil}=0.100\mu$m \\ 

$a_{\xi}$=50\AA, $q_{\rm gra}=0.01$ 
	&$\alpha_{\rm carb}=-1.54$
	&$\alpha_{\rm sil}=-2.21$\\

$E_{6.2}=3$, $E_{7.7}=E_{8.6}=2$\tablenotemark{b}
	&$\beta_{\rm carb}=-0.165$
	&$\beta_{\rm sil}=0.30$ \\

H/C from eq.\ (\ref{eq:nh2nc})
	& $C_{\rm carb}=9.99\times 10^{-12}$
	& $C_{\rm sil}=1.00\times 10^{-13}$ \\

\hline
\end{tabular}
\tablenotetext{a}{Taken from Weingartner \& Draine (2001a).}
\tablenotetext{b}{See Table \ref{tab:drude_parameters}.}
\end{table}

\subsection{Observational Constraints\label{sec:obs_constraints}}

We will compare our model results with observational data for 
3 regions: the average emission per H observed by 
DIRBE (Arendt et al.\ 1998) and 
FIRAS (Finkbeiner et al.\ 1999) for 
high-Galactic-latitude regions; 
the DIRBE flux and the 4.5--11.7$\mu$m IRTS spectrum 
(Onaka et al.\ 1996) for the ``MIRS'' region;
and the DIRBE flux and the 2.8--3.9$\mu$m
IRTS spectrum (Tanaka et al.\ 1996) for the ``NIRS'' region.
For the MIRS and NIRS regions the DIRBE fluxes were obtained
from the DIRBE data set, obtained from the National Space Science
Data Center (NSSDC).
In Table \ref{tab:obs_data} we list the DIRBE fluxes for these
three regions.
 
\begin{table}
\caption[]{DIRBE results for high galactic latitudes and NIRS and MIRS regions
\label{tab:obs_data}.}
\begin{tabular}{cccccccccc}
\hline \hline
Item
	& 2.2$\micron$ & 3.5$\micron$ & 4.9$\micron$ & 12$\micron$ & 25$\micron$ & 60$\micron$ & 100$\micron$ & 140$\micron$ & 240$\micron$ \\
\hline
$\lambda_{\rm effective}$ ($\micron$)
	& 2.22 & 3.53 & 4.88 & 12.3 & 20.8 & 56.0 & 97.7 & 148 & 248 \\
FWHM ($\micron$)
	& 0.37 & 0.97 & 0.67 &  8.1 &  8.9 & 27.3 & 31.0 &  35 & 101 \\
HGL\tablenotemark{a}       
	& - & 0.97 & 1.11 & 7.16 & 3.57 & 5.30 & 18.6 
                   & 22.5 & 10.1 \\ 
MIRS region\tablenotemark{b}
	& 6.65 & 3.47 & 1.68 & 3.79
               & 2.10 & 4.85 & 11.5 & 19.4 & 8.20\\
NIRS region\tablenotemark{c}  
	& 7.13 & 3.47 & 1.55 & 3.01 
               & 1.58 & 3.63 & 9.01 & 14.9 & 6.27\\ 
\hline
\end{tabular}
\tablenotetext{a}{$\Delta (\lambda I_\lambda)/\Delta \NH$ 
	for $|b|\simgt 25^{\rm o}$ 
             ($10^{-26} \erg \s^{-1} \sr^{-1} {\rm H}^{-1}$)
	(Arendt et al.\ 1998).}  
\tablenotetext{b}{$\lambda I_\lambda$ 
		($10^{-3} \erg \s^{-1} \cm^{-2} \sr^{-1}$)
		for $44^{\rm o}\le l \le 44^{\rm o}40^{'}$, 
		$-0^{\rm o}40^{'}\le b \le 0^{\rm o}$
		(Hauser et al.\ 1998).}
\tablenotetext{c}{$\lambda I_\lambda$
		($10^{-3} \erg \s^{-1} \cm^{-2} \sr^{-1}$)
		for $47^{\rm o}30^{'}\le l \le 48^{\rm o}$, 
		$|b|\le 15^{'}$	(Hauser et al.\ 1998).}
\end{table}

\subsection{Model Spectra: High-Galactic-Latitude Emission
		\label{sec:model_spectra:HGL}}

The high galactic latitude regions are optically thin to starlight, 
so the starlight illuminating the grains is approximately uniform,
with intensity $\chiMMP\approx 1$.
We adopt our standard grain model, with parameters from 
Table \ref{tab:model_parameters}.

\begin{figure}			
\epsfig{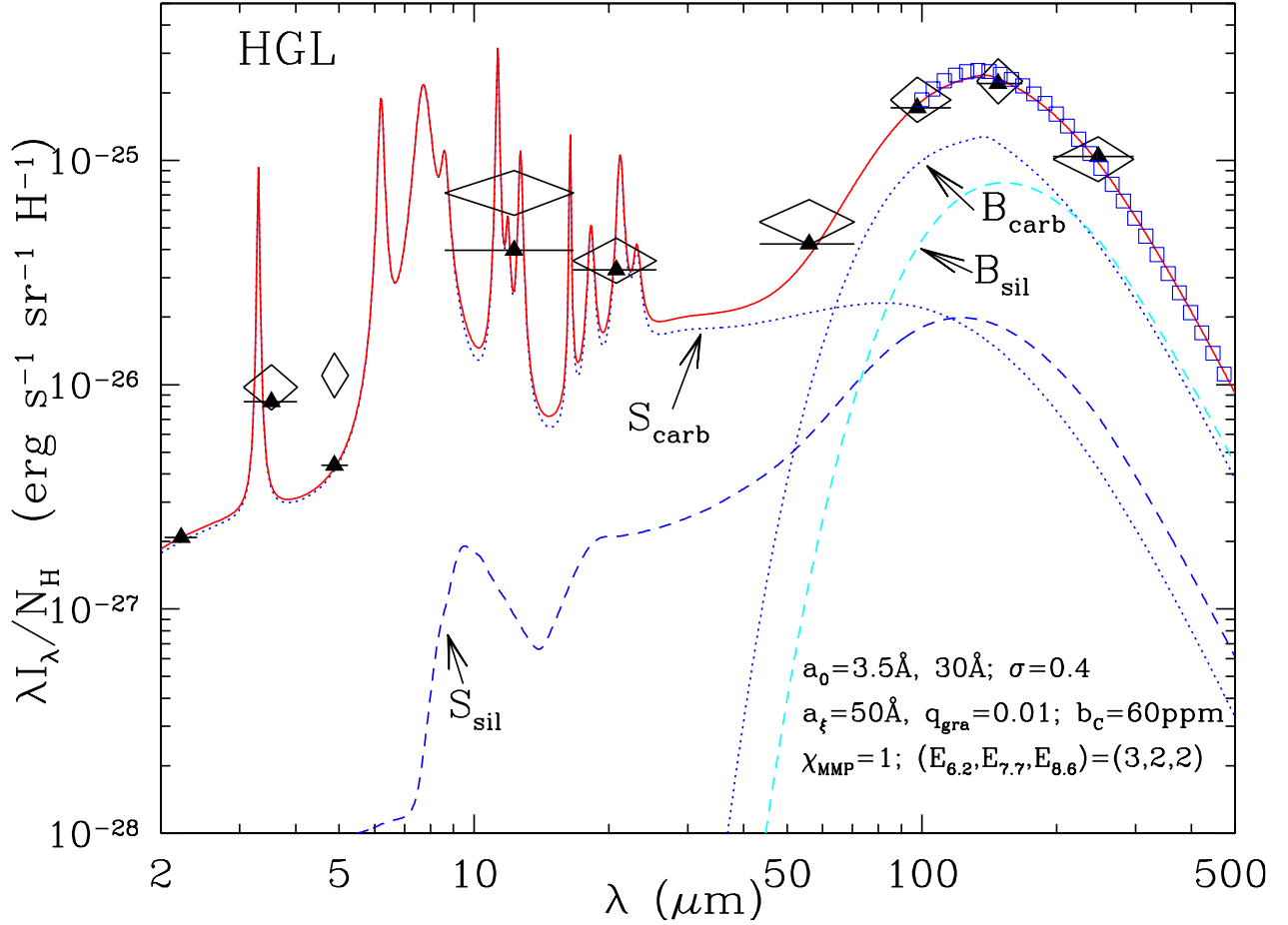}
\caption{
	\footnotesize
        \label{fig:HGL}
        Comparison of the model to the observed
        emission from the diffuse ISM at
        high galactic latitudes ($|b|\ge 25^{\rm o}$).
	Curves labelled B$_{\rm sil}$ and B$_{\rm carb}$ show emission
	from ``big'' ($a\geq 250\Angstrom$) silicate and carbonaceous 
	grains; curves labelled S$_{\rm sil}$ and S$_{\rm carb}$ show
	emission from ``small'' ($a<250$\Angstrom) silicate and
	carbonaceous grains (including PAHs). 
        Triangles show the model spectrum (solid curve)
	convolved with the DIRBE filters.
        Observational data are from 
        DIRBE (diamonds; Arendt et al.\ 1998), 
	and FIRAS (squares; Finkbeiner et al.\ 1999).
        }
\end{figure}

Arendt et al.\ (1998) have determined $\Delta I_\lambda/\Delta \NH$, 
which we compare with $\emiss(\chi)$ from eq.\ (\ref{eq:I_thin_dust}). 
In Figure \ref{fig:HGL} we show 
$\lambda I_\lambda/\NH$ for our dust model for $\chiMMP=1$.
The model is in excellent agreement with the observed
dust emission per H nucleon for $\lambda \geq 100\micron$.
At $60\micron$ the model emission is $\sim18\%$ low, which is 
probably within the uncertainties in the DIRBE fluxes\footnote{
	Dwek et al.\ [1997] have estimated that the DIRBE fluxes have
	1-$\sigma$ uncertainties of $\pm20\%$.}
and the uncertainties in subtraction
of zodiacal emission in high galactic latitude regions (Arendt et al.\ 1998).
There is a significant discrepancy between model and observation at
12$\micron$, where the model flux is only 55\% of the observed flux.
In view of the difficulties in 
correction for zodiacal emission, which is relatively strong at 12$\micron$,
the shortfall at 12$\micron$ may not be real.
However, agreement at 12$\micron$ could be improved 
if the PAH ionization fraction
$\fion(a)$ were reduced for $6\ltsim a \ltsim 10\Angstrom$,
since in this size range the strong $7.7\micron$ feature in
ionized PAHs radiates away energy at the expense 
of the 11.3, 11.9, and 12.7$\micron$ C-H out-of-plane bending modes.
Unfortunately, there are as yet no observational data regarding the strength
of the 6.2, 7.7, and 8.6$\micron$ emission from high galactic latitudes.

The model emission also appears to fall short at 5$\micron$,
where the model flux is only 40\% of the observed value.
At these shorter wavelengths correction for zodiacal emission 
is increasingly difficult. However, if the reported 
$\Delta I_\lambda/\Delta N_{\rm H}$ is 
correct it would require additional opacity at this wavelength for the 
ultrasmall grain component. Such additional opacity could, for 
example, be contributed by carbon chain molecules with a carbon 
abundance of $\simlt 1$ppm (Allamandola et al.\ 1999b).

\begin{figure}			
\epsfig{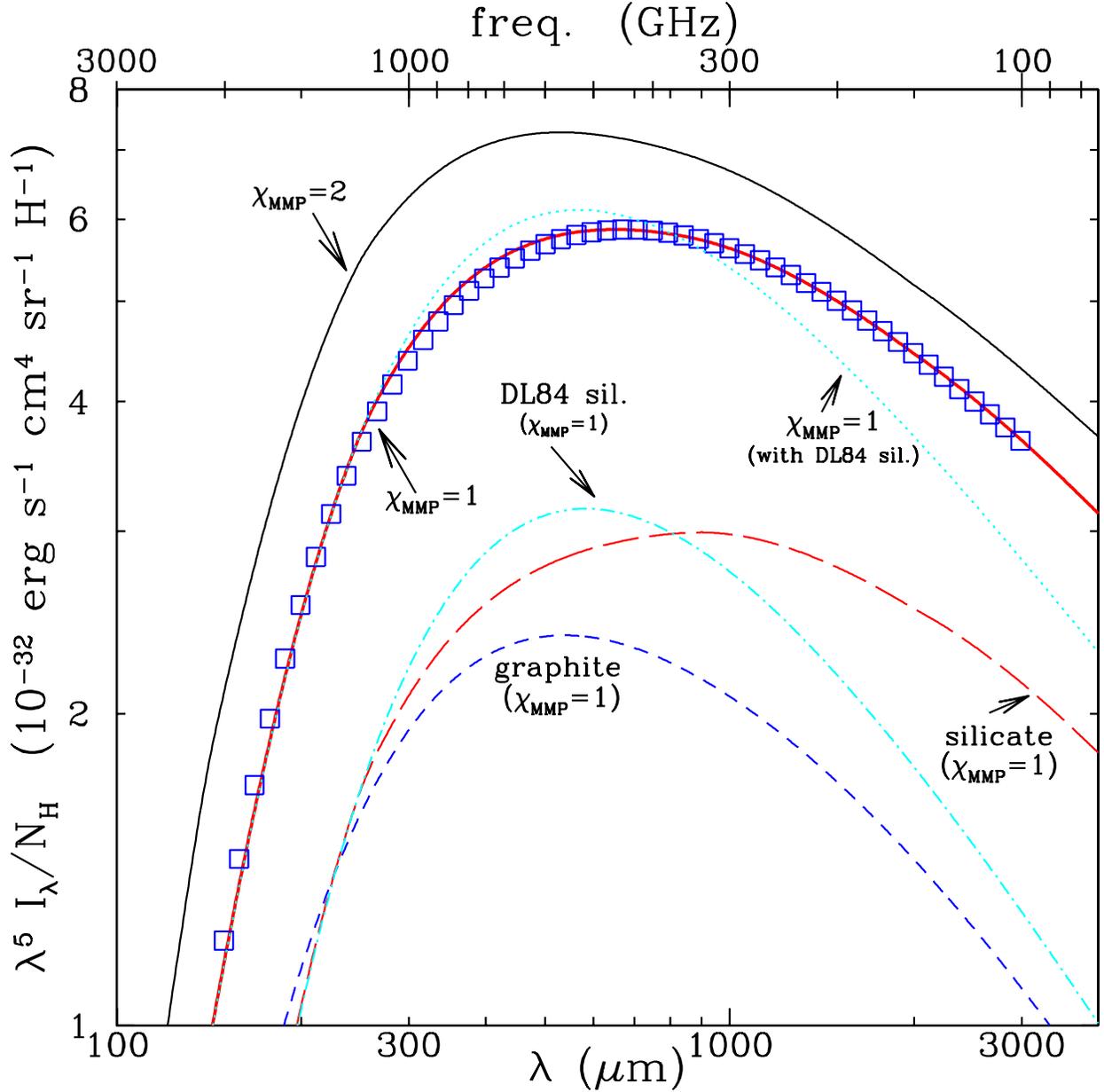}
\caption{
	\footnotesize
        \label{fig:far_IR}
        The far-IR spectrum per H nucleon.
	Squares are the FIRAS data represented by a ``two-temperature'' 
	fit (Finkbeiner et al.\ 1999). 
	Dotted line shows model spectra for $\chiMMP=1$, calculated using
	the silicate dielectric function from DL84.
	Solid lines are model spectra for $\chiMMP=1$ and $2$, labelled
	by the value of $\chiMMP$, calculated using the modified silicate
	dielectric function from the present work (see \S\ref{sec:optc_sil}).
	The model for $\chiMMP=1$ closely reproduces the observed
	FIRAS emission	(as discussed in \S\ref{sec:optc_sil}, the 
	silicate dielectric
	function at $\lambda >250\mu$m has been modified slightly to
	improve the agreement).
	Broken lines show the contribution of carbonaceous grains and
	silicate grains to the emission for $\chiMMP=1$
	(dot-dashed line shows emission calculated using DL84 silicate
	dielectric function).
        }
\end{figure}

The far-IR--submillimeter emission from dust at high galactic latitudes
has been measured by FIRAS (Wright et al.\ 1991; Reach et al.\ 1995; 
Finkbeiner et al.\ 1999). In Figure \ref{fig:far_IR} we show the 
average FIRAS spectrum from Finkbeiner et al.\ (1999) together with 
the far-IR emission calculated for our grain model, for grains 
illuminated by $\chiMMP=1$. Even using the original DL84 dielectric
functions for graphite and silicate, the model is in very good quantitative
agreement with the FIRAS measurements for $\lambda < 1100\micron$.
The agreement is improved if the silicate emissivity is
modified for $\lambda>250\micron$, as described in 
\S\ref{sec:optc_sil}.
The resulting emission spectrum 
is then in excellent agreement with the observations.
Finkbeiner et al.\ (1999) had argued that the FIRAS spectrum was 
indicative of two types of dust grains, one of which was quite cold.
However, we see here that the ``graphite-silicate'' grain model,
with the silicate emission modified by no more than $\pm12\%$ for
$\lambda < 1100\micron$, gives essentially perfect agreement with
the FIRAS spectra.

\subsection{Model Spectra: Galactic Plane Regions
	\label{sec:model_spectra:IRTS}}

The DIRBE short wavelength channels are 
dominated by starlight. We model the DIRBE spectra by taking
\begin{equation}
\label{eq:stars_plus_dust}
I_\lambda = (I_\lambda)_{\rm dust} +
A 
\left[\sum_{i=2}^4 W_i B_\lambda(T_i)\right] 
\frac{1-{\rm exp}\left[-\NH \Sigma_{\rm abs}(\lambda)\right]}
{\Sigma_{\rm abs}(\lambda)},
\end{equation}
where $(I_\lambda)_{\rm dust}$ is
obtained from eq.\ (\ref{eq:I_dust}),
the multiplicative factor $A$ determines the strength of the
starlight, and the $W_i$ and $T_i$ are the same as for the ISRF
(see \S\ref{sec:isrf}).
Equation (\ref{eq:stars_plus_dust}) would be exact if the stellar
density was proportional to the dust density and the starlight was 
isotropic at all points along the line of sight 
(so that scattering does not alter the intensity).
While not expected to be highly accurate, this approximation should give
a reasonable approximation to the overall spectrum.

We adopt our ``standard'' grain model (see
Table \ref{tab:model_parameters}).
This assumes that the dust in the 6 kpc molecular ring (where
most of the interstellar matter in the MIRS and NIRS fields is located)
has the same size distribution as dust in the local diffuse ISM;
this assumption is likely to be incorrect in detail, but is the
best we can do at this time.

\begin{figure}			
\epsfig{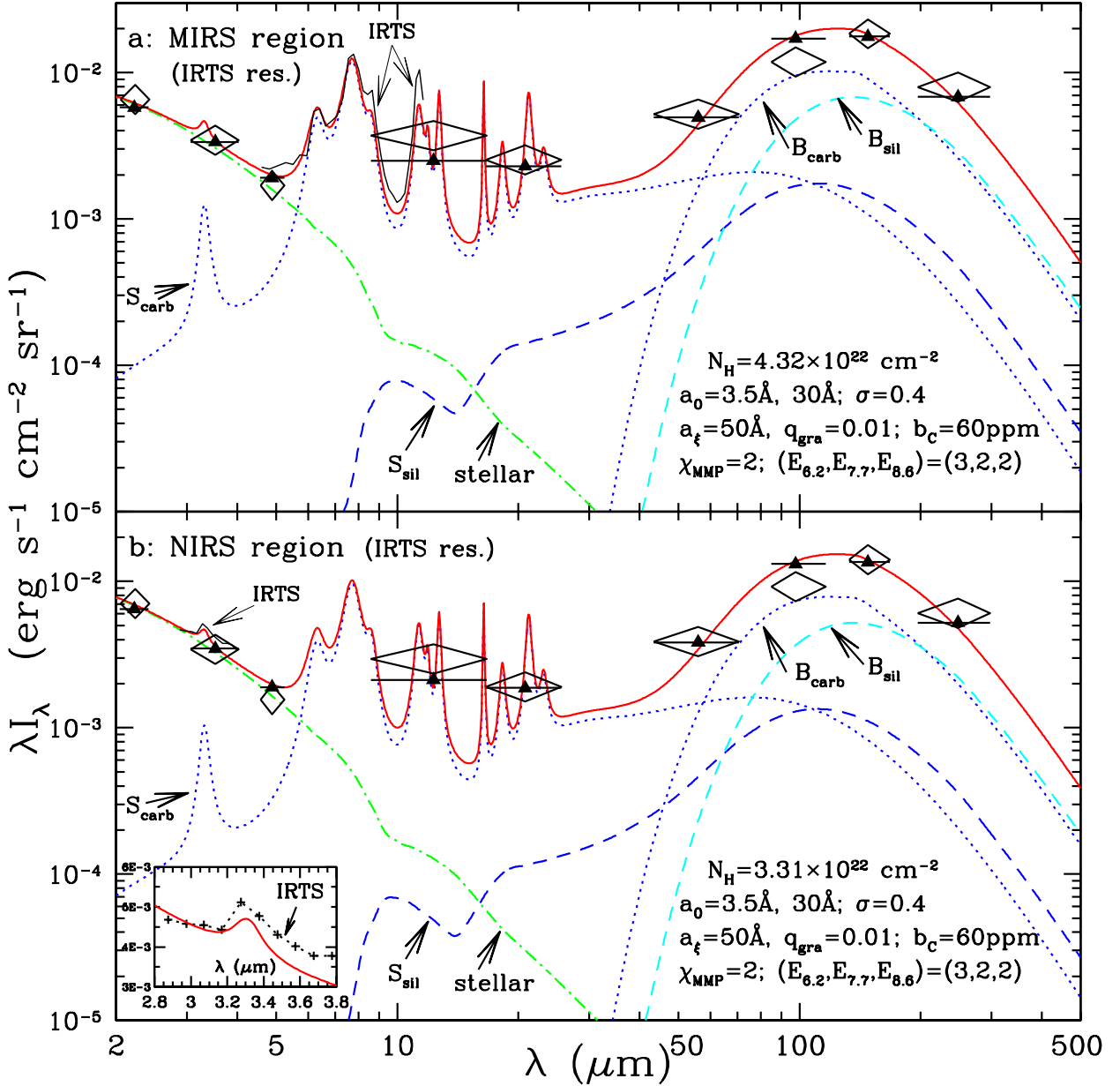}
\caption{
	\footnotesize
        \label{fig:irts_chi2}
	Infrared emission from dust plus 
	starlight (see text) for two regions
	in the Galactic plane: (a) the MIRS region
        ($44^{\rm o}\le l \le 44^{\rm o}40^\prime$, 
        $-0^{\rm o}40^\prime\le b \le 0^{\rm o}$), and
	(b) the NIRS region
	($47^{\rm o}30^\prime\le l \le 48^{\rm o}$, $|b|\le 15^\prime$).
	For both regions we use our standard dust parameters,
	with the PAH ionization fraction $\fion(a)$ taken to be
	the weighted mean for the local ISM
	(43\% CNM, 43\% WNM, 14\% WIM).
	The starlight intensity heating the dust 
	has been taken to be $\chiMMP=2$.
	The solid curve shows the overall model spectrum, degraded to
	the IRTS resolution; triangles show the model 
        spectrum convolved with the DIRBE filters.
	DIRBE observations are shown as diamonds.
	For the MIRS field we show the IRTS MIRS 5--12$\micron$ spectrum 
	(thin solid line).
	For the NIRS field we show the IRTS NIRS 2.8-3.9$\micron$ spectrum
	(thin solid line, also shown as cross-dotted curve in inset).
	Curves labelled B$_{\rm sil}$ and B$_{\rm carb}$ show emission
	from ``big'' ($a\geq 250\Angstrom$) silicate and carbonaceous 
	grains; curves labelled S$_{\rm sil}$ and S$_{\rm carb}$ show
	emission from ``small'' ($a<250\Angstrom$) silicate and
	carbonaceous grains (including PAHs).
        }
\end{figure}

\subsubsection{Single-$\chi$ Models 
	\label{sec:single-chi}}

The grains in the MIRS and NIRS Galactic plane regions
will be illuminated by a range of radiation fields, since each
$0.5^\circ\!\times\!0.5^\circ$ field (52~pc$\times$52~pc @ 6 kpc) will include
diffuse clouds, molecular clouds and star-forming regions.
The simplest approximation is to assume that the dust is 
uniformly distributed in the beam and heated by a uniform 
radiation field with intensity $\chiMMP$ relative to the 
local ISRF. For a given $\chiMMP$, the column density $\NH$ 
is adjusted to match the DIRBE 140$\micron$ intensity.
The constant $A$ in eq.\ (\ref{eq:stars_plus_dust})
determining the strength of the stellar contribution
has been adjusted so that the spectrum agrees with
the DIRBE 3.5$\micron$ measurement. We find that the 
$\chiMMP$=2 model provides a rough overall fit to the 
IRTS spectra and DIRBE photometry. 

In Figure \ref{fig:irts_chi2} we show the emission spectrum 
calculated for the ``standard'' dust model, for grains heated 
by $\chiMMP=2$, for both the MIRS and NIRS regions. Note that 
the model spectra have been smoothed to the resolution 
of the NIRS and MIRS instruments on IRTS for comparison with the 
IRTS spectra. Also note that the column densities $\NH$ are 
very large, corresponding to $A_V = 23$ and 17 mag for the MIRS 
and NIRS regions, respectively. These column densities are large 
enough so that self-absorption is significant at 
$\lambda\ltsim 15\micron$. In particular, the 3.3,
6.2, and 11--12$\micron$ features are noticeably attenuated 
by the dust absorption.

This simple model provides fair agreement with the DIRBE observations 
of these two fields. The largest discrepancy is at $100\micron$, where 
the observed flux is only 70\% of the model value.

The model is in reasonably good agreement with the IRTS 5--12$\micron$
spectrum in the MIRS region. 
The model is $\sim$40\% below 
the 11.3$\micron$ peak observed by IRTS, 
and $\sim$32\% below the DIRBE 12$\micron$ measurement, but
note that 
1) the claimed absolute calibration accuracy of the MIRS instrument is
only $\pm30\%$ (Onaka et al.\ 1996),
2) the DIRBE photometric accuracy is estimated to be $\pm20\%$
(Dwek et al.\ 1997).

The model is also in reasonably good agreement with the IRTS NIRS 
2.8--3.9$\micron$ spectrum in the NIRS region. Recall that the 
stellar contribution has been adjusted to match the DIRBE 3.5$\micron$
photometry; with this choice of stellar continuum, the model falls 
$\simlt$15\% below the NIRS spectrophotometry for $\lambda<3.7\mu$m.
This agreement is satisfactory, since the difference of the overall 
slope of the 2.8--3.9$\micron$ NIRS spectrum and the slope indicated 
by the DIRBE 2.2, 3.5, and 4.9$\micron$ photometry suggests that the 
uncertainties in the NIRS spectrophotometry may exceed the claimed 
$\pm5\%$ absolute calibration uncertainty (Tanaka et al.\ 1996).
The model prediction for the strength of the 3.3$\micron$ emission feature
is in good agreement with the NIRS photometry 
(see the inset in Figure \ref{fig:irts_chi2}b).

In the context of the single-$\chi$ model, dust models with 
either weaker or stronger radiation fields are less successful 
than the $\chiMMP$=2 model in reproducing the observed spectra.
The $\chiMMP$=1 model does not emit sufficiently at 5--12$\mu$m
due to self-absorption by the very large column density of dust; 
on the other hand, the $\chiMMP$=3 model, although providing 
a reasonably good fit to the 5--12$\mu$m spectrum, emits too much 
(by $\simgt 75\%$) at 100$\mu$m.

\subsubsection{Multi-$\chi$ Models \label{sec:multichi}}

\begin{figure}			
\epsfig{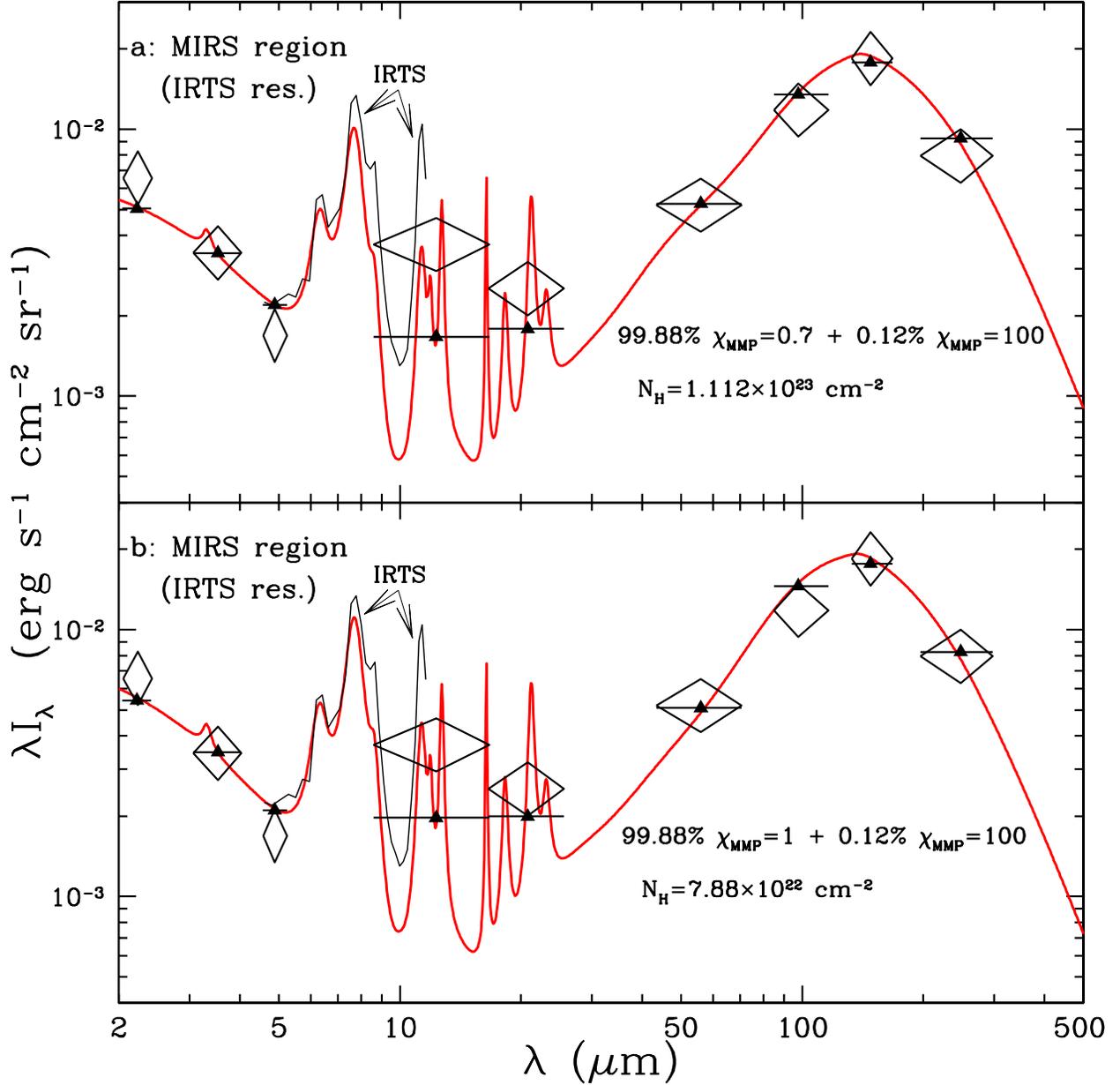}
\caption{\footnotesize
	\label{fig:irts_multi}
	Same as Figure \ref{fig:irts_chi2}a, but for multi-$\chi$ models.
	Upper panel: a model where the far-IR emission is provided
	by dust with $\chiMMP=0.7$, with additional emission from dust
	heated by $\chiMMP=100$.  While providing good agreement with
	the 60--240$\micron$ DIRBE fluxes, the model is a factor of
	two low in the 12$\micron$ region, partly due to the absorption
	contributed by the very large column density of $\chiMMP=0.7$ dust.
	Lower panel: a model where the far-IR emission is provided by
	dust heated by $\chiMMP=1$, with additional emission from dust
	with $\chiMMP=100$. There is less dust absorption,
        but the model still falls short in 
        the 12$\micron$ region.
	}
\end{figure}

The overprediction of the $100\micron$ flux in 
Figure \ref{fig:irts_chi2} could be due in large part to the
simplification of assuming that all of the dust is heated by
a single radiation field. Consider instead a model where
a fraction $f_j$ of the dust is heated by starlight with 
intensity $\chi_j$. If the regions of different $\chi_j$ are 
randomly located on scales small compared to the beamsize, 
then we may approximate the different dust emissivities as 
uniformly mixed, so that the emergent intensity becomes
\begin{equation}
\label{eq:I_dust,multi}
(I_\lambda)_{\rm dust} =
\left[\sum_{j} f_j \emiss(\chi_{j})\right]
\frac{1-{\rm exp}\left[-\NH \Sigma_{\rm abs}(\lambda)\right]}
{\Sigma_{\rm abs}(\lambda)} ~.
\end{equation}
The $\chiMMP=2$ model of Figure \ref{fig:irts_chi2} exceeds the 
observed 100$\micron$ emission by about 40\%. This might indicate
that the dust used to reproduce the 140$\micron$ 
emission is somewhat too hot. We can try to 
remedy this by using cold dust to provide the 140 and 240$\micron$
emission, plus additional warmer dust to provide sufficient 
60$\micron$ emission. Two examples of multi-$\chi$ models for 
the MIRS region are shown here. In Figure \ref{fig:irts_multi}a
we show a multi-$\chi$ model where the cooler dust is heated by
$\chiMMP=0.7$, and the warmer dust by $\chiMMP=100$.
This model reproduces the 60, 100, 140, and 240$\micron$ photometry,
but the mid-IR absorption by the large amount of cool
dust causes the 11--12$\micron$ emission to be too low by more than
a factor of two. If the cooler dust is taken to be heated by 
$\chiMMP=1$, we obtain Figure \ref{fig:irts_multi}b.  This model 
now exceeds the observed 100$\micron$ flux by $\sim$20\%. 
The total dust column density has been reduced by $\sim$30\% 
compared to Figure \ref{fig:irts_multi}a, but is still $\sim$80\% 
greater than in the single-$\chi$ model of Figure \ref{fig:irts_chi2}a --
there is sufficient absorption at 12$\micron$ that this model falls 
short of the observed flux by $\sim 30\%$.
We have also carried out calculations for models 
combining three or more $\chi$ values
and found no significant improvement.
We conclude that multi-$\chi$ models offer no significant improvement
over the simple single-$\chi$ model of \S\ref{sec:single-chi}.

\subsection{Emission Spectra for Different Starlight Intensities
	\label{sec:vary_starlight}}

\begin{figure}			
\epsfig{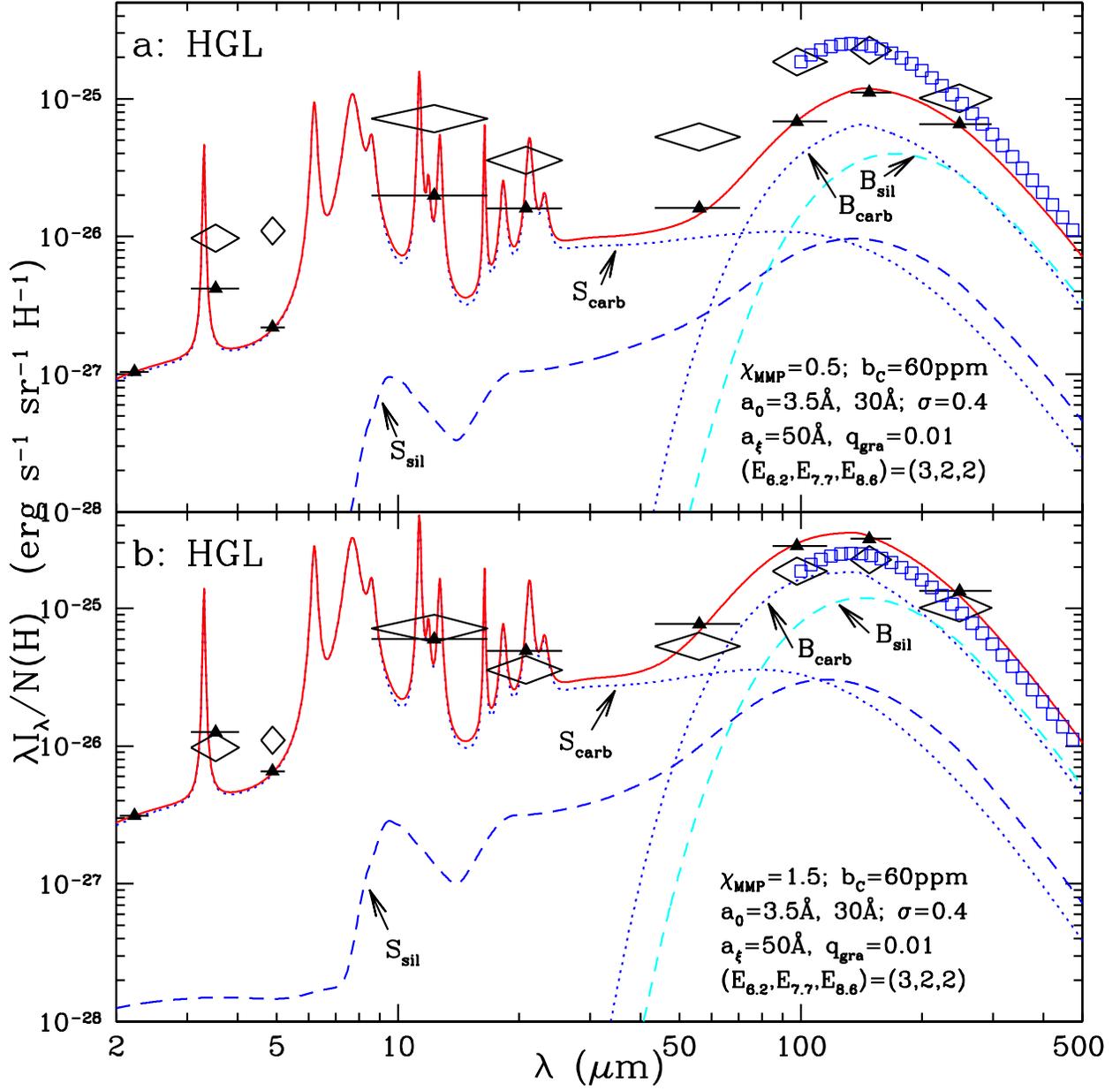}
\caption{
	\footnotesize
	\label{fig:HGL0515}
	As in Figure \ref{fig:HGL}, but for
	dust illuminated by 
	starlight intensities $\chiMMP=0.5$ (upper panel)
	or $\chiMMP=1.5$ (lower panel).
	}
\end{figure}

In \S\ref{sec:model_spectra:HGL} we have shown models for 
optically-thin emission from our standard model heated 
by starlight with $\chiMMP=1$ for the high-latitude dust.
To study the sensitivity to the assumed starlight intensity,
in Figure \ref{fig:HGL0515}
we show spectra for $\chiMMP=0.5$ and 1.5. 
As expected from Figure \ref{fig:HGL}, the $\chiMMP=0.5$ 
model yields too little emission (by $\approx 50\%$) over 
the entire spectral range; the $\chiMMP$=1.5 model emits 
too much at $\lambda\simgt 50\mu$m.

\begin{figure}			
\epsfig{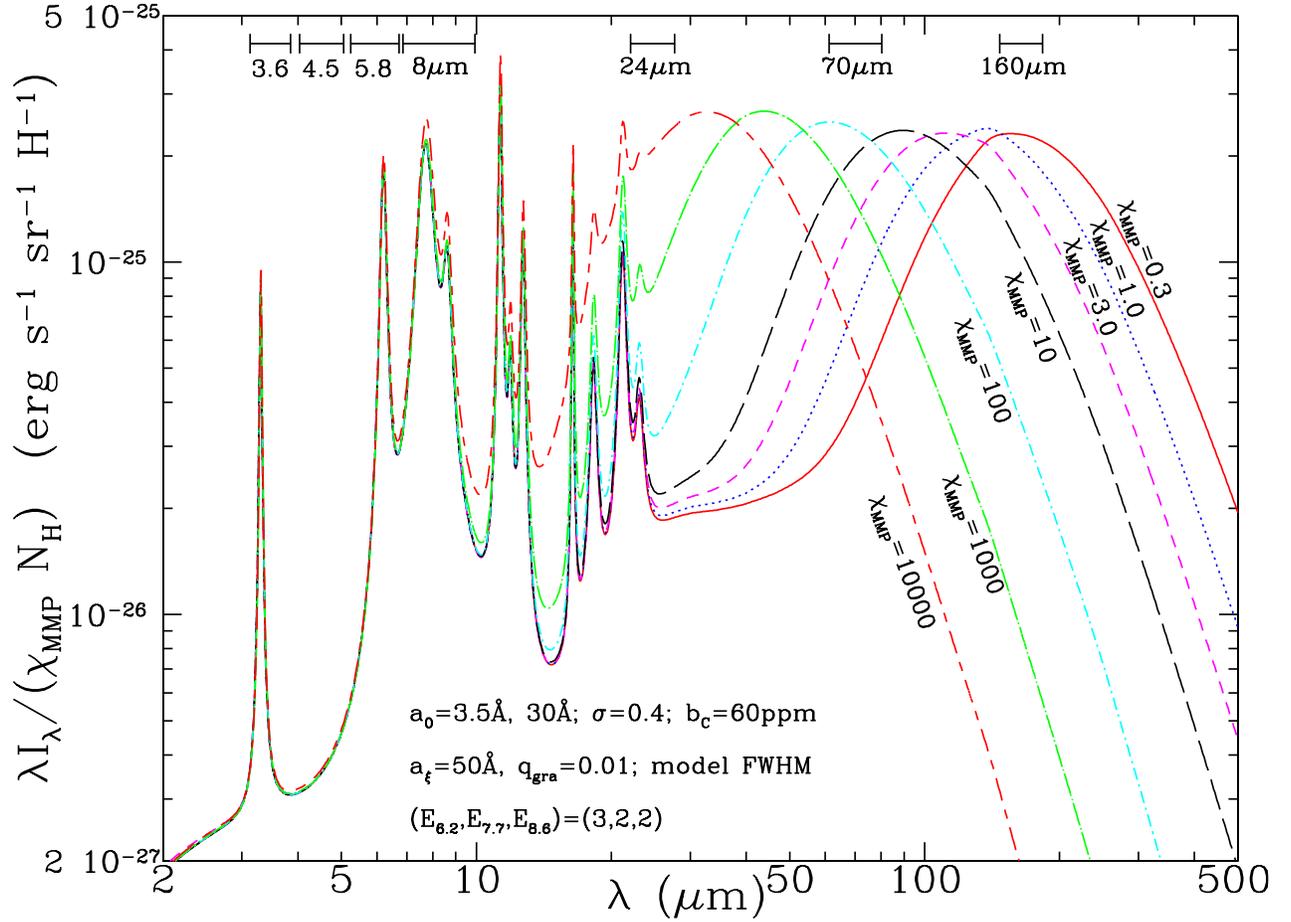}
\caption{
	\footnotesize
        \label{fig:HGL_varychi}
        IR spectra (scaled by the starlight intensity)
        for our standard grain model for different illuminating 
        radiation fields. The overall shape for $\lambda \simlt 15\mu$m 
        is almost unchanged for regions with starlight intensities 
        varying by about five orders of magnitude. Note we have 
        assumed the same ionization fraction ($\fion$) as in the 
        local diffuse ISM.
	We also show the wavelength coverage of the 4 IRAC bands and
	3 MIPS bands on SIRTF.
        }
\end{figure}

Finally, in Figure \ref{fig:HGL_varychi} we show the emission 
from dust heated by radiation fields with $\chiMMP=0.3, 1, 3, 
10, 100, 10^3, 10^4$ (we assume all have the same $\fion$ 
as the local ISM).
We see that the fraction 
of the infrared emission emerging at $\lambda < 30\micron$ is 
relatively independent of $\chiMMP$ (over this range), 
since this part of the 
spectrum is dominated by single-photon heating. This naturally 
explains why the observed PAH emission features show little 
spectral change in various regions with UV intensities differing
by five orders of magnitude (e.g., see Boulanger 1999). However, 
as expected, the emission at $\lambda \gtsim 50\micron$ shows 
a shift from longer to shorter wavelengths as $\chiMMP$ is 
increased from 0.3 to $10^4$.

The {\it Space Infrared Telescope Facility} (SIRTF) will be
capable of sensitive imaging using the {\it Infrared Array Camera} 
(IRAC) at 3.6, 4.5, 5.8, and 8.0$\micron$, and using the 
{\it Multiband Imaging Photometer} (MIPS) at 24, 70, and 160$\micron$.
In Table \ref{tab:sirtf} we show the band-averaged intensities
for our standard dust model heated by starlight radiation fields 
of various intensities.
\begin{table}[h]
\caption[]{$I_\nu/N_{\rm H}$ ($\MJy \sr^{-1}/10^{20}\cm^{-2}$)
	Averaged over SIRTF Bands for Standard Dust Model 
        Illuminated by Starlight Intensity $\chiMMP$\label{tab:sirtf}.}
\begin{tabular}{cccccccc}
\hline
 & IRAC & IRAC & IRAC & IRAC & MIPS & MIPS &MIPS \\
$\chiMMP$	& 3.6$\micron$ 
		& 4.5$\micron$ 
		& 5.8$\micron$ 
		& 8.0$\micron$ 
		& 24$\micron$
		& 70$\micron$
		& 160$\micron$ \\
\hline
0.3	&3.49$\times 10^{-4}$
	&1.63$\times 10^{-4}$ 
	&2.02$\times 10^{-3}$
	&7.30$\times 10^{-3}$ 
        &8.88$\times 10^{-3}$
	&0.0311
	&0.373\\
1	&1.16$\times 10^{-3}$
	&5.45$\times 10^{-4}$ 
	&6.75$\times 10^{-3}$
	&0.0243
	&0.0302
	&0.188
	&1.200\\
3	&3.49$\times 10^{-3}$
	&1.63$\times 10^{-3}$ 
	&0.0202
	&0.0726
	&0.0937
	&0.932
	&2.96\\
10	&0.0116
	&5.45$\times 10^{-3}$ 
	&0.0675
	&0.243
	&0.335
	&4.51
	&6.85\\
100	&0.116
	&0.0545
	&0.677
	&2.44
	&4.35
	&52.4
	&24.2\\
10$^3$  &1.17
	&0.548
	&6.82
	&25.1
	&81.5
	&331
	&64.3\\
10$^4$  &11.8
	&5.61
	&71.7 
	&290
	&1690
	&1330
	&143\\
\hline
\end{tabular}
\end{table}

\subsection{PAH Feature Strengths and the H/C Ratio\label{sec:h2c}}

\begin{figure}			
\epsfig{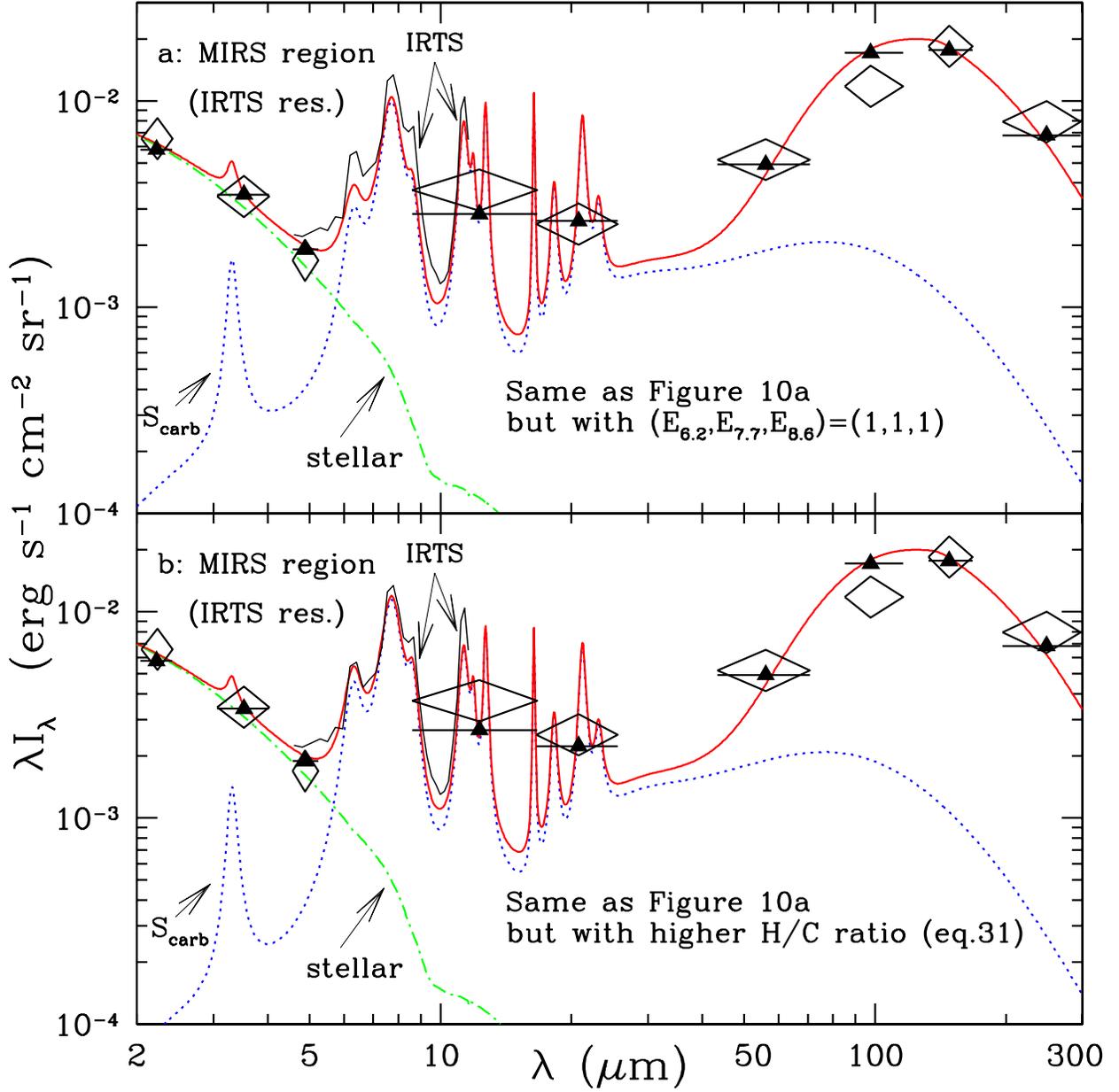}
\caption{
	\footnotesize
        \label{fig:irts_e1_h2c}
        (a): Same as Figure \ref{fig:irts_chi2}a
	     but with $(E_{6.2},E_{7.7},E_{8.6})=(1,1,1)$.
	(b): Same as Figure \ref{fig:irts_chi2}a
	but with a higher H/C
        ratio (eq.\ [\ref{eq:high_nh2nc}]).
        }
\end{figure}

Our standard model assumes enhancements $E_{6.2,7.7,8.6}=(3,2,2)$
of the PAH band strengths.
To demonstrate the need for such enhancement,
in Figure \ref{fig:irts_e1_h2c}a we plot the best-fitting model 
using the {\it laboratory-measured} PAH feature strengths
[i.e. $E_{6.2,7.7,8.6}=(1,1,1)$]. Although the fit to the 
11.3$\mu$m peak becomes a little better than our ``standard'' 
model (Figure \ref{fig:irts_chi2}a), the fit to the 6.2, 
7.7, 8.6$\mu$m region is not as good as the standard model. 
We also note that the fit does not necessarily get improved 
by increasing the PAH feature strengths by a {\it larger}
factor as expected from energy conservation consideration.

Previous sections all assume a H/C ratio representing compact, 
symmetric PAHs (eq.\ [\ref{eq:nh2nc}], \S\ref{sec:optc_pah}). 
For PAHs with open, uneven structures (e.g., elongated PAHs), 
the H/C ratio is expected to be higher. With this in mind, 
we consider a dust model with a relatively higher H/C ratio
\begin{equation}\label{eq:high_nh2nc}
{\rm H/C} = 
\left\{\begin{array}{lr} 
0.5, & \numC \le 60,\\
0.5/\sqrt{\numC/60}, & 60 \le \numC \le 240,\\
0.25, & \numC \ge 240.\\
\end{array}\right.
\end{equation}
The corresponding spectrum is shown in Figure \ref{fig:irts_e1_h2c}b.
The C-H features (3.3, 8.6, 11.3$\mu$m) become slightly stronger,
and the C-C features (6.2, 7.7$\mu$m) become slightly weaker (as
expected from energy conservation), but the entire spectrum is 
nearly indistinguishable from the model with a lower H/C ratio 
(eq.\ [\ref{eq:nh2nc}]; Figure \ref{fig:irts_chi2}a). 

The dust model described in this work assumes essentially
full hydrogen coverage for PAHs. Dehydrogenation becomes 
significant only for the smallest PAHs. Upon absorption of an 
energetic photon, large PAHs will not have CH bond rupture 
since the absorbed energy will promptly be redistributed 
among many vibrational modes. Tielens et al.\ (1987) concluded 
that small PAHs ($\numC \simlt 25$) would be partially 
dehydrogenated in regions with intense UV fields, while large 
PAHs ($\numC \simgt 25$) would be completely hydrogenated in 
those regions. Allain, Leach \& Sedlmayr (1996) reexamined this 
and reached similar conclusions -- as shown in their Figure 10, 
PAHs with $\numC \simgt 30$ in diffuse clouds are almost 100\% 
hydrogenated. Therefore, it is reasonable to assume full hydrogen 
coverage. Figure \ref{fig:irts_e1_h2c}b shows that the infrared 
emission spectrum is insensitive to modest changes in the H/C 
ratio (except for possible effects on the wavelengths and strengths
of the C-H, C-C bands upon dehydrogenation [e.g., see Pauzat, Talbi, 
\& Ellinger 1997]). 

In regions very rich in H atoms, a PAH molecule may acquire
excess H atoms attached to the peripheral C atoms and 
become ``superhydrogenated''. The addition of extra H atoms
to PAHs results in a decrease of the level of ``aromaticity''
and an increase of the level of ``aliphaticity'' by
weakening the strength of the aromatic C-H stretching 
band and creating new aliphatic C-H stretching bands
(Bernstein, Sandford, \& Allamandola 1996). The 3.40, 3.46, 
3.51, and 3.56$\mu$m subfeatures superposed on the shoulder
of the 3.3$\mu$m feature seen in many objects may be explained 
by such ``superhydrogenation'' (Schutte et al.\ 1993; Bernstein 
et al.\ 1996). The very strong 8.6$\mu$m band seen in the compact
HII region IRAS 18434-0242 (Roelfsema et al.\ 1996) may be due to 
the extra H atoms bonded to internal C atoms, although the presence 
of non-compact PAH ions can also enhance the 8.6$\mu$m band 
(Roelfsema et al.\ 1996). Due to the lack of sufficient experimental 
data and quantum chemical calculations, we have not included possible
``superhydrogenation'' in our model. 

\subsection {Total Infrared Emission and Optical-UV Albedos}

For $\chiMMP=1$ our model for the high galactic latitude emission
(Figure \ref{fig:HGL}) yields a total infrared intensity 
$4\pi \int \emiss d\lambda = 
{\rm 4.36\times 10^{-24}\ ergs\ s^{-1}\ H^{-1}}$
(the fractional contributions of S$_{\rm carb}$, B$_{\rm carb}$, 
and silicate grains [S$_{\rm sil}$+B$_{\rm sil}$], are approximately 
33\%, 37\%, and 30\%, respectively). As seen from Figures \ref{fig:HGL} 
and \ref{fig:far_IR}, our model appears to be in excellent agreement 
with the overall power radiated by interstellar dust.

By contrast, the model of Dwek et al.\ (1997) radiates $\sim$37\% more 
power per H nucleon ($6.0\times10^{-24}\erg\s^{-1}\ {\rm H}^{-1}$),
and the model of D\'{e}sert et al.\ (1990) radiates $\sim$59\% more power 
per H nucleon ($6.92\times10^{-24}\erg\s^{-1}\ {\rm H}^{-1}$).
These differences are due to the fact that
1) the model of Dwek et al.\ (1997) has a UV extinction which is
higher than the average Galactic extinction curve (see their 
Figure 4) and probably has lower optical-UV albedos than ours; 
2) the albedos predicted by the D\'{e}sert et al.\ (1990) model 
are significantly lower than those of the current model. 

\begin{figure}			
\epsfig{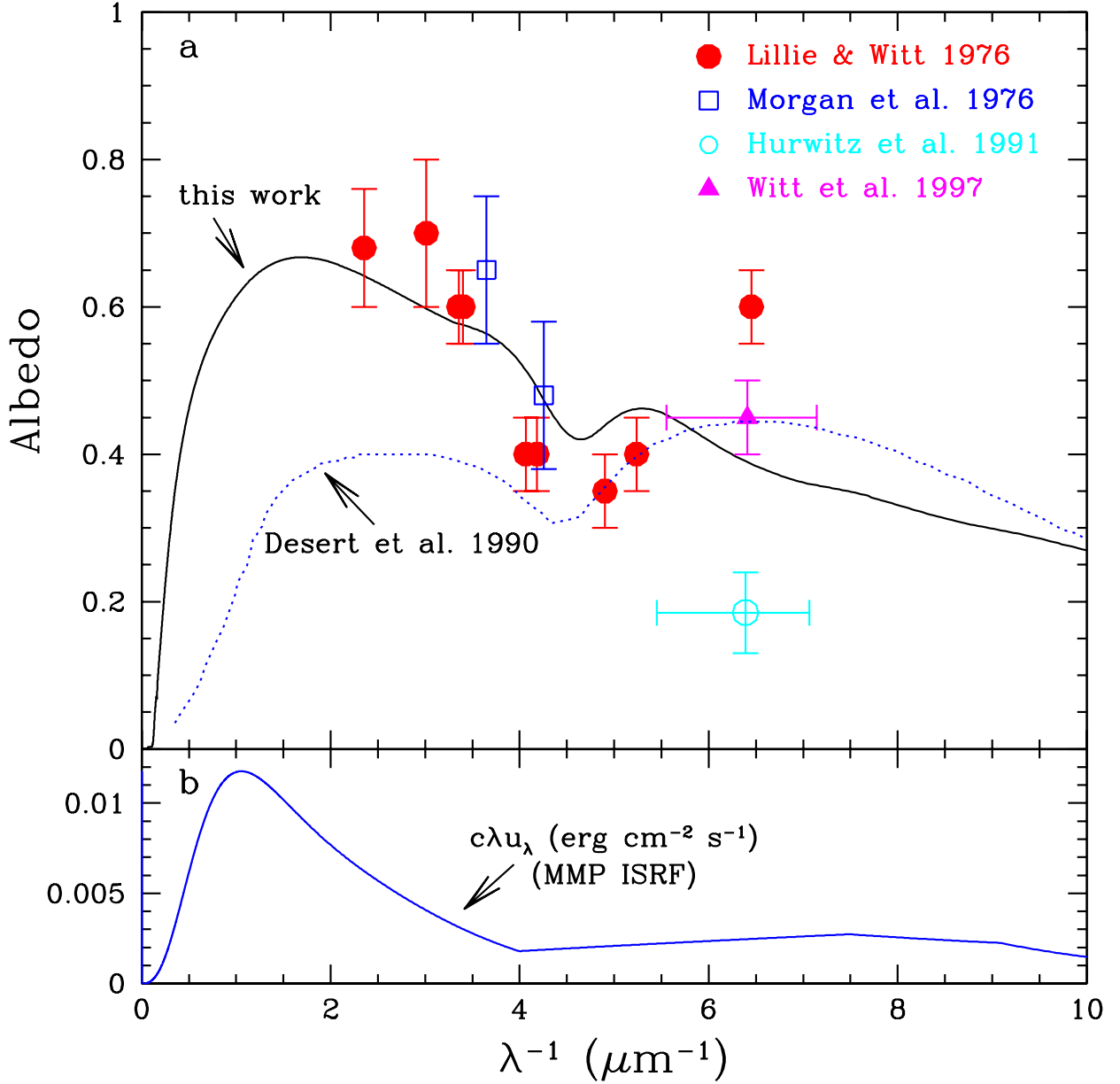}
\caption{
	\footnotesize
        \label{fig:albedo}
        (a): Albedo calculated for the present dust 
        model (solid; also see Figure 15 in Weingartner \& Draine 2001a) 
        and for the D\'{e}sert et al.\ (1990) model (dotted;
        see their Figure 6) together with observational data for
        the diffuse Galactic light
        (filled circles -- Lillie \& Witt 1976; 
        open squares -- Morgan et al.\ 1976;
        open circles -- Hurwitz et al.\ 1991;
        triangles -- Witt et al.\ 1997).
        (b): The solar neighbourhood ($\chiMMP=1$) 
        interstellar radiation field (Mathis et al.\ 1983).
        The D\'esert et al.\ grain model is much more strongly absorbing
	at $\lambda^{-1} < 4\micron^{-1}$ where the ISRF energy 
	is concentrated.
        }
\end{figure}

In Figure \ref{fig:albedo} we show the albedos calculated for the 
present model (also see Figure 15 in Weingartner \& Draine [2001a])
together with the dust albedos estimated from observations of the 
diffuse Galactic light (Lillie \& Witt 1976; Morgan, Nandy, \& 
Thompson 1976; Hurwitz, Bowyer, \& Martin 1991; Witt, Friedmann, 
\& Sasseen 1997). The current model appears to be in good agreement 
with the observationally-determined albedos. For comparison, 
Figure \ref{fig:albedo} also shows the albedo for the
grain model of D\'{e}sert et al.\ (1990), which evidently
has an absorptivity about 20--50\% higher than both the  
current model and observations at $\lambda^{-1} < 4\micron^{-1}$
where the bulk of the starlight energy is located.

\begin{figure}			
\epsfig{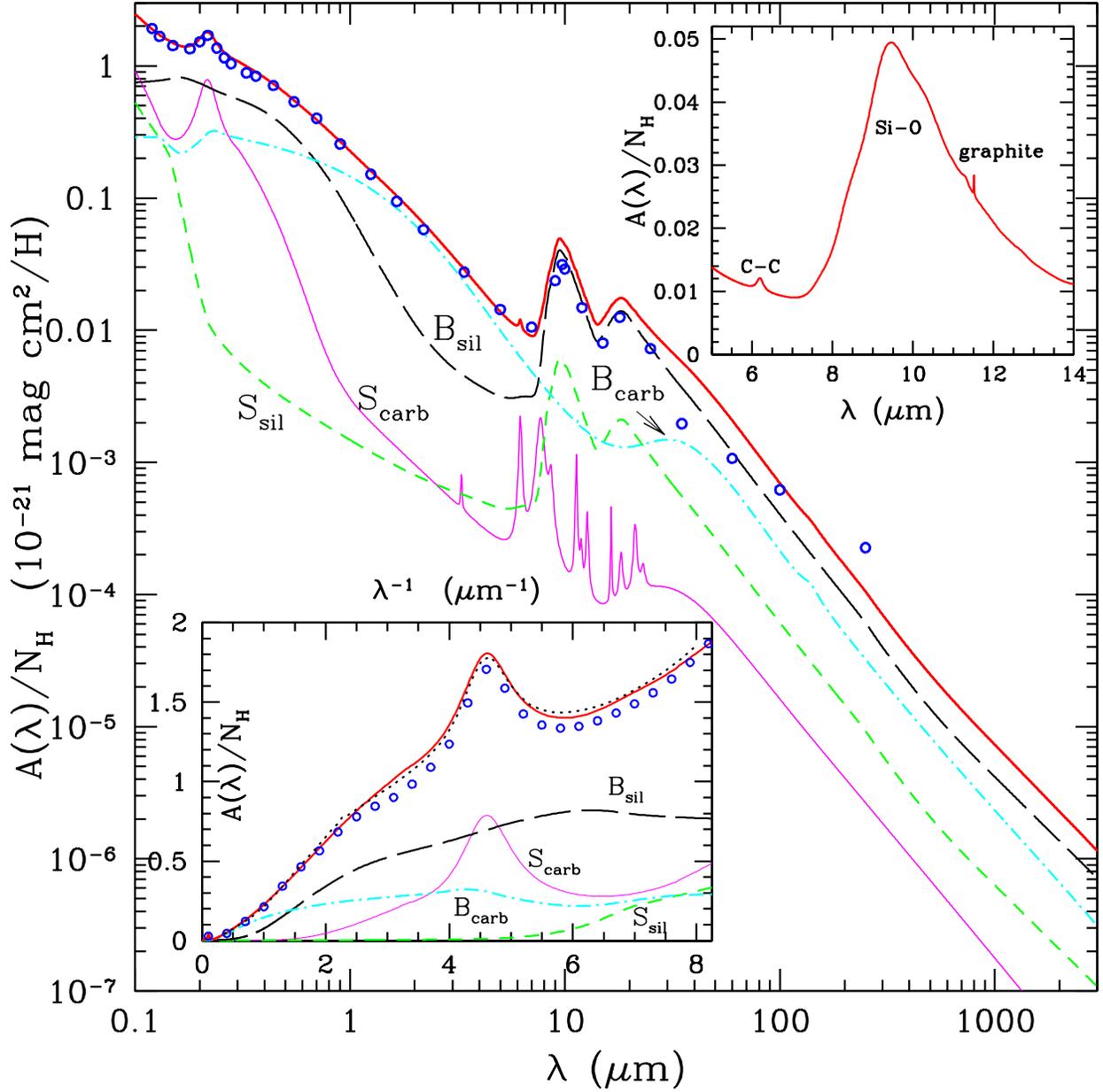}
\caption{
       \label{fig:extcurve}
        The calculated extinction for the 
        present grain model (heavy solid line) and the observed 
        average ($R_V\approx 3.1$) interstellar extinction 
        curve (dotted line: Fitzpatrick 1999; open circles: Mathis 1990). 
        Model results are the sum of 4 components:
      ``${\rm B_{sil}}$'' (silicate, $a\ge 250$\AA; long-dashed line);
      ``${\rm S_{sil}}$'' (silicate, $3.5\le a \le 250$\AA; 
        short-dashed line); 
      ``${\rm B_{carb}}$'' (carbonaceous, $a\ge 250$\AA; dot-dashed line); 
        and ``${\rm S_{carb}}$'' (carbonaceous, $3.5\le a \le 250$\AA, 
        including PAHs; thin solid line). 
        The upper-right inset illustrates the aromatic 
        6.2$\mu$m absorption feature and the 9.7$\mu$m silicate 
        band. The narrow feature at 11.5$\mu$m is due to a lattice 
        mode of crystalline graphite (Draine 1984) and would
        likely be smoothed out in an imperfect polycrystalline 
        sample. The lower-left inset plots the extinction curve 
        against $\lambda^{-1}$ ($\mu {\rm m}^{-1}$).
	}         
\end{figure}

\section{Discussion\label{sec:discussion}}

Figure \ref{fig:extcurve} compares the extinction curve obtained 
by our dust model with $\bc=60$ppm ($\bcone=45$ppm, $\bctwo=15$ppm)
with the average interstellar extinction curve and also 
shows the contributions of the individual
dust components.
It can be seen that PAHs, the dominant 
contributor to the 2175\AA\ hump, have a relatively flat behavior 
for $5 < \lambda^{-1} < 8\micron^{-1}$ 
and thus do not dominate the far-UV extinction. 
This is consistent with Greenberg \& Chlewicki (1983) in which the
2175\AA\ hump was shown to correlate poorly with the far-UV extinction. 

Our dust model implies a total dust mass per H nucleon 
$m_{\rm dust}\approx 1.89\times 10^{-26}\ {\rm g\ H^{-1}}$, or a gas-to-dust
ratio $1.4m_{\rm H}/m_{\rm dust}=124.$
Let $\Sigma_{\rm ext}(\lambda)\equiv 
N_{\rm H}^{-1}\int (dn/da) C_{\rm ext}(a,\lambda)da$ 
be the total extinction cross section per H nucleon. 
The mass absorption coefficient for the dust is just
$\kappa_{\rm abs}(\lambda) = (1-\albedo)\Sigma_{\rm ext}(\lambda)/m_{\rm dust}
= 0.4\ {\rm ln}10\ (1-\albedo) [A(\lambda)/N_{\rm H}]/m_{\rm dust} =
4.88\times 10^{25} (1-\albedo) [A(\lambda)/N_{\rm H}]\ {\rm cm^2\ g^{-1}}$. 
In Table \ref{tab:kext} we tabulate $\kappa_{\rm abs}$ at 
selected wavelengths.
Note that these numbers are for dust in the average diffuse ISM,
with $R_V \equiv A_V/E(B-V)\approx 3.1$. 
In diffuse regions with $R_V > 3.1$ the extinction
per H is reduced at $\lambda < 0.9\micron$ due to changes in the
grain size distribution (see WD01), but the extinction does not
appear to be affected at $\lambda \gtsim 0.9\micron$
(Cardelli, Clayton, \& Mathis 1989).
In dark clouds ice mantle formation adds additional absorption features,
most prominently the 3.1$\micron$ O-H stretch in H$_2$O, but for
$\lambda \gtsim 10\micron$ 
the opacities in Table \ref{tab:kext} probably remain a good approximation
even for molecular regions with densities up to $\sim 10^6 \cm^{-3}$.

\begin{deluxetable}{rccccc}
\tabletypesize{\footnotesize}
\tablewidth{0pt}
\tablecaption{Extinction and absorption properties for
		dust in the diffuse ISM.\tablenotemark{a}\newline
	NOTE: values of $\Sigma_{\rm ext}$ and $\kappa_{\rm abs}$ 
	in ApJ Table 6 are too small by a factor of 1.086 .
		\label{tab:kext}}
\tablehead{
	\colhead{band}&
	\colhead{$\lambda$}&
	\colhead{albedo}&
        \colhead{$g\equiv\langle\cos\theta\rangle$ \tablenotemark{b}}&
	\colhead{$\Sigma_{\rm ext}$ \tablenotemark{c}}&
	\colhead{$\kappa_{\rm abs}$ \tablenotemark{d}} \\
	\colhead{}&
	\colhead{($\mu$m)}&
	\colhead{}&
	\colhead{}&
	\colhead{($\cm^{2}\ {\rm H}^{-1}$)}&
	\colhead{($\cm^{2}\g^{-1}$)}
	}
\startdata
  & 0.0100	& 0.47	& 1.00 &$7.11\times 10^{-22}$ & $2.00\times 10^{4}$ \\
  & 0.0488	& 0.30	& 0.86 &$1.52\times 10^{-21}$ & $5.66\times 10^{4}$ \\
Ly edge  &0.0912& 0.24	& 0.73 &$2.51\times 10^{-21}$ & $1.01\times 10^{5}$ \\
  & 0.100	& 0.27	& 0.72 &$2.28\times 10^{-21}$ & $8.83\times 10^{4}$ \\
Ly$\alpha$&0.1215&0.32	& 0.73 &$1.73\times 10^{-21}$  & $6.20\times 10^{4}$ \\
  & 0.150	& 0.38	& 0.70 &$1.38\times 10^{-21}$  & $4.56\times 10^{4}$ \\
  & 0.220	& 0.42	& 0.56 &$1.65\times 10^{-21}$  & $5.05\times 10^{4}$ \\
  & 0.300	& 0.58	& 0.57 &$1.00\times 10^{-21}$ & $2.25\times 10^{4}$ \\

U & 0.365	& 0.62	& 0.58 &$8.28\times 10^{-22}$ & $1.68\times 10^{4}$ \\
B & 0.440	& 0.65	& 0.57 &$6.73\times 10^{-22}$ & $1.26\times 10^{4}$ \\
V & 0.550	& 0.67	& 0.54 & $5.06\times 10^{-22}$ & 8960 \\
R & 0.700	& 0.66	& 0.48 &$3.59\times 10^{-22}$  & 6460 \\
I & 0.900	& 0.63	& 0.40 &$2.47\times 10^{-22}$  & 4820 \\
J & 1.22	& 0.58  & 0.29 &$1.53\times 10^{-22}$  & 3440\\
H & 1.63	& 0.51  & 0.21 &$9.76\times 10^{-23}$  & 2510\\
K & 2.20	& 0.43  & 0.13 &$5.92\times 10^{-23}$  & 1780\\
L & 3.45	& 0.28  & 0.0051  &$2.57\times 10^{-23}$ & 984\\
  & 3.60	& 0.26  & -0.0037 &$2.37\times 10^{-23}$ & 926\\
  & 4.50	& 0.18  & -0.037 &$1.54\times 10^{-23}$	& 670\\ 
M & 4.80 	& 0.16  & -0.043 &$1.37\times 10^{-23}$	& 611\\ 
  & 5.80	& 0.099 & -0.052 &$1.02\times 10^{-23}$ & 488\\ 
  & 8.00	& 0.016 & -0.050 &$1.54\times 10^{-23}$	& 803\\ 
  & 9.70	& 0.003 & -0.036 &$4.39\times 10^{-23}$	& 2320\\
N & 10.6	& 0.003 & -0.032 &$3.36\times 10^{-23}$	& 1770\\
  & 12.0	& 0.003 & -0.029 &$1.93\times 10^{-23}$	& 1020\\ 
Q & 21.0	& $5\times 10^{-4}$ & -0.024 &$1.30\times 10^{-23}$ & 693\\ 
  & 25.0	& $3\times 10^{-4}$ & -0.024 &$9.29\times 10^{-24}$ & 493\\ 
  & 60.0	& $1\times 10^{-4}$ &$< 0.01$ &$1.91\times 10^{-24}$ & 101\\
  & 70.0	& $1\times 10^{-4}$ &$< 0.01$ &$1.38\times 10^{-24}$ & 73.1\\
  & 100		& $< 10^{-4}$ & $< 0.01$ &$6.40\times 10^{-25}$ & 34.0 \\
  & 140		& $< 10^{-4}$ & $< 0.01$ &$3.31\times 10^{-25}$ & 17.6 \\
  & 160		& $< 10^{-4}$ & $< 0.01$ &$2.44\times 10^{-25}$ & 13.0 \\
  & 240		& $< 10^{-4}$ & $< 0.01$ &$1.06\times 10^{-25}$ & 5.61 \\
  & 450		& $< 10^{-4}$ & $< 0.01$ &$2.80\times 10^{-26}$ & 1.49 \\
  & 850		& $< 10^{-4}$ & $< 0.01$ &$8.83\times 10^{-27}$ & 0.468 \\
  & 1350	& $< 10^{-4}$ & $< 0.01$ &$4.05\times 10^{-27}$ & 0.215 \\
  & 2000	& $< 10^{-4}$ & $< 0.01$ &$2.10\times 10^{-27}$ & 0.111 \\
  & 3000	& $< 10^{-4}$ & $< 0.01$ &$1.05\times 10^{-27}$ & 0.0559 \\
\hline
\enddata
\vspace*{-0.3cm}
\tablenotetext{a}{An expanded version of this table is available at
	{\tt http://www.astro.princeton.edu/$\sim$draine/dust/dustmix.html}}
\tablenotetext{b}{Average of $\cos\theta$ for scattered light, where
	$\theta$ is the scattering angle.}
\tablenotetext{c}{$\Sigma_{\rm ext}$ is the dust extinction cross section
	per H nucleon.}
\tablenotetext{d}{$\kappa_{\rm abs}$ is the absorption opacity of the dust 
	alone;
	the opacity of the gas+dust mixture is
	$\kappa_{\rm abs}/125$.
	To within $\pm$10\%, 
       $\kappa_{\rm abs}\approx 3.17\times 10^5
       \left(\lambda/\mu {\rm m}\right)^{-2} ({\rm cm^2\ g^{-1}})$  
	for $20 \!<\! \lambda \!<\! 700\micron$, and
       $\kappa_{\rm abs} \approx 3.89\times 10^4\ 
       \left(\lambda/\mu {\rm m}\right)^{-1.68} {\rm cm^2\ g^{-1}}$
	for $700 \!<\! \lambda \!<\! 10^4\micron$.}
\end{deluxetable}


Our dust model requires a total of 254ppm of C and 48 ppm of Si
in grains, both of which are significantly higher than the 
maximum available for the dust according to the recent work
of Snow \& Witt (1995), who suggest that the elemental abundances
in the ISM are just 60\%--70\% of the solar
value\footnote{Although the present model uses more silicon 
than other existing models, the predicted silicate optical depth
$\Delta \tau_{\rm 9.7\mu m}/N_{\rm H} 
\approx 3.3\times 10^{-23}\ {\rm cm^2/H}$ is not inconsistent  
with observations ($\Delta \tau_{\rm 9.7\mu m}/N_{\rm H} 
\approx 2.8-5.9\times 10^{-23}\ {\rm cm^2/H}$; see Draine 1989b,
Mathis 1998).}. 
However, Weingartner \& Draine (2001a) have argued that it is difficult to
see how the observed interstellar extinction can be explained if
interstellar abundances are appreciably subsolar;
they argue that the case for subsolar abundances is not compelling.
Dwek (1997) has argued that the composite dust model (Mathis 1996) --
which aims at making economical use of dust materials -- is still short 
of carbon when PAHs are included to account for the mid-IR emission. 
 
Our model has an absorption feature at 6.2$\mu$m due to 
the PAH C-C stretching mode (see Figure \ref{fig:extcurve}). 
The integrated optical depth of the 
6.2$\mu$m feature relative to the silicate optical depth is 
$(\int \Delta\tau d\lambda^{-1})_{6.2}
/\Delta\tau_{9.7} \approx 2.9\cm^{-1}$.
The aromatic 6.2$\mu$m absorption feature has
been detected in several objects including both local 
sources and Galactic center sources (Schutte et al.\ 1998;
Chiar et al.\ 2000). On average, the observed integrated 
strength of the 6.2$\mu$m band per unit silicate band depth is 
about 2.2cm$^{-1}$
with no significant difference between the Galactic sources and 
the local sources (Schutte et al.\ 1998). 
Our model is in good agreement with this determination.

In addition to the 6.2$\mu$m feature, our model also predicts 
a weak narrow absorption feature at 3.3$\mu$m with 
$\int \Delta\tau_{3.3} d\lambda^{-1}/\Delta\tau_{9.7} 
\approx 0.50\cm^{-1}$.
This feature has been detected in the Galactic center source GCS 3 
with $(\int \Delta\tau d\lambda^{-1})_{3.3}/\Delta\tau_{9.7} 
\approx 0.34\cm^{-1}$ (Chiar et al.\ 2000).
It has also been seen in absorption in some heavily extincted 
molecular cloud sight lines (Sellgren et al.\ 1995; Brooke, Sellgren, 
\& Geballe 1999). The 7.7, 8.6, and 11.3$\mu$m features 
are hidden by the much stronger 9.7$\mu$m 
silicate feature, and therefore will be difficult to observe as
absorption features.

The aromatic absorption bands allow us to place constraints on 
the PAH abundance if the PAH band strengths are known.
On the other hand, one can also infer information on the PAH band 
strengths from the aromatic absorption bands given knowledge 
of the PAH abundance derived from other considerations such as 
the 2175\AA\ hump and/or the 3--12$\mu$m PAH emission bands.  
For example, for $\bc$=60ppm in PAHs (the maximum allowed by the
2175\AA\ hump), 
and assuming $a_{\xi}\approx50\Angstrom$,
enhancement of the 6.2$\micron$ band by
a factor $E_{6.2} > 3$ would result in a 6.2$\mu$m absorption 
feature which would be stronger than observed;
conversely, if we had not enhanced the 6.2$\mu$m feature strength 
by a factor $E_{6.2}$=3, then the 6.2$\mu$m absorption feature 
strength in the model would be weaker than observed. However,
this depends on the poorly-constrained value of $a_{\xi}$,
which determines the radius where carbonaceous grains make 
the transition from being PAH-like to becoming graphite-like.
It is clear that $(\int \Delta\tau d\lambda^{-1})_{6.2}$ would 
increase (decrease) as $a_{\xi}$ is increased (decreased) 
(e.g., the $a_{\xi}=100$\AA\ model results in 
$(\int \Delta\tau d\lambda^{-1})_{6.2}
/\Delta\tau_{9.7} \approx 4.2\cm^{-1}$).
Therefore, until we have better understanding
of the PAH-graphite transition, we should not over-state the 
constraints of the aromatic absorption bands on the PAH band strengths.
However, the fact that our model is consistent with the 3.3, 6.2$\mu$m
absorption bands, the 2175\AA\ hump, and the 3--12$\mu$m emission 
indicates that the choice of $a_{\xi}$=50\AA\ is reasonable. 

We must also stress that the strengths of these PAH absorption 
bands could vary with physical conditions due to changes in the 
PAH ion fraction $\fion(a)$. In regions with increased 
$\fion$, the 6.2$\mu$m absorption feature would be strengthened
and the 3.3$\mu$m feature would be weakened 
(see Table \ref{tab:drude_parameters}).
The fact that there is no significant variation between the
Galactic center sources and the local sources for the 6.2$\mu$m
absorption strength relative to the silicate optical depth
(Schutte et al.\ 1998) suggests that these regions probably have 
similar radiation field-to-electron density ratios. This is also 
indicated by the $P(7.7\mu {\rm m})/P(11.3\mu {\rm m})$ data of 
Chan et al.\ (2001) (see their Figure 2c). 

Significant variations in the relative abundances of PAHs to large 
grains are expected in different environments due to depletion 
(e.g., accretion of PAHs into grain mantles, agglomeration of 
PAHs and other very small grains to form composite grains),
production (e.g. fragmentation of graphitic grains, hydrogenated 
amorphous carbon [Scott, Duley, \& Pinho 1997] or organic refractory
materials [Greenberg et al.\ 2000]),
or drift of large grains relative to small grains due to
anisotropic starlight (Weingartner \& Draine 2001c).
Such local variations in PAH abundance are indicated by the large 
cloud-to-cloud variations in mid-IR emission (Verter et al.\ 2000)
as well as the lower UV extinction in dense regions.

Ultrasmall silicate grains are excluded as a major component 
since the IRTS and ISO spectroscopic observations of the
diffuse ISM show no evidence of the Si-O 9.7$\mu$m
emission feature which would be expected for transiently-heated
tiny silicate dust (Mattila et al.\ 1996; Onaka et al.\ 1996).
However, the presence of these particles cannot be ruled out 
since a weak 9.7$\mu$m emission feature may be hidden by the 
strong PAH features at 8.6$\micron$ and 11.3$\micron$. 
On the basis of the observed IR emission spectrum for 
the diffuse ISM, the observed ultraviolet extinction curve,
and the 10$\mu$m silicate absorption profile, Li \& Draine (2001) 
have obtained upper limits on the abundances of ultrasmall 
($a\ltsim15\Angstrom$) amorphous and crystalline 
silicate grains. It is found that, contrary to previous work
($\simlt 1\%$ of solar silicon abundance [D\'esert et al.\ 1986]), 
as much as $\sim$10\% of interstellar Si could be in 
$a\ltsim 15\Angstrom$ silicate grains without violating 
observational constraints. Not more than $\sim$5\% of the Si 
can be in crystalline silicates (of any size).

\section{Summary
	\label{sec:summary}}

We have modeled the infrared emission spectra from 
dust in the diffuse ISM of the Milky Way,
including the prominent emission features at 3.3, 6.2, 7.7, 
8.6 and 11.3$\mu$m and the continuum 
emission from near-IR to submillimeter.
We have compared our model to 
the observed emission from high galactic latitude regions as well as
from regions in the Galactic plane. 

The dust model adopted here consists
of a mixture of amorphous silicate grains and 
carbonaceous grains, both having a wide size
distribution ranging from
large grains $\sim$1$\micron$ in diameter.
down to molecules containing 
tens of atoms.
We assume that the carbonaceous grains
have graphitic properties at radii 
$a \gtrsim 50$\AA, and polycyclic 
aromatic hydrocarbon (PAH)-like properties at the very small 
sizes (see \S\ref{sec:optc_gra}) which account for the 
3.3, 6.2, 7.7, 8.6 and 11.3$\mu$m emission features 
seen in a wide range of objects. 

Based on recent progress made in both laboratory and 
theoretical studies of PAH absorption spectra, and guided 
by the astronomical observations, we have constructed 
``astronomical'' absorption cross sections for both neutral 
and ionized PAHs from the far ultraviolet to the infrared 
(see \S\ref{sec:optc_pah}, Table \ref{tab:drude_parameters},
and Appendix \ref{sec:appendix}) including dependence on
grain size (which affects the onset of the absorption edge 
at long wavelength [see \S\ref{app:visible_to_uv}]), 
H/C ratio, and charge state. Detailed modelling indicates 
that the feature strengths of the 7.7, 8.6$\mu$m bands need 
to be increased by a factor $\sim$2, and a factor $\sim$3 for 
the 6.2$\mu$m band (see \S\ref{sec:optc_pah} and 
compare Figures \ref{fig:irts_chi2}a and \ref{fig:irts_e1_h2c}a). 
In addition, a small amount of continuum absorption 
needs to be added to the PAH absorption 
spectrum to account for the near-IR continuum emission at 
1--5$\mu$m underlying the 3.3$\mu$m C-H stretching feature 
detected in various sources (see \S\ref{sec:pahircont}).
We represent this continuum opacity by adding about 1\% of the
opacity of graphite.
We have also slightly modified the far-infrared and submillimeter emissivity
of ``astronomical silicate'' in order to explicitly demonstrate that the
observed emission from the interstellar medium can be reproduced
using reasonable properties for carbonaceous and silicate grains.

We model the heat capacities of carbonaceous and silicate grains
using vibrational models developed by DL01.
The thermal-discrete 
method developed in DL01 is used to
calculate the energy (temperature) distribution functions 
for transiently heated very small grains (see \S\ref{sec:T_spike}). 
We find that for grains heated by starlight with $\chiMMP=1$,
the energy distribution function is quite broad even for grains
as large as $a=200\Angstrom$, resulting in stronger
emission at short wavelengths than would be computed using the
``equilibrium temperature'' $\bar{T}$.
The PAH charging computation is performed using the technique 
developed by Weingartner \& Draine (2001b) (see \S\ref{sec:charging}).

We adopt the size distributions for large grains 
(see \S\ref{sec:sizedb}) derived from modeling the 
interstellar extinction curve (Weingartner \& Draine 2001a).
We find that if the very small carbonaceous grain population includes
two log-normal size distributions characterized by 
$a_{01}=3.5$\AA, $a_{02}=30$\AA, and widths 
$\sigma=0.4$ for the PAH size distributions (with 
about 60ppm of C [relative to H] contained in 
PAHs; 45ppm in the former, and 15ppm in the latter), then the overall
grain model is able
to closely reproduce the observed emission of the diffuse 
ISM from near-IR to submillimeter 
(see \S\ref{sec:results}) while at the same time reproducing
the observed extinction curve.
Our dust model 
predicts a pronounced absorption feature at 6.2$\mu$m 
and a much weaker feature at 3.3$\mu$m due to PAHs with 
optical depths consistent with observations 
(see \S\ref{sec:discussion}). 

We will show elsewhere (Draine \& Li 2000b) that
the very same ultrasmall 
grain component (i.e. PAHs) invoked here to account for the 
3--60$\mu$m {\it vibrational} emission 
will produce {\it ``rotational''} emission which
can quite naturally 
explain the observed 10--100GHz ``anomalous'' microwave emission from
the interstellar medium,
confirming the original estimates by Draine \& Lazarian (1998a,b).

While real interstellar grains are no doubt more complex than the
simple model presented here, this extension of the original 
graphite-silicate grain model (Mathis, Rumpl, \& Nordsieck 1977; 
Draine \& Lee 1984) to explicitly include a PAH population leads 
to a well-defined grain model which is in excellent agreement with 
available observational constraints, ranging from the observed 
extinction curve to the observed infrared emission.

\acknowledgments
We thank L.J. Allamandola, R. Arendt, J.M. Greenberg, and J.C. Weingartner
for helpful discussions. 
We are grateful to the anonymous referee for helpful comments.
We thank D.M. Hudgins and L.J. Allamandola
for providing us with access to their PAH database. 
We thank F. Boulanger, D. Cesarsky, E. Sturm, K.I. Uchida, 
and L. Verstraete for providing us with observational data. 
We thank D.P. Finkbeiner and D.J. Schlegel for help in obtaining 
the DIRBE fluxes for regions in the Galactic plane. 
We thank R.H. Lupton for the availability of the SM plotting package.
This research was supported in part by 
NASA grant NAG5-7030 and NSF grants AST-9619429 and AST-9988126.

\appendix

\section{Optical Properties of Polycyclic Aromatic Hydrocarbons
	\label{sec:appendix}
	}

\subsection{The UV and Far-UV Range\label{app:UV_to_FUV}}

Donn (1968) originally proposed that PAHs 
could contribute significantly to the interstellar UV 
extinction. Although a single PAH species has strong UV absorption 
features (UV Atlas 1966), a cosmic mixture of many individual
molecules, radicals, and ions, with many overlapping absorption
bands, may effectively produce an absorption continuum, except
perhaps in the visible, where some narrow features may stand out
and account for some of the diffuse interstellar bands, and in 
the 2000--2400\AA\ region, where a concentration of strong
absorption features may blend together to produce the 2175\AA\
extinction feature.

Efforts over the past two decades to measure the absorption 
spectrum of PAHs have been primarily for individual ionized/neutral 
PAH molecules in hopes of identifying carriers of the diffuse
interstellar bands (see Salama et al.\ 1996 and references therein).
To the best of our knowledge, the only high quality 
absorption spectra published for natural mixtures of PAHs from the visible
to far-UV are those of L\'{e}ger et al.\ (1989b) 
and Joblin et al.\ (1992).
For ${\rm 3.3\,\mu m^{-1} < \lambda^{-1} < 10\,\mu m^{-1}}$,
we adopt their experimental spectra as a starting point. 
All these spectra have a strong UV feature around 2175\AA\ (but
relatively broader than the interstellar feature). We use the method
of Fitzpatrick \& Massa (1990) to fit the {\it averaged} spectra 
of L\'{e}ger et al.\ (1989b)\footnote{%
	We adopted the PAH mixtures 
	obtained from a coal pitch extract evaporated at 380K (averaged 
	over those of 381K and 385K), 410K, 440K 
	(averaged over 436K and 440K), 
	and 476K. In these mixtures over 300 molecules were identified with 
	molecular mass ranging from ${\rm \sim 150 -350\,amu}$. 
	The measurements were performed only for 
	${\rm \lambda^{-1} < 7.7\,\mu m^{-1}}$, beyond which 
	extrapolations were made.} 
and of Joblin et al.\ (1992)\footnote{%
	These mixtures were from 
	a coal pitch extract evaporated at 570K, 600K and 630K, with
	molecular masses ranging from 170 to 550 amu.}
by a Drude profile plus a linear continuum and a nonlinear 
far-ultraviolet ($\lambda^{-1} > 5.9\,\mu$m$^{-1}$) rise. 
To be consistent with the observed extinction, we then 
adjust the peak position of the near-UV hump (due to the 
$\pi$-$\pi^{*}$ transition) to be at ${\rm \lambda_{\pi-\pi^{*}}^{-1}
= 4.6\,\mu m^{-1}}$ and the band width FWHM to be 
${\rm \gamma \lambda_{\pi-\pi^{*}}^{-1} = 1.0\,\mu m^{-1}}$, 
with the integrated cross section conserved.\footnote{%
	The carrier of the 2175\AA\ extinction hump remains
	unidentified more than 35 years after its first detection 
	(Stecher 1965). It is commonly attributed to some sort of 
	graphitic carbonaceous material (see Draine 1989a). 
	A correlation between 
	the 2175\AA\ hump and the IRAS 12$\mu$m emission  
	(dominated by PAHs) was found by Boulanger, Pr\'{e}vot, \& Gry (1994) 
	in the Chamaeleon cloud, suggesting a common carrier. 
	Arguments against PAHs as the 2175\AA\ hump carrier can be 
	found in Mathis (1998) where it was argued that the hump peak 
	is sensitive to the PAH shape (but also see Duley \& Seahra 1998)
	and the hump strength is senstive to the states of ionization 
	(e.g. Lee \& Wdowiak [1993] found that the 2175\AA\ hump is weakened 
	in cations). In the dust model presented here the both the
	2175\AA\ extinction hum and the 12$\micron$ emision are
	mainly contributed by PAHs, but we note that PAHs are not the major 
	contributor to the far-UV ($\lambda < 1800$\AA) extinction. 
	This is consistent with the observation
	that the far-UV extinction does not correlate with
	either the 2175\AA\ 
	hump (Greenberg \& Chlewicki 1983)
	or the IRAS 12$\mu$m emission (Boulanger et al.\ 1994).
	}
We thus obtain eq.\ (\ref{eq:cabs_pah_6}), 
for $3.3 < \lambda^{-1} < 5.9\micron^{-1}$.

For $\lambda^{-1} > 10\,\mu$m$^{-1}$, no experimental data for 
PAH mixtures are available. We adopt that of coronene 
C$_{24}$H$_{12}$, characterized by a strong increase
with $\lambda^{-1}$ out to a strong peak at $\approx 700$\AA\ 
(due to the $\sigma$-$\sigma^{*}$ transition). 
Other simpler aromatic molecules such as naphthalene (C$_{10}$H$_{8}$)
(Koch, Otto, \& Radler 1972), benzene (C$_{6}$H$_{6}$)
and its derivatives (Robin 1975) display similar spectral behaviour 
in the range of 10--30eV.
For ${\rm 10\,\mu m^{-1} < \lambda^{-1} < 15\,\mu m^{-1}}$,
the laboratory measurements of coronene (L\'{e}ger et al.\ 1989b) 
are fitted by a Drude profile plus a linear continuum to obtain
eq.\ (\ref{eq:cabs_pah_3}).

For $\lambda^{-1} > 17.25\,\mu$m$^{-1}$, 
we adopt the absorption cross sections per C atom calculated from 
the refractive indices of graphite (DL84);
see eq.\ (\ref{eq:cabs_pah_1}).
A smooth transition from the absorption spectrum of coronene 
to that of graphite is made by extrapolating the coronene data to 
$\lambda^{-1} \approx 17.25\,\mu$m$^{-1}$,
resulting in eq.\ (\ref{eq:cabs_pah_2}).

To smoothly join our eq.\ (\ref{eq:cabs_pah_3}) 
for $\lambda^{-1}>10\micron^{-1}$ to
eq.\ (\ref{eq:cabs_pah_5}) for $\lambda^{-1} < 7.7\micron^{-1}$,
for $7.7 < \lambda^{-1} < 10\micron^{-1}$
we adopt the polynomial eq.(\ref{eq:cabs_pah_4}),
which provides a good fit to the experimental data for PAH mixtures.

\subsection{The Visible and Near-UV Range \label{app:visible_to_uv}}

The visible and near UV absorption properties of PAHs have been extensively
studied (Clar 1964; Salama et al.\ 1996 and references therein). However,
there are no high quality absorption spectra for PAH mixtures available
in the literature. Therefore we adopt that of circumanthracene 
(${\rm C_{40}H_{16}}$), the largest symmetric PAH with the visible and 
near UV spectrum measured to date. The laboratory absorption spectrum of 
circumanthracene (Clar 1964) can be 
approximated by
\begin{equation}
\label{eq:c40h16_abs}
\left[C_{\rm abs}(\lambda)/\numC\right]_{\rm C_{40}H_{16}} = 2.54\times 
10^{-17 - 3.431\left(\lambda/{\rm \mu m}\right)}\ {\rm cm^2/C}.
\end{equation} 
We adopt the cutoff function of D\'{e}sert et al.\ (1990),
\begin{equation}
\label{eq:pah_cutoff_func}
{\rm cutoff}(\lambda, \lambda_c) = \frac{1}{\pi} 
\arctan\left[\frac{10^3 (y-1)^3} {y}\right] + \frac{1}{2}, ~~~ 
y=\lambda_c/\lambda,
\end{equation}
where the cutoff wavelength ${\rm \lambda_c}$ -- 
the visual absorption edge -- is derived from the 
extensive discussion of Salama et al.\ (1996):
\begin{equation}
\label{eq:pah_cutoff_wave}
{\rm
\left[\frac{\lambda_c}{\mu m}\right] = \left\{\begin{array}{lr}
   1/\left(3.804M^{-0.5} + 1.052\right), & {\rm for\ neutral\ PAHs};\\
   1/\left(2.282M^{-0.5} + 0.889\right), & {\rm for\ PAH\ cations};\\
\end{array}\right.}
\end{equation}
where $M$, the number of fused benzenoid rings, is approximately 
$M\approx 0.4 \numC$ for $\numC >40$, 
$M\approx 0.3 \numC$ for $\numC <40$ 
($\numC$ is the number of carbon atoms). Upon ionization, 
the visual absorption edge $\lambda_{\rm c}$ shifts to longer wavelength;
$\lambda_{\rm c}$ also shifts to longer wavelength as the PAH
size increases. 

To smoothly join the visual-near UV spectrum with that of 
${\rm \lambda^{-1} > 3.3\,\mu m^{-1}}$ (\S\ref{app:UV_to_FUV}), 
we multiply the 
absorption cross section (\ref{eq:c40h16_abs})
of circumanthracene by a factor 1.36,
to obtain eq.\ (\ref{eq:cabs_pah_7}) for 
$1\ltsim \lambda^{-1}\ltsim 3.3\micron^{-1}$.

\renewcommand{\tabcolsep}{0.05cm}
\begin{deluxetable}{lccccccclll}
\tabletypesize{\scriptsize}
\tablewidth{0pt}
\tablecaption{Observed PAH band widths (FWHM, $\mu$m) in selected
sources\label{tab:pah_fwhm}.}
\tablehead{
\colhead{Objects}&
\colhead{3.3$\mu$m}&
\colhead{6.2$\mu$m}&
\colhead{7.7$\mu$m}&
\colhead{8.6$\mu$m}&
\colhead{11.3$\mu$m}&
\colhead{11.9$\mu$m}&
\colhead{12.7$\mu$m}&
\colhead{Reported res.}&
\colhead{Method\tablenotemark{a}}&
\colhead{Spectrum sources}
}

\startdata
~~~ diffuse ISM & 0.20\tablenotemark{b} & 0.67 & 0.91 & 0.55 & 0.63 & - & - 
            & $\Delta \lambda=0.3\mu$m & Drude& Onaka et al.\ 1996\\
~~~ vdB 133 (little UV flux) & - & 0.30 & 0.78 & 0.50 & 0.60 & - & 0.85 
          &$\lambda/\Delta \lambda=40$ &Drude &Uchida et al.\ 1998\\
~~~ M17 (PDR) & - & 0.30 & 0.68& 0.42& 0.60 & - & 0.80 
          &$\lambda/\Delta \lambda=40$ &Drude &Cesarsky et al.\ 1996a\\
~~~ NGC 7023 (reflection nebula) & 0.04\tablenotemark{c} & 0.28 & 0.60 
          &0.42 &0.43 & - & 0.80 &$\lambda/\Delta \lambda=40$ &Drude 
          &Cesarsky et al.\ 1996b\\
~~~ NGC 7023 (reflection nebula) & 0.04\tablenotemark{c} & 0.17 & 0.72 
          &0.28 &0.22 & - & 0.33 &$\lambda/\Delta \lambda=500$ &Eye
          &Moutou et al.\ 1999b\\
~~~ NGC 1333 (reflection nebula)\tablenotemark{d} & 0.04\tablenotemark{e} 
          & 0.35 & 0.70 &0.43 & 0.40\tablenotemark{f} & - & - 
          & $\lambda/\Delta \lambda=40$ &Lorentzian &Uchida et al.\ 2000\\
~~~ $\rho$ Oph (molecular cloud) & - & 0.27 & 0.70 & 0.42 & 0.43 & - & 1.18 
          & $\lambda/\Delta \lambda=40$ &Lorentzian 
          &Boulanger et al.\ 1996\\
~~~ SMC & - & 0.20 & 0.53 & 0.36 & 0.54 & - & -
          & $\lambda/\Delta \lambda=40$ & Lorentzian &Reach et al.\ 2000\\
~~~ Diffuse ISM & - & 0.20 & 0.71 & 0.42 & 0.27 & - & - 
            & $\Delta \lambda=0.1\mu$m &Cauchy &Mattila et al.\ 1996\\
~~~ NGC 891 (external galaxy) & - & 0.22 & 0.62 & 0.30 & 0.30 & - & - 
           & $\Delta \lambda=0.1\mu$m &Cauchy &Mattila et al.\ 1999\\
~~~ 28 normal galaxies (averaged) & - & 0.36 & 0.77 & 0.45 & 0.42 & - 
           & - & $\Delta \lambda=0.1\mu$m &Drude &Helou et al.\ 2000\\
~~~ Orion ionization front & 0.04\tablenotemark{g} & 0.24\tablenotemark{h}
          & 0.80\tablenotemark{h}& 0.28\tablenotemark{h}& 0.14 & -
          & 0.27 & $\Delta \lambda=0.09\mu$m &Gaussian &Roche et al.\ 1989\\ 
~~~ NGC 2023 (reflection nebula) & 0.04\tablenotemark{e}& 0.20 & 0.74 &0.38 
          &0.36 & - & - &$\lambda/\Delta \lambda=500$ &Drude
          &Verstraete et al.\ 2001\\
~~~ Red Rectangle (post-AGB) & 0.035\tablenotemark{g}& 0.16 & 0.65 
          & 0.36 & 0.21 & 0.30 & 0.42 & $\lambda/\Delta \lambda\approx 
1000$ 
          &Eye &Tielens et al.\ 1999\\ 
~~~ 30 Dor (LMC)\tablenotemark{i}& 0.036 & 0.30 & 0.60 & 0.40 & 0.20 & - & -
          & $\lambda/\Delta \lambda=1500$ &Drude &Sturm et al.\ 2000\\
~~~ M82 (starburst galaxy) & 0.04 & 0.20 & 0.71 & 0.42 & 0.19 & - & - 
          & $\lambda/\Delta \lambda=1500$ &Drude &Sturm et al.\ 2000\\
~~~ NGC 253 (starburst galaxy) & 0.042 & 0.18 & 0.78 &0.45 & 0.19 & - & - 
          & $\lambda/\Delta \lambda=1500$ &Drude &Sturm et al.\ 2000\\
~~~ Circinus(Seyfert 2 galaxy) & 0.045 & 0.16 & 0.82 &0.43 & 0.20 & - & - 
          & $\lambda/\Delta \lambda=1500$ &Drude &Sturm et al.\ 2000\\
\hline
~~~ {\bf Present Model}& 0.04\tablenotemark{j} & 0.20 & 0.70 & 0.40 
          & 0.20\tablenotemark{k} & 0.30 &0.30\tablenotemark{k}& - 
          &- &this work\\
\hline
\enddata
\tablenotetext{a}{FWHM values are determined by 
                       fitting the PAH features either 
                       with a set of Drude profiles, 
                       or Cauchy, Gaussian, Lorentzian profiles,
                       or simply estimated by reading off 
                       the source spectra (hereafter ``Eye'');
                       those marked with ``Cauchy'', ``Gaussian'',
                       or ``Lorentzian'' are taken from the source
                       references; those marked with ``Drude'' or 
                       ``Eye'' are obtained in this work. 
			}
\tablenotetext{b}{Tanaka et al.\ 1996 (resolution $\Delta \lambda=0.13\mu$m).}
\tablenotetext{c}{ Moutou et al.\ 1999a 
                 (resolution $\lambda/\Delta \lambda=500$).}
\tablenotetext{d}{Uchida et al.\ (2000) have shown that the FWHM of 
               the 6.2, 7.7, 8.6$\mu$m bands remains constant over 
               different nebular positions with $G_0 \simgt 200$; 
               however, the 7.7$\mu$m band appears to be broader 
               at low $G_0$.}
\tablenotetext{e}{Joblin et al.\ 1996a 
		(resolution $\Delta \lambda=0.008\mu$m).}
\tablenotetext{f}{Joblin et al.\ 1996b 
                   (resolution $\lambda/\Delta \lambda=60$).}
\tablenotetext{g}{Geballe et al.\ 1989 
                   (resolution $\Delta \lambda=0.009\mu$m).}
\tablenotetext{h}{Bregman et al.\ 1989 
                  (resolution $\Delta \lambda=0.12\mu$m).}
\tablenotetext{i}{Except the 3.3 and 11.3$\mu$m bands, the signal-to-noise 
		ratio is very low for the
                30 Dor spectrum
                 (Sturm et al.\ 2000).}
\tablenotetext{j}{In a high resolution ($\lambda/\Delta \lambda\approx 1400$) 
             survey of eight sources, Tokunaga et al.\ (1991) found
             that the 3.3$\mu$m feature in extended objects such as
             planetary nebulae and HII regions has a band width of
             $\approx 0.042\mu$m; this was further confirmed by 
             Roche et al.\ (1996) who found 18 planetary nebulae all 
             have a band width of $\approx 0.04\mu$m for the 
             3.3$\mu$m feature ($\lambda/\Delta \lambda=500$).}
\tablenotetext{k}{On average, the 10--15$\mu$m spectra of 
                  16 HII regions, YSOs, and evolved stars 
                  (with resolution $\lambda/\Delta \lambda=500-1500$)
                  show a band width of $\approx 0.20, 0.29\mu$m for 
                  the 11.3, 12.7$\mu$m band, respectively 
                  (see Hony et al.\ 2001).}
\end{deluxetable}

\subsection{The Near-IR, Mid-IR Range \label{app:NIR_to_MIR}}

During the last few years, a great deal of progress has been 
made by experimentalists and theorists in the study of the 
infrared properties of both neutral and ionized PAHs 
(Allamandola et al.\ 1999a; Hudgins \& Allamandola 
1999a, 1999b; Langhoff 1996; and references therein). 
The infrared spectra of PAHs are characterized by the 
C-C and C-H stretching/bending vibrational features. 
Observationally, the feature positions and widths are 
approximately constant in a wide variety of objects.
Table \ref{tab:pah_fwhm} summarizes the band widths of the 
major PAH features observed in some well-observed astronomical sources.
We represent the PAH infrared 
absorption (actually emission) cross sections by a set of 
Drude profiles with peak wavelengths 
$\lambda_j$ and FWHM $\gamma_j\lambda_j$ taken from the 
observations\footnote{
We rely on the highest resolution observations to indicate the
FWHM of the narrower PAH features.}
(see Table \ref{tab:pah_fwhm}), respectively. 
Although the laboratory spectra of various PAH 
species are not exactly at the same wavelengths 
as the astronomical observations,
we note that large symmetric compact 
PAHs are in close agreement with interstellar 
observations (see Langhoff 1996). 

Although weaker than the
11.3$\mu$m resonance, other C-H out-of-plane 
(oop) bending modes\footnote{The C-H oop bending modes are 
characteristic of the edge structure of PAHs. They occur at 
different wavelengths, depending on the number of neighbouring 
hydrogen atoms on one aromatic ring: for neutral PAHs, 
isolated C-H oop band (solo-CH) occurs at $\sim$ 11.3$\mu$m, 
doubly adjacent CH (duet-CH) at $\sim$ 11.9$\mu$m, 
triply adjacent CH (trio-CH) at $\sim$ 12.7$\mu$m, 
quadruply adjacent CH at $\sim$ 13.6$\mu$m. 
Upon ionization, the solo and duet CH modes shift to 
shorter wavelength (see Hudgins \& Allamandola 1999b).},
have also been detected in various regions (Witteborn et al.\ 1989; 
Hony et al.\ 2001). From the point of view of stability, 
compact, symmetric structures are favored for interstellar PAHs 
(van der Zwet \& Allamandola 1985; Omont 1986; Allamandola et al.\ 1989;
Jochims et al.\ 1994). Thus one expects a much lower
quartet-CH (13.6$\mu$m) abundance than solo-, duet-, and trio-CH for 
interstellar PAHs, as has been recently confirmed by ISO
observations (Hony et al.\ 2001).
For large symmetric compact PAHs, it is reasonable to 
assume equal single, double, and triple adjacent CH units 
(Stein \& Brown 1991). 
We therefore
include Drude profiles with central wavelengths 
$\lambda_j = 11.9, 12.7\mu$m and FWHM $\gamma_j\lambda_j = 0.30, 0.30\mu$m 
in our PAH model spectrum for the duet-CH, trio-CH out-of-plane 
bending modes, respectively.

Based on recent laboratory and theoretical studies of 
neutral and ionized PAHs (deFrees et al.\ 1993; 
Szczepanski \& Vala 1993; Hudgins, Allamandola, \& Sandford 1994; 
Hudgins \& Allamandola 1995a, 1995b, 1997; Langhoff 1996; 
Pauzat, Talbi, \& Ellinger 1997; Hudgins \& Sandford 1998a, 1998b, 1998c) 
we estimate the integrated cross section of the 
C-C, C-H bands for neutral PAHs and PAH cations (normalized by the 
number of C atoms for C-C modes and by the number 
of H atoms for C-H modes). The results are presented in 
Table \ref{tab:drude_parameters}. 
We have considered only two charge states -- neutral PAHs and 
PAH cations -- since no experimental measurements have been made for
multiply-ionized PAHs. It has been shown that
the IR properties of PAH anions closely resemble those of PAH 
cations (e.g., see Szczepanski, Wehlburg, \& Vala 1995; Langhoff 1996; 
Hudgins et al.\ 2000) except for the very strong 3.3$\mu$m C-H stretch
enhancement in the anion\footnote{%
	The 3.3$\mu$m C-H stretching
	mode is about 3 times more intense in anions 
	than the corresponding mode 
	in neutral PAHs; however, no cation features have been observed
	in this mode (Hudgins et al.\ 2000).}.

\subsection{The Far-IR Range\label{app:FIR}} 

The far-IR properties of PAHs are poorly known.
Moutou et al.\ (1996) have carried out measurements of 40 PAH 
species in the range of 14--75$\mu$m. They found four prominent
features at 16.2, 18.3, 21.2 and 23.1$\mu$m (with band widths 
FWHM = 10, 20, 18, 20\ cm$^{-1}$, respectively) and a weak continuum 
beyond 40$\mu$m\footnote{%
	A 16.4$\mu$m emission feature was detected 
	in the ISO-SWS spectra of NGC 7023, M17 and the Orion Bar 
	by Moutou et al.\ (2000). Therefore we adopt 16.4$\mu$m rather than
	16.2$\mu$m.}.     
To include these far-IR features, we adopt 4 Drude profiles for the 
16.4, 18.3, 21.2 and 23.1$\mu$m bands and one Drude profile for the 
far-IR continuum (${\rm \lambda > 24 \mu m}$). Band widths 
are taken from Moutou et al.\ (1996) except for the 
16.4$\mu$m feature, for which we adopt 
$\sim 0.16\mu$m (Moutou et al.\ 2000). 
The integrated band strengths for the 16.4, 18.3, 21.2, and 23.1~$\mu$m bands
were estimated from the 40 species in Table 3 of
Moutou et al.\ (1996) by 
(1) summing the band strengths per C atom of features within $\pm20\cm^{-1}$
of the nominal band centers,
(2) dividing by 40, and
(3) multiplying the resulting band strength per C atoms by a factor of 2 
(a) since many of the species studied by Moutou et al.\ 
(e.g., hexabenzocoronene C$_{42}$H$_{18}$) are highly symmetric and
show only weak absorption in the 14--40$\micron$ region and
(b) in order to reproduce the observed strengthof the 16.4$\micron$ feature
relative to shorter wavelength features (e.g., 11.3$\micron$) in the spectrum
of NGC 7023 (Moutou et al.\ 1999a).

The parameters for the FIR continuum Drude profile
(${\rm \sigma_{int,14}}$, $\lambda_{14}$, $\gamma_{14}$; 
see Table \ref{tab:drude_parameters}) were chosen 
by setting
$\sigma_{{\rm int},14} \approx 
      \int_{14\micron}^{\infty}N_{\rm C}^{-1}
	C_{\rm abs}^{\rm PAH}(\lambda)d\lambda^{-1}
	-\sum_{j=10}^{13}\sigma_{{\rm int},j}$ 
       averaged over the laboratory spectra of 40 PAH species 
       (Moutou et al.\ 1996),
and choosing $\lambda_{14}$, $\gamma_{14}$ to
(1) minimize effects on the near-IR ($\lambda < 14\,\mu {\rm m}$) 
              spectrum; 
(2) avoid a noticeable hump in the far infrared 
              ($\lambda > 24 \mu {\rm m}$) spectrum;
(3) minimize absorption at $\lambda > \lambda_{\rm max}$,
              where $\lambda_{\rm max}$
              corresponds to the lowest vibrational mode 
              (see DL01);              
(4) have the far infrared continuum 
               dropping approximately as ${\rm \propto \lambda^{-2}}$
for ${\rm \lambda > 24 \mu m}$.

\end{document}